\documentclass[aps,prd,onecolumn,10pt]{revtex4-1}

\usepackage{graphicx}
\usepackage{epstopdf}
\usepackage{amsmath}

\newcommand{\be}{\begin{equation}}
\newcommand{\ee}{\end{equation}}
\newcommand{\bea}{\begin{eqnarray}}
\newcommand{\eea}{\end{eqnarray}}
\newcommand{\nn}{\nonumber}
\newcommand{\bdm}{\begin{displaymath}}
\newcommand{\edm}{\end{displaymath}}

\begin{document}

\title{Binary dynamics from spin1-spin2 coupling at fourth post-Newtonian order}
\author{Michele Levi}
\affiliation{Racah Institute of Physics, Hebrew University, Jerusalem 91904, Israel}
\date{\today}

\begin{abstract} 
We calculate via the effective field theory (EFT) approach the next-to-next-to-leading order (NNLO) spin1-spin2 conservative potential for a binary. Hereby, we first demonstrate the ability of the EFT approach to go at NNLO in post-Newtonian (PN) corrections from spin effects. The NNLO spin1-spin2 interaction is evaluated at fourth PN order for a binary of maximally rotating compact objects. This sector includes contributions from diagrams, which are not pure spin1-spin2 diagrams, as they contribute through the leading-order spin accelerations and precessions, that should be first taken into account here. The fact that the spin is derivative-coupled adds significantly to the complexity of computations. In particular, for the irreducible two-loop diagrams, which are the most complicated to evaluate in this sector, irreducible two-loop tensor integrals up to order 4 are required. The EFT calculation is carried out in terms of the nonrelativistic gravitational (NRG) fields. However, not all of the benefits of the NRG fields apply to spin interactions, as all possible diagram topologies are realized at each order of $G$ included. Still, the NRG fields remain advantageous, and thus there was no use of automated computations in this work. Our final result can be reduced, and a corresponding Hamiltonian may be derived. 
\end{abstract}

\maketitle

\section{Introduction}\label{sec:intro}

Though predictions of General Relativity (GR) have been confirmed in all observations and experiments to date, one of its most essential predictions, namely, that of gravitational waves (GW), has not yet been directly observed. Worldwide efforts are undertaken in order to observe such signals either with ground-based \cite{ligo, virgo, geo600, lcgt}, or future space-based \cite{lisa} detectors. Inspiralling compact binaries, which are promising candidate sources for such signals, can be described analytically by the post-Newtonian (PN) approximation of GR \cite{Blanchet:2002av}. It turns out that even relatively high-order PN corrections for the binary inspiral, as high as the fourth order PN (4PN) correction, have a phenomenological impact on the theoretical waveform templates, required for successful detection (see \cite{Foffa:2011ub} and references therein). Moreover, astrophysical objects are expected to have significant spins, for which spin effects have a big impact on the event rates expected in GW detectors \cite{Reisswig:2009vc}. It is therefore desirable to have PN corrections involving spin effects to the same high orders as for the nonspinning case. 

A novel effective field theory (EFT) approach was suggested recently by Goldberger and Rothstein for the treatment of the binary inspiral problem \cite{Goldberger:2004jt, Goldberger:2007hy}. The EFT approach is very advantageous in applying the efficient standard tools of quantum field theory to GR, such as Feynman diagrams and dimensional regularization. Subsequently, PN corrections of conservative dynamics have been reproduced: 1PN and 2PN corrections for binaries \cite{Goldberger:2004jt, Gilmore:2008gq}, 2PN correction for the n-body problem \cite{Chu:2008xm}, first promoting the use of automated computations in the EFT approach, and recently even the 3PN correction \cite{Foffa:2011ub}. The EFT approach was further extended to include spin effects \cite{Porto:2005ac}, and PN  corrections involving spin were tackled as well: the next-to-leading (NLO) spin1-spin2 \cite{Porto:2006bt, Porto:2007tt, Levi:2008nh, Porto:2008tb} (complete results also in \cite{Steinhoff:2007mb}) and NLO spin-squared interactions \cite{Porto:2008jj} (complete results also in \cite{Hergt:2008jn, Steinhoff:2008ji, Hergt:2010pa}) at 3PN order were computed. The more complex NLO spin-orbit interaction at 2.5PN order was also computed \cite{Perrodin:2010dy, Porto:2010tr, Levi:2010zu} (first obtained in \cite{Tagoshi:2000zg, Faye:2006gx, Blanchet:2006gy, Damour:2007nc}). We note that following \cite{Steinhoff:2009ei, Steinhoff:2009hx} in more traditional methods, the NLO spin-orbit and spin(a)-spin(b) interactions for the n-body problem \cite{Hartung:2010jg} and the next-to-next-to-leading order spin-orbit interaction for a binary \cite{Hartung:2011te} were also obtained. Following the EFT treatment of the radiation sector in the nonspinning case \cite{Goldberger:2009qd}, the spin-orbit, spin1-spin2, and spin-squared components of multipole moments were computed to NLO in \cite{Porto:2010zg}, where spin-orbit radiative effects at NLO were already obtained in \cite{Blanchet:2006gy}, and have been pushed further to include tail effects in \cite{Blanchet:2011zv}. Recently, the EFT formalism has been extended to incorporate radiation reaction \cite{Galley:2009px}, and similar EFT approaches were developed to investigate the extreme mass ratio inspiral problem \cite{Galley:2008ih, Galley:2010xn, Galley:2011te}, which is also relevant for GW detection, and to treat weak ultra relativistic scattering \cite{Kol:2011pj}. 

A major improvement for the obtainment of higher-order PN corrections via the EFT approach was presented in \cite{Kol:2007bc}. There, a reduction over the time dimension of the metric \`a la Kaluza-Klein was made, and was demonstrated to improve the 1PN order EFT computation. This nonrelativistic parametrization of the metric defines a set of new nonrelativistic gravitational (NRG) fields, which were later used to reproduce, e.g.~the 2PN and 3PN order corrections via EFT as was already noted. The advantages of the NRG fields are numerous. First, there is the physical interpretation of the different field components and the clear coupling hierarchy to the mass and spin. Second, the derivation of the self-gravitational vertices is simple since there is a full explicit expression for the pure gravitational action \cite{Kol:2010si}, as well as a simple expression for its stationary part which contributes at leading orders. Further, simple propagators are obtained, there are no mixed 2-point functions, and the derivation of the mass couplings is also simple and immediate. Because of the structure of vertices and mass couplings to the worldline, not all possible diagram topologies are realized at each order, and the number of diagrams, and, in particular, the more complicated ones, is reduced. Those are pushed to higher orders, for example, with respect to the standard Lorentz covariant parametrization, where a one-loop diagram is eliminated at 1PN \cite{Kol:2007bc}, or to the Arnowitt-Deser-Misner (ADM) parametrization, where mass couplings are eliminated at 2PN \cite{Kol:2010ze}.

In this paper, we calculate the next-to-next-to-leading order (NNLO) spin1-spin2 conservative potential for a binary of compact spinning objects at the 4PN order, via the EFT approach in terms of the NRG fields. With this result, we demonstrate for the first time the ability of the EFT approach to go beyond the NLO in PN corrections involving spin. The NNLO spin1-spin2 interaction sector includes contributions from 56 diagrams, of which 47 are pure spin1-spin2 diagrams, while further 9 arise from other sectors, that contribute through their SSC dependent parts, and/or through the LO spin equations of motion (EOM), that should first be taken into account here. Of the pure spin1-spin2 diagrams, there are 41 new diagrams here, while 6 others already appeared at the NLO spin1-spin2 sector, though they include new ingredients in the worldline couplings. In particular, there are 7 two-loop diagrams contributing. We note that, unfortunately, not all of the benefits of the NRG fields apply for spin interactions, as all possible diagram topologies are realized at each order of $G$ included, which was already illustrated in the NLO spin interactions. Still, the NRG fields remain advantageous, and thus there was no use of automated computations in this work. Our final result can be reduced, and a corresponding Hamiltonian may be derived. An alternative derivation of the NNLO spin1-spin2 Hamiltonian can be found in \cite{Hartung:2011ea}. 

Throughout this paper, we use $c\equiv1$, $\eta_{\mu\nu}\equiv diag[1,-1,-1,-1]$, and the convention for the Riemann tensor is 
$R^\mu_{~\nu\alpha\beta}\equiv\partial_\alpha\Gamma^\mu_{\nu\beta}-\partial_\beta\Gamma^\mu_{\nu\alpha}+\Gamma^\mu_{\lambda\alpha}\Gamma^\lambda_{\nu\beta}-\Gamma^\mu_{\lambda\beta}\Gamma^\lambda_{\nu\alpha}$. 
Greek letters denote indices in the global coordinate frame, while lowercase Latin letters from the beginning of the alphabet denote indices in the local Lorentz frame. All indices run from 0 to 3, while spatial tensor indices from 1 to 3, are denoted with lowercase Latin letters from the middle of the alphabet. The notation $\int_{\bf{k}} \equiv \int \frac{d^d{\bf{k}}}{(2\pi)^d}$ is used for abbreviation (boldface characters denote d-dimensional vectors). The scalar triple product appears here with no brackets, i.e.~$\vec{a}\times\vec{b}\cdot\vec{c}\equiv(\vec{a}\times\vec{b})\cdot\vec{c}$ (as there is in fact no ambiguity regarding the order in which the product can be performed).

The paper is organized as follows. In Sec.~\ref{sec:eft}, we briefly review the EFT approach for the binary inspiral with spinning objects, and present the Feynman rules required for the EFT computation with the NRG fields. In Sec.~\ref{sec:calc}, we present the evaluation of the NNLO spin1-spin2 Lagrangian/Routhian, going over all contributing Feynman diagrams, and giving the value of each diagram. In Sec.~\ref{sec:res}, we present the NNLO spin1-spin2 Lagrangian/Routhian EFT result, and explain how to derive from it the NNLO spin1-spin2 Hamiltonian. In Sec.~\ref{sec:conc}, we summarize our main conclusions. Finally, in Appendix \ref{app:a}, we include the tensor Fourier, one-loop and two-loop reduced integrals required here, whereas in Appendix \ref{app:b}, we give the LO spin EOM that contribute here. 

\section{EFT approach for binary inspiral with spinning objects}\label{sec:eft} 

In this section, we present the ingredients required in order to perform the EFT calculation of the NNLO spin1-spin2 interaction in terms of NRG fields, namely, the Feynman rules and the effective action from which they are derived. Here, we review briefly and build on Secs.~II and III of \cite{Levi:2010zu} and references therein, following similar notations and conventions as those that were used there. 

First, we parametrize the metric in a nonrelativistic form according to the Kaluza-Klein ansatz
\be \label{eq:kka}
d\tau^2 = g_{\mu\nu}dx^{\mu}dx^{\nu} \equiv e^{2 \phi}(dt - A_i\, dx^i)^2 -e^{-2 \phi} \gamma_{ij}dx^i dx^j~,
\ee
defining the set of nonrelativistic gravitational (NRG) fields ${\phi, A_i,\gamma_{ij}\equiv\delta_{ij}+\sigma_{ij}}$.  
Then, in terms of the NRG fields the metric reads
\be \label{eq:gkk}
g_{\mu\nu}=
\left(\begin{array}{cc} 
e^{2\phi}      & -e^{2\phi} A_j \\
-e^{2\phi} A_i & -e^{-2\phi}\gamma_{ij}+e^{2\phi} A_i A_j
\end{array}\right)\simeq
\left(\begin{array}{cc} 
1+2\phi+2\phi^2 & -A_j-2A_j\phi-2A_j\phi^2 \\
-A_i-2A_i\phi-2A_j\phi^2   & -\delta_{ij}+2\phi\delta_{ij}-\sigma_{ij}-2\phi^2\delta_{ij}+2\phi\sigma_{ij}+A_iA_j
\end{array}\right),
\ee
where we have written the approximation for the metric in the weak-field limit up to the orders in the fields that are required for this work. 

The action describing the dynamics of the binary system is given by
\be
S=S_g+S_{pp},
\ee
where $S_g$ is the pure gravitational action, and $S_{pp}$ is the worldline point-particle action for each of the two particles in the binary. 

We consider first the purely gravitational action. It is the usual Einstein-Hilbert action plus a gauge-fixing term, which we take as the fully harmonic gauge, i.e.
\be
S_g = S_{EH} + S_{GF} = -\frac{1}{16\pi G} \int d^4x \sqrt{g} \,R + \frac{1}{32\pi G} \int d^4x\sqrt{g}
\,g_{\mu\nu}\Gamma^\mu\Gamma^\nu, 
\ee
where $\Gamma^\mu\equiv\Gamma^\mu_{\rho\sigma}g^{\rho\sigma}$. The full explicit expression for the Einstein-Hilbert action and for the fully harmonic gauge fixing in terms of NRG fields was given in \cite{Kol:2010si}, where it was obtained using Cartan's method of 2-forms. Thus, there is no need to expand for each required ingredient specifically, and the propagators and self-gravitational vertices can be obtained readily from the action. 

The NRG scalar, vector, and 2-tensor field propagators in the harmonic gauge are then given by 
\begin{align}
\label{eq:prphi} \parbox{18mm}{\includegraphics{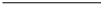}}
 & = \langle{~\phi(x_1)}~~{\phi(x_2)~}\rangle = ~~~~4\pi G~~~ \int_{\bf{k}} \frac{e^{i{\bf k}\cdot\left({\bf x}_1 - {\bf x}_2\right)}}{{\bf k}^2}~\delta(t_1-t_2),\\ 
\label{eq:prA} \parbox{18mm}{\includegraphics{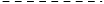}}
 & = \langle{A_i(x_1)}~{A_j(x_2)}\rangle = -16\pi G~\delta_{ij} \int_{\bf{k}} \frac{e^{i{\bf k}\cdot\left({\bf x}_1 - {\bf x}_2\right)}}{{\bf k}^2}~\delta(t_1-t_2),\\ 
\label{eq:prsigma} \parbox{18mm}{\includegraphics{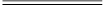}}
 & = \langle{\sigma_{ij}(x_1)}{\sigma_{kl}(x_2)}\rangle = ~~32\pi G~P_{ij;kl} \int_{\bf{k}}\frac{e^{i{\bf k}\cdot\left({\bf x}_1 - {\bf x}_2\right)}}{{\bf k}^2}~\delta(t_1-t_2),
\end{align}
where $P_{ij;kl}\equiv\frac{1}{2}\left(\delta_{ik}\delta_{jl}+\delta_{il}\delta_{jk}-2\delta_{ij}\delta_{kl}\right)$. Here and henceforth, the Feynman rules are presented in position space. This makes more sense considering the nature of the binary inspiral problem, in which the external positions of the particles are given, rather than the usual external momenta in quantum field theory. 
 
There are also time-dependent quadratic vertices, which result from the fact that the propagators are actually relativistic rather than instantaneous. The Weyl rescaling present in the NRG parametrization eliminates undesired mixed quadratic vertices, so that the 2-point functions between the three different fields are 0: $\langle \phi A_i \rangle = \langle \phi~\sigma_{jk} \rangle = \langle A_i\sigma_{jk} \rangle=0$. Thus, the Feynman rules for the propagator correction vertices are given by
\begin{align}
\label{eq:prtphi}  \parbox{18mm}{\includegraphics{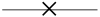}}
 & = ~~\frac{1}{8\pi G}~~\int d^4x~\left(\partial_t\phi\right)^2, \\ 
\label{eq:prtA}   \parbox{18mm}{\includegraphics{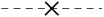}}
 & = -\frac{1}{32\pi G} \int d^4x~\left(\partial_tA_i\right)^2, \\ 
\label{eq:prtsigma} \parbox{18mm}{\includegraphics{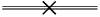}}
 & = \frac{1}{128\pi G} \int d^4x~\left[2(\partial_t\sigma_{ij})^2-(\partial_t\sigma_{ii})^2\right], 
\end{align}
where the crosses represent the self-gravitational quadratic vertices, which contain two time derivatives.

There are also contributions from three-graviton vertices of cubic gravitational self-interaction. The Feynman rules for the three-graviton vertices required for the NNLO of the spin1-spin2 interaction are given by
\begin{align}
\label{eq:phiA2} \parbox{18mm}{\includegraphics{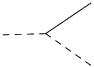}}
 & = \frac{1}{8\pi G}\int d^4x~\phi\left(\partial_iA_j\left(\partial_iA_j-\partial_jA_i\right)+\left(\partial_iA_i\right)^2\right), \\ 
\label{eq:A2sigma} \parbox{18mm}{\includegraphics{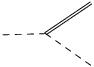}}
 & = -\frac{1}{64\pi G}\int d^4x\left[2\sigma_{ij}\left(\partial_iA_k\partial_jA_k+\partial_kA_i\partial_kA_j-2\partial_kA_i\partial_jA_k+2\partial_iA_j\partial_kA_k\right)\right.\nn \\ 
 &  \,\,\,\,\,\,\,\,\,\,\,\,\,\,\,\,\,\,\,\,\,\,\,\,\,\,\,\,\,\,\,\,\,\,\,\,\,\,\,\,\,\, \left. -\sigma_{kk}\left(\partial_iA_j\left(\partial_iA_j-\partial_jA_i\right)+\left(\partial_iA_i\right)^2\right)\right], \\ 
\label{eq:sigmaphi2} \parbox{18mm}{\includegraphics{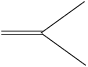}}
 & = ~ \frac{1}{16\pi G}\int  d^4x~\left(2\sigma_{ij}\partial_i\phi\partial_j\phi-\sigma_{jj}\partial_i\phi\partial_i\phi\right), \\
\label{eq:Aphi2dt} \parbox{18mm}{\includegraphics{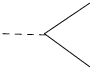}}
 & = -\frac{1}{4\pi G}\int d^4x~A_i\partial_i\phi\partial_t\phi,\\
\label{eq:phiAsigmadt} \parbox{18mm}{\includegraphics{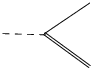}}
 & = ~ \frac{1}{8\pi G}\int d^4x~\left[2\sigma_{ij}\left(\partial_i\phi\partial_tA_j-\partial_t\phi\partial_iA_j\right)-\sigma_{jj}\left(\partial_i\phi\partial_tA_i-\partial_t\phi\partial_iA_i\right)\right],\\ 
\label{eq:A3t} \parbox{18mm}{\includegraphics{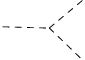}}
 & = \frac{1}{16\pi G}\int d^4x~A_i\partial_iA_j\partial_tA_j.
\end{align}
where the first three vertices are stationary, and can be read off the stationary Kaluza-Klein part of the EH action. The three last vertices are time dependent, and each contain a single time derivative.

In the NNLO spin1-spin2 interaction, contributions from four-graviton vertices of quartic gravitational self-interaction also appear. The Feynman rule for the four-graviton vertex required to the order considered here is given by
\begin{align}
\label{eq:phi^2A^2} \parbox{18mm}{\includegraphics{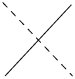}}
 & = \frac{1}{4\pi G}\int d^4x~\phi^2\left(\partial_iA_j\left(\partial_iA_j-\partial_jA_i\right)+\left(\partial_iA_i\right)^2\right),
\end{align}
where this vertex is again stationary.

Now, we consider the worldline point-particle action. This is given by
\be
S_{pp} = S_{pp(m)} + S_{pp(\bf{S})},  
\ee
where $S_{pp(m)}$ denotes the coupling of the point particles to gravity without the inclusion of spin, and $S_{pp(\bf{S})}$ the coupling of the spin degrees of freedom of the particles to gravity. Considering the gravitational coupling to the two massive compact objects, we take the worldline action of a point particle for each of the objects, so that we have
\be
S_{pp(m)} = -\sum_{i=1}^2 m_i \int d\lambda_i,
\ee
where finite-size effects are not taken into account here as their contribution enters at higher orders \cite{Goldberger:2004jt}. We parametrize the worldline using the coordinate time $t=x^0$, i.e.~$\lambda=t$, so that we have for $u^\mu\equiv dx^\mu/d\lambda$: $u^0=1$, $u^i=dx^i/dt\equiv v^i$. Thus, the Feynman rules for the one-graviton couplings to the worldline mass required for the NNLO of the spin1-spin2 interaction are given by
\begin{align}
\label{eq:mphi} \parbox{12mm}{\includegraphics{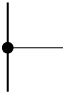}}
 & = - m \int dt~\phi~\left(1+\frac{3}{2}v^2+\cdots\right), \\ 
\label{eq:mA} \parbox{12mm}{\includegraphics{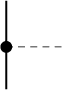}}
 & = ~m \int dt~A_iv^i~\left(1+\cdots\right), \\ 
\label{eq:msigma} \parbox{12mm}{\includegraphics{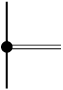}}
 & = \frac{m}{2} \int dt~\sigma_{ij}v^iv^j~\left(1+\cdots\right),
\end{align}
where the heavy solid lines represent the worldlines and the spherical black blobs represent the particles masses on the worldline. The ellipsis denotes higher orders in $v$, beyond the order considered here. 

For the two-graviton couplings to the worldline mass required at this order, we have the following Feynman rule:   
\begin{align}
\label{eq:mphi^2}  \parbox{12mm}{\includegraphics{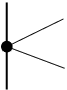}}
 & = -\frac{m}{2} \int dt~\phi^2~\left(1+\cdots\right). 
\end{align}
As expected, mass couplings do not play a major role in the spin1-spin2 interaction.

Next, we consider the gravitational coupling to the two spinning compact objects. Here, we are not concerned with finite-size effects of spin, which are quadratic in the individual spins, and hence are not relevant for the spin1-spin2 interaction (see \cite{Porto:2008jj} for spin1-spin1 effects). Thus we consider here only the part of the point particle action, which is linear in the spin of the particles given by 
\be 
S_{pp(\bf{S})} = -\frac{1}{2} \sum_{i=1}^2 \int d\lambda_i\, S^{ab}_i\omega_{\mu ab} u_i^\mu,
\ee 
where $\omega_\mu^{ab}\equiv e^{b\nu}D_\mu e^a_\nu$ are the Ricci rotation coefficients, and this form for the couplings was introduced in \cite{Porto:2008tb}, following the Routhian in \cite{Yee:1993ya}. Since the spin degrees of freedom are naturally formulated in terms of tetrads, here it is more convenient to start from the standard Lorentz covariant parametrization $g_{\mu\nu} \equiv \eta_{\mu\nu} + h_{\mu\nu}$, and the derivation of the couplings to the worldline spins is not so immediate as that of the couplings to the masses. Our background reference tetrad, expanded in terms of $h_{\mu\nu}$, is given by 
\be
e^a_\mu = \delta^a_\mu + \frac{1}{2}h^a_\mu - \frac{1}{8}h^a_\rho h^\rho_\mu + \frac{1}{16}h^a_\rho h^\rho_\kappa  h^\kappa_\mu + \cdots.
\ee
Using this tetrad, and expanding up to the third order in $h_{\mu\nu}$, which is the order required for the NNLO spin1-spin2 interaction, we obtain the following Lagrangian: 
\bea
L_{pp(\bf{S})} = &&\frac{1}{2}S^{ab}h_{a\mu,b} u^\mu + \frac{1}{4}S^{ab} h^\nu_b\left(\frac{1}{2}h_{a\nu,\mu}+h_{\nu\mu,a}-h_{a\mu,\nu}\right)u^\mu \nn\\
&& + \frac{1}{8}S^{ab} h_{a\sigma}\left(h^\sigma_\rho h^\rho_{b,\mu} + h^\nu_b h^\sigma_{\mu,\nu} + \frac{3}{2}h^\sigma_\nu \left(h^\nu_{\mu,b}-{h_{\mu b,}}^\nu\right)\right)u^\mu + \cdots.
\eea
Now we should transform from the Lorentz covariant parametrization to the NRG fields using Eq.~(\ref{eq:gkk}) in order to obtain the couplings to the worldline spin.

The Feynman rules for the one-graviton couplings to the worldline spin are thus given by
\begin{align}
\label{eq:sA}  \parbox{12mm}{\includegraphics{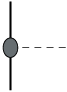}}
 & = \int dt \,\,\frac{1}{2}\left(S^{ij}\partial_iA_j - S^{0i}\partial_iA_jv^j + S^{0i}\partial_0A_i\right), \\ 
\label{eq:sphi}   \parbox{12mm}{\includegraphics{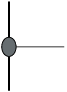}}
 & = \int dt \,\,\left(S^{ij}\partial_j\phi v^i + S^{0i}\partial_i\phi - S^{0i}\partial_0\phi v^i \right), \\  
\label{eq:ssigma}   \parbox{12mm}{\includegraphics{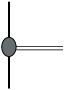}}
 & = \int dt \,\,\frac{1}{2}\left(S^{ij}\partial_i\sigma_{jk} v^k + S^{0i}\partial_0\sigma_{ij}v^j\right),
\end{align}
where the (gray) oval blobs represent the spins on the worldlines. Note that here the full expressions for the one-graviton spin couplings should be considered.  

For two-graviton couplings to the worldline spin, the Feynman rules required for the NNLO spin1-spin2 interaction are: 
\begin{align}
\label{eq:sphiA}  \parbox{12mm}{\includegraphics{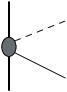}}
 & = \int dt \left[2S^{ij}\partial_iA_j\phi + \frac{1}{2}S^{ij}A_j\partial_i\phi + \frac{1}{2}S^{ij}A_j\partial_0\phi v^i \right. \nn \\
 & \left. \,\,\,\,\,\,\,\,\,\,\,\,\,\,\,\,\,\,\,\,\,\,
+ \frac{1}{2}S^{0i} \left(A_j\partial_j\phi v^i-3A_j\partial_i\phi v^j-A_i\partial_j\phi v^j +\partial_jA_i\phi v^j-2\partial_iA_j\phi v^j+A_i\partial_0\phi+3\partial_0A_i\phi \right)\right], \\ 
\label{eq:ssigmaA}  \parbox{12mm}{\includegraphics{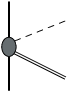}}
 & = \int dt~\frac{1}{4}S^{ij}\left(\partial_jA_k-\partial_kA_j\right)\sigma_{ik} + \cdots, \\ 
\label{eq:sA^2}   \parbox{12mm}{\includegraphics{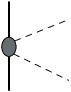}}
 & = \int dt \left[\frac{1}{2}S^{ij} \left(A_k\partial_jA_iv^k +  \frac{1}{2}A_i\partial_jA_kv^k + \frac{1}{4} A_j\partial_kA_iv^k + \frac{1}{4}A_i\partial_0A_j\right) + \frac{1}{4}S^{0i}A_j\left(\partial_iA_j - \partial_jA_i\right) + \cdots \right],\\ 
\label{eq:sphi^2}   \parbox{12mm}{\includegraphics{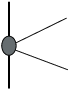}}
 & = \int dt\left(2S^{0i} \phi \partial_i\phi  + \cdots\right) , \\
\label{eq:sphisigma}   \parbox{12mm}{\includegraphics{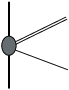}}
 & = \int dt \,\,\frac{1}{2}\left[S^{ij}\sigma_{ik} \left(\partial_j\phi v^k+\partial_k\phi v^j\right) - S^{0i}\sigma_{ij}\partial_j\phi  + \cdots \right],  
\end{align}
where the ellipsis denotes higher orders in $v$, beyond the order considered here. The first coupling here was already encountered at the NLO spin1-spin2 interaction, but as it is extended here to a higher PN order, it becomes much more complicated. Further, new couplings arise here, which are also complex.

At the NNLO spin1-spin2 interaction, we also have to include three-graviton spin couplings, that are encountered for the first time. For three-graviton couplings to the worldline spin, the Feynman rule required here is 
\begin{align}
\label{eq:sphi^2A}  \parbox{12mm}{\includegraphics{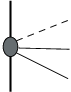}}
 & = \int dt \left(4S^{ij}\phi^2\partial_iA_j + \frac{3}{2}S^{ij}\phi\,\partial_i\phi\, A_j + \cdots \right), 
\end{align}
where naturally the gravito-magnetic vector is involved in this high-order field coupling LO contribution.

Two important features of the spin couplings can be noticed very early on and should be stressed. First is the fact that the spin, formally being a tensor, is derivative-coupled unlike the scalar mass. This fact translates into the requirement of higher-order tensor expressions for all the integrals involved in the calculations (see Appendix \ref{app:a} for more details), which adds significantly to the complexity of computations. Moreover, the derivative-coupling also allows for time derivatives in the worldline couplings, which are an additional complication in the spin computations. The time derivatives also make the corresponding terms scale at higher PN orders. The second notable feature is the fact that the spin couplings contain $S^{i0}$ entries, which represent the redundant unphysical degrees of freedom related with the spin tensor. These are taken at this stage as independent degrees of freedom, yet eventually, possibly even after the obtainment of the EOM, they are reduced from the final result using some spin supplementary condition (SSC). These will also yield contributions of higher PN orders with respect to the $S^{ij}$ spin tensor components. Both features make the PN order of the spin couplings implicit. This makes the power counting, which is essential in the EFT approach, more difficult with respect to the nonspinning case.

\section{spin1-spin2 interaction at fourth post-Newtonian order}\label{sec:calc}

In this section, we evaluate the relevant two-body effective action by its diagrammatic expansion. For the nPN order in spin interactions, we need to consider Feynman diagrams up to the $G^{\lceil n-1\rceil}$ order, where $\lceil n \rceil$ is the ceiling value of n. Thus in the NNLO spin1-spin2 potential, which is evaluated at 4PN, we have diagram contributions up to order $G^3$, coming from all 12 possible topologies appearing at these orders, as displayed in Figs.~1--3 of \cite{Gilmore:2008gq}: one topology at $O(G)$, two at $O(G^2)$, and nine topologies at $O(G^3)$. Unfortunately, not all benefits of the NRG fields are present in spin interactions, where in general all topologies are realized at each new order of $G$ included, unlike the nonspinning case, where there is a reduction in the number of topologies and diagrams. Hence, using the NRG fields, the NLO spin interactions, for example, include the one-loop diagrams, which are omitted from the 1PN potential. Similarly, the NNLO spin1-spin2 here includes all $G^3$ topologies, such as those with a single cubic vertex or a quartic vertex, unlike the NNLO nonspinning case -- the 2PN potential computed in \cite{Gilmore:2008gq}. 

For the construction of the Feynman diagrams, we use the Feynman rules from Sec.~\ref{sec:eft}, which we PN expand, see Secs.~IV and V of \cite{Levi:2010zu} for more detail. All in all, we have 56 diagrams contributing to the NNLO spin1-spin2 interaction, 47 of which are pure spin1-spin2 diagrams. Six other diagrams contribute through the LO EOM from spin interactions, as both of the objects are considered spinning: three appeared as spin-orbit diagrams at NLO, and three are orbital interaction diagrams, which appeared at 2PN. Three further diagrams from the spin-orbit sector contribute through their SSC dependent parts only, thus we do not repeat them here. Of the pure spin1-spin2 diagrams, 41 are new diagrams, while 6 others already appeared at the NLO spin1-spin2 sector, although they include new terms in the worldline couplings. 

Here we denote ${\vec{r}}\equiv{\vec{x}}_1(t)-{\vec{x}}_2(t)$, $r\equiv\left|\vec{r}\right|$, and ${\vec{n}}\equiv\frac{{\vec{r}}} {r}$. The spin is represented by a 3-vector defined by $S^{ij} \equiv \epsilon^{ijk}S^{k}$. The labels 1 and 2 are used for the left and right worldlines, respectively. The [$1\leftrightarrow2$] notation stands for a similar term, whose value is obtained under the interchange of particles labels. Note that under this exchange, $\vec{n}\to-\vec{n}$. Finally, a multiplicative factor of $\int dt$ is suppressed and omitted from all diagram values.
 
\subsection{Order $G$ Feynman diagrams}

\begin{figure}[ht]
\includegraphics{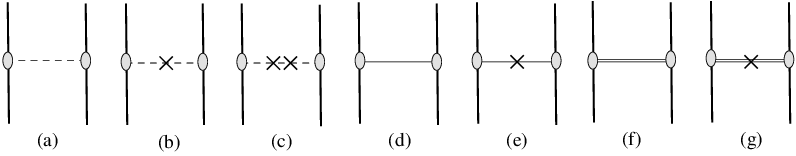}
\caption{NNLO spin1-spin2 Feynman diagrams of order $G$: One-graviton exchange. The solid, dashed, and double lines represent the $\phi$, $A_i$, and $\sigma_{ij}$ fields, respectively.}
\label{fig:4PNG}
\end{figure}

\begin{figure}[ht]
\includegraphics{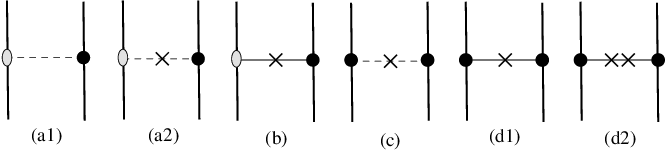}
\caption{Feynman diagrams of order $G$ that contribute to the NNLO spin1-spin2 interaction: One-graviton exchange of spin-orbit and orbital interactions, which yield acceleration and precession terms. Since both objects are considered to be spinning here, these diagrams contribute through the substitution of the LO spin-orbit and spin1-spin2 EOM, see Appendix \ref{app:b}. Diagrams (a1), (a2), and (b) should be included together with their mirror images.} 
\label{fig:4PNGeom}
\end{figure}

For the NNLO spin1-spin2 interaction, we have 13 one-graviton exchange diagrams to evaluate as shown in Figs.~\ref{fig:4PNG} and \ref{fig:4PNGeom}. Figure \ref{fig:4PNG} contains seven pure spin1-spin2 diagrams, whereas Fig.~\ref{fig:4PNGeom} contains diagrams which appeared in the NLO spin-orbit sector or the 2PN orbital interaction sector. Diagrams (a) and (b) in Fig.~\ref{fig:4PNG}, appeared already in the NLO spin1-spin2 evaluation, and they correspond to diagrams (b) and (c), respectively, of Fig.~2 in \cite{Levi:2008nh}. The NLO evaluation yielded spin-precession terms which were then omitted, but do contribute at this order at the substitution of the LO spin-orbit precession, see Appendix \ref{app:b}. We note that the substitution of lower-order EOM in higher-order PN Lagrangians and Hamiltonians is a well-founded procedure, see, e.g.~\cite{Schafer:1984,Damour:1985,Damour:1990jh}. In diagram (b), double-precession terms arise, but these will only contribute from the next PN order, so they are dropped here. In principle, new diagrams are added in this sector by just inserting further propagator correction vertices.  Thus there are three new diagrams (as diagrams (d) and (f) here also correspond to (a1) and (a2), respectively, of Fig.~2 in \cite{Levi:2008nh}). We note that unlike the diagrams of nonspinning interactions, which require a tensor Fourier integral of order $2n$ for $n$ propagator correction vertices, here a tensor of order $2n+2$ is required, due to the derivative-coupling of spin, which makes the computations heavier. In particular, we note diagram (c), which requires Fourier tensor integrals of orders 5 and 6, see, e.g.~Appendix \ref{app:a} for these Fourier integral tensors up to order 4. The expressions for the orders 5 and 6 are too lengthy to be included here, as they have 26 and 76 generic terms, respectively. Again, new precession terms which arise at the evaluation at this order are dropped at the use of the LO EOM of spin. We stress that at the NNLO spin1-spin2 level, accelerations, and precession terms are inevitable. Finally, we recall that there are several ways to evaluate the diagrams including time derivatives, differing by just total time derivatives. 

The values of the one-graviton exchange diagrams are then given in the following: 
\bea
Fig.~1(a)&=& -\frac{G}{r}\dot{S}_1^{0i}\dot{S}_2^{0i}+\frac{G}{r^2}\left[\dot{S}_1^{0i}\left((\vec{S}_2\times\vec{n})^i-S_2^{0i}\vec{v}_2\cdot\vec{n}-v_2^iS_2^{0j}n^j\right)\right.\nn\\
&&\left.~~~~~-\dot{S}_2^{0i}\left((\vec{S}_1\times\vec{n})^i-S_1^{0i}\vec{v}_1\cdot\vec{n}-v_1^iS_1^{0j}n^j\right)\right]-\frac{G}{r^3}\left[S_1^{0i}S_2^{0i}\left(2\vec{v}_1\cdot\vec{v}_2-3\vec{v}_1\cdot\vec{n}\vec{v}_2\cdot\vec{n}\right)\right.\nn\\
&&\left.~~~~~+S_1^{0i}S_2^{0j}\left(2v_2^iv_1^j-3v_2^in^j\vec{v}_1\cdot\vec{n}-3n^iv_1^j\vec{v}_2\cdot\vec{n}-3n^in^j\vec{v}_1\cdot\vec{v}_2\right)\right],\\
Fig.~1(b)&=& \frac{G}{2r}\left[\dot{S}_1^{0i}\left((\vec{S}_2\times\vec{a}_2)^i-(\vec{S}_2\times\vec{n})^i\vec{a}_2\cdot\vec{n}\right)+\dot{S}_2^{0i}\left((\vec{S}_1\times\vec{a}_1)^i-(\vec{S}_1\times\vec{n})^i\vec{a}_1\cdot\vec{n}\right)\right]\nn\\
&&-\frac{G}{2r^2}\left[S_1^{0i}\left((\vec{S}_2\times\vec{v}_1)^i\vec{a}_2\cdot\vec{n}+(\vec{S}_2\times\vec{n})^i\vec{a}_2\cdot\vec{v}_1+(\vec{S}_2\times\vec{a}_2)^i\vec{v}_1\cdot\vec{n}+(\vec{S}_2\times\vec{a}_1)^i\vec{v}_2\cdot\vec{n}\right.\right.\nn\\
&&\left.~~~~~+v_2^i\vec{S}_2\times\vec{a}_1\cdot\vec{n}-n^i\vec{S}_2\times\vec{v}_2\cdot\vec{a}_1-3(\vec{S}_2\times\vec{n})^i\vec{v}_1\cdot\vec{n}\vec{a}_2\cdot\vec{n}-3n^i\vec{S}_2\times\vec{a}_1\cdot\vec{n}\vec{v}_2\cdot\vec{n}\right)\nn\\
&&~~~~~+\dot{S}_1^{0i}\left((\vec{S}_2\times\vec{v}_1)^i\vec{v}_2\cdot\vec{n}-2(\vec{S}_2\times\vec{v}_2)^i\vec{v}_2\cdot\vec{n}-(\vec{S}_2\times\vec{n})^iv_2^2+v_2^i\vec{S}_2\times\vec{v}_1\cdot\vec{n}-n^i\vec{S}_2\times\vec{v}_2\cdot\vec{v}_1\right.\nn\\
&&\left.~~~~~+3(\vec{S}_2\times\vec{n})^i(\vec{v}_2\cdot\vec{n})^2-3n^i\vec{S}_2\times\vec{v}_1\cdot\vec{n}\vec{v}_2\cdot\vec{n}\right)\nn\\
&&~~~~~+\dot{\vec{S}}_1\cdot\vec{S}_2\vec{v}_2\cdot\vec{n}-\dot{\vec{S}}_1\cdot\vec{v}_2\vec{S}_2\cdot\vec{n}-\dot{\vec{S}}_1\cdot\vec{n}\vec{S}_2\cdot\vec{v}_2+3\dot{\vec{S}}_1\cdot\vec{n}\vec{S}_2\cdot\vec{n}\vec{v}_2\cdot\vec{n}\nn\\
&&~~~~-S_2^{0i}\left((\vec{S}_1\times\vec{v}_2)^i\vec{a}_1\cdot\vec{n}+(\vec{S}_1\times\vec{n})^i\vec{a}_1\cdot\vec{v}_2+(\vec{S}_1\times\vec{a}_1)^i\vec{v}_2\cdot\vec{n}+(\vec{S}_1\times\vec{a}_2)^i\vec{v}_1\cdot\vec{n}\right.\nn\\
&&\left.~~~~~+v_1^i\vec{S}_1\times\vec{a}_2\cdot\vec{n}-n^i\vec{S}_1\times\vec{v}_1\cdot\vec{a}_2-3(\vec{S}_1\times\vec{n})^i\vec{v}_2\cdot\vec{n}\vec{a}_1\cdot\vec{n}-3n^i\vec{S}_1\times\vec{a}_2\cdot\vec{n}\vec{v}_1\cdot\vec{n}\right)\nn\\
&&~~~~~-\dot{S}_2^{0i}\left((\vec{S}_1\times\vec{v}_2)^i\vec{v}_1\cdot\vec{n}-2(\vec{S}_1\times\vec{v}_1)^i\vec{v}_1\cdot\vec{n}-(\vec{S}_1\times\vec{n})^iv_1^2+v_1^i\vec{S}_1\times\vec{v}_2\cdot\vec{n}-n^i\vec{S}_1\times\vec{v}_1\cdot\vec{v}_2\right.\nn\\
&&\left.~~~~~+3(\vec{S}_1\times\vec{n})^i(\vec{v}_1\cdot\vec{n})^2-3n^i\vec{S}_1\times\vec{v}_2\cdot\vec{n}\vec{v}_1\cdot\vec{n}\right)\nn\\
&&\left.~~~~~-\dot{\vec{S}}_2\cdot\vec{S}_1\vec{v}_1\cdot\vec{n}+\dot{\vec{S}}_2\cdot\vec{v}_1\vec{S}_1\cdot\vec{n}+\dot{\vec{S}}_2\cdot\vec{n}\vec{S}_1\cdot\vec{v}_1-3\dot{\vec{S}}_2\cdot\vec{n}\vec{S}_1\cdot\vec{n}\vec{v}_1\cdot\vec{n}\right]\nn\\
&&-\frac{G}{2r^3}\left[S_1^{0i}\left((\vec{S}_2\times\vec{v}_1)^i\vec{v}_1\cdot\vec{v}_2-(\vec{S}_2\times\vec{v}_1)^iv_2^2-2(\vec{S}_2\times\vec{v}_2)^i\vec{v}_1\cdot\vec{v}_2-v_1^i(\vec{S}_2\times\vec{v}_2)\cdot\vec{v}_1\right.\right.\nn\\
&&~~~~~-3(\vec{S}_2\times\vec{v}_1)^i\vec{v}_1\cdot\vec{n}\vec{v}_2\cdot\vec{n}+3(\vec{S}_2\times\vec{v}_1)^i(\vec{v}_2\cdot\vec{n})^2+6(\vec{S}_2\times\vec{v}_2)^i\vec{v}_1\cdot\vec{n}\vec{v}_2\cdot\vec{n}+6(\vec{S}_2\times\vec{n})^i\vec{v}_1\cdot\vec{v}_2\vec{v}_2\cdot\vec{n}\nn\\
&&~~~~~+3(\vec{S}_2\times\vec{n})^i\vec{v}_1\cdot\vec{n}v_2^2-3v_1^i\vec{S}_2\times\vec{v}_1\cdot\vec{n}\vec{v}_2\cdot\vec{n}-3v_2^i\vec{S}_2\times\vec{v}_1\cdot\vec{n}\vec{v}_1\cdot\vec{n}+3n^i\vec{S}_2\times\vec{v}_2\cdot\vec{v}_1\vec{v}_1\cdot\vec{n}\nn\\
&&\left.~~~~~-3n^i\vec{S}_2\times\vec{v}_1\cdot\vec{n}\vec{v}_1\cdot\vec{v}_2-15(\vec{S}_2\times\vec{n})^i\vec{v}_1\cdot\vec{n}(\vec{v}_2\cdot\vec{n})^2+15n^i\vec{S}_2\times\vec{v}_1\cdot\vec{n}\vec{v}_1\cdot\vec{n}\vec{v}_2\cdot\vec{n}\right)\nn\\
&&~~~~~+S_2^{0i}\left((\vec{S}_1\times\vec{v}_2)^i\vec{v}_1\cdot\vec{v}_2-(\vec{S}_1\times\vec{v}_2)^iv_1^2-2(\vec{S}_1\times\vec{v}_1)^i\vec{v}_1\cdot\vec{v}_2-v_2^i(\vec{S}_1\times\vec{v}_1)\cdot\vec{v}_2\right.\nn\\
&&~~~~~-3(\vec{S}_1\times\vec{v}_2)^i\vec{v}_1\cdot\vec{n}\vec{v}_2\cdot\vec{n}+3(\vec{S}_1\times\vec{v}_2)^i(\vec{v}_1\cdot\vec{n})^2+6(\vec{S}_1\times\vec{v}_1)^i\vec{v}_1\cdot\vec{n}\vec{v}_2\cdot\vec{n}+6(\vec{S}_1\times\vec{n})^i\vec{v}_1\cdot\vec{v}_2\vec{v}_1\cdot\vec{n}\nn\\
&&~~~~~+3(\vec{S}_1\times\vec{n})^i\vec{v}_2\cdot\vec{n}v_1^2-3v_2^i\vec{S}_1\times\vec{v}_2\cdot\vec{n}\vec{v}_1\cdot\vec{n}-3v_1^i\vec{S}_1\times\vec{v}_2\cdot\vec{n}\vec{v}_2\cdot\vec{n}+3n^i\vec{S}_1\times\vec{v}_1\cdot\vec{v}_2\vec{v}_2\cdot\vec{n}\nn\\
&&\left.\left.~~~~~-3n^i\vec{S}_1\times\vec{v}_2\cdot\vec{n}\vec{v}_1\cdot\vec{v}_2-15(\vec{S}_1\times\vec{n})^i\vec{v}_2\cdot\vec{n}(\vec{v}_1\cdot\vec{n})^2+15n^i\vec{S}_1\times\vec{v}_2\cdot\vec{n}\vec{v}_1\cdot\vec{n}\vec{v}_2\cdot\vec{n}\right)\right],
\eea

\bea
Fig.~1(c)&=&
-\frac{G}{8r}\left[3\vec{S}_1\cdot\vec{S}_2\vec{a}_1\cdot\vec{a}_2-\vec{S}_1\cdot\vec{a}_1\vec{S}_2\cdot\vec{a}_2-\vec{S}_1\cdot\vec{a}_2\vec{S}_2\cdot\vec{a}_1-3\vec{S}_1\cdot\vec{S}_2\vec{a}_1\cdot\vec{n}\vec{a}_2\cdot\vec{n}+\vec{S}_1\cdot\vec{a}_1\vec{S}_2\cdot\vec{n}\vec{a}_2\cdot\vec{n}\right.\nn\\
&&~~~~~+\vec{S}_1\cdot\vec{a}_2\vec{S}_2\cdot\vec{n}\vec{a}_1\cdot\vec{n}+\vec{S}_1\cdot\vec{n}\vec{S}_2\cdot\vec{a}_1\vec{a}_2\cdot\vec{n}+\vec{S}_1\cdot\vec{n}\vec{S}_2\cdot\vec{a}_2\vec{a}_1\cdot\vec{n}+\vec{S}_1\cdot\vec{n}\vec{S}_2\cdot\vec{n}\vec{a}_1\cdot\vec{a}_2\nn\\
&&\left.~~~~~-3\vec{S}_1\cdot\vec{n}\vec{S}_2\cdot\vec{n}\vec{a}_1\cdot\vec{n}\vec{a}_2\cdot\vec{n}\right]\nn\\ &&+\frac{G}{8r^2}\left[3\vec{S}_1\cdot\vec{S}_2\vec{a}_2\cdot\vec{n}v_1^2+6\vec{S}_1\cdot\vec{S}_2\vec{a}_2\cdot\vec{v}_1\vec{v}_1\cdot\vec{n}-2\vec{S}_1\cdot\vec{v}_1\vec{S}_2\cdot\vec{v}_1\vec{a}_2\cdot\vec{n}-2\vec{S}_1\cdot\vec{v}_1\vec{S}_2\cdot\vec{a}_2\vec{v}_1\cdot\vec{n}\right.\nn\\
&&~~~~~-2\vec{S}_1\cdot\vec{v}_1\vec{S}_2\cdot\vec{n}\vec{a}_2\cdot\vec{v}_1-2\vec{S}_1\cdot\vec{n}\vec{S}_2\cdot\vec{v}_1\vec{a}_2\cdot\vec{v}_1-\vec{S}_1\cdot\vec{n}\vec{S}_2\cdot\vec{a}_2v_1^2-2\vec{S}_1\cdot\vec{a}_2\vec{S}_2\cdot\vec{v}_1\vec{v}_1\cdot\vec{n}\nn\\
&&~~~~~-\vec{S}_1\cdot\vec{a}_2\vec{S}_2\cdot\vec{n}v_1^2-9\vec{S}_1\cdot\vec{S}_2(\vec{v}_1\cdot\vec{n})^2\vec{a}_2\cdot\vec{n}+6\vec{S}_1\cdot\vec{v}_1\vec{S}_2\cdot\vec{n}\vec{v}_1\cdot\vec{n}\vec{a}_2\cdot\vec{n}\nn\\
&&~~~~~+6\vec{S}_1\cdot\vec{n}\vec{S}_2\cdot\vec{v}_1\vec{v}_1\cdot\vec{n}\vec{a}_2\cdot\vec{n}+3\vec{S}_1\cdot\vec{a}_2\vec{S}_2\cdot\vec{n}(\vec{v}_1\cdot\vec{n})^2+3\vec{S}_1\cdot\vec{n}\vec{S}_2\cdot\vec{a}_2(\vec{v}_1\cdot\vec{n})^2\nn\\
&&~~~~~+3\vec{S}_1\cdot\vec{n}\vec{S}_2\cdot\vec{n}\vec{a}_2\cdot\vec{n}v_1^2+6\vec{S}_1\cdot\vec{n}\vec{S}_2\cdot\vec{n}\vec{a}_2\cdot\vec{v}_1\vec{v}_1\cdot\vec{n}-15\vec{S}_1\cdot\vec{n}\vec{S}_2\cdot\vec{n}(\vec{v}_1\cdot\vec{n})^2\vec{a}_2\cdot\vec{n}\nonumber\\
&&~~~~~-3\vec{S}_1\cdot\vec{S}_2\vec{a}_1\cdot\vec{n}v_2^2-6\vec{S}_1\cdot\vec{S}_2\vec{a}_1\cdot\vec{v}_2\vec{v}_2\cdot\vec{n}+2\vec{S}_1\cdot\vec{v}_2\vec{S}_2\cdot\vec{v}_2\vec{a}_1\cdot\vec{n}+2\vec{S}_1\cdot\vec{a}_1\vec{S}_2\cdot\vec{v}_2\vec{v}_2\cdot\vec{n}\nn\\
&&~~~~~+2\vec{S}_1\cdot\vec{n}\vec{S}_2\cdot\vec{v}_2\vec{a}_1\cdot\vec{v}_2+2\vec{S}_1\cdot\vec{v}_2\vec{S}_2\cdot\vec{n}\vec{a}_1\cdot\vec{v}_2+\vec{S}_1\cdot\vec{a}_1\vec{S}_2\cdot\vec{n}v_2^2+2\vec{S}_1\cdot\vec{v}_2\vec{S}_2\cdot\vec{a}_1\vec{v}_2\cdot\vec{n}\nn\\
&&~~~~~+\vec{S}_1\cdot\vec{n}\vec{S}_2\cdot\vec{a}_1v_2^2+9\vec{S}_1\cdot\vec{S}_2(\vec{v}_2\cdot\vec{n})^2\vec{a}_1\cdot\vec{n}-6\vec{S}_1\cdot\vec{n}\vec{S}_2\cdot\vec{v}_2\vec{v}_2\cdot\vec{n}\vec{a}_1\cdot\vec{n}\nn\\
&&~~~~~-6\vec{S}_1\cdot\vec{v}_2\vec{S}_2\cdot\vec{n}\vec{v}_2\cdot\vec{n}\vec{a}_1\cdot\vec{n}-3\vec{S}_1\cdot\vec{n}\vec{S}_2\cdot\vec{a}_1(\vec{v}_2\cdot\vec{n})^2-3\vec{S}_1\cdot\vec{a}_1\vec{S}_2\cdot\vec{n}(\vec{v}_2\cdot\vec{n})^2\nn\\
&&\left.~~~~~-3\vec{S}_1\cdot\vec{n}\vec{S}_2\cdot\vec{n}\vec{a}_1\cdot\vec{n}v_2^2-6\vec{S}_1\cdot\vec{n}\vec{S}_2\cdot\vec{n}\vec{a}_1\cdot\vec{v}_2\vec{v}_2\cdot\vec{n}+15\vec{S}_1\cdot\vec{n}\vec{S}_2\cdot\vec{n}(\vec{v}_2\cdot\vec{n})^2\vec{a}_1\cdot\vec{n}\right]\nn\\
&&-\frac{G}{8r^3}\left[3\vec{S}_1\cdot\vec{S}_2v_1^2v_2^2+6\vec{S}_1\cdot\vec{S}_2(\vec{v}_1\cdot\vec{v}_2)^2-2\vec{S}_1\cdot\vec{v}_1\vec{S}_2\cdot\vec{v}_1v_2^2-4\vec{S}_1\cdot\vec{v}_1\vec{S}_2\cdot\vec{v}_2\vec{v}_1\cdot\vec{v}_2\right.\nn\\
&&~~~~~-4\vec{S}_1\cdot\vec{v}_2\vec{S}_2\cdot\vec{v}_1\vec{v}_1\cdot\vec{v}_2-2\vec{S}_1\cdot\vec{v}_2\vec{S}_2\cdot\vec{v}_2v_1^2-9\vec{S}_1\cdot\vec{S}_2(\vec{v}_1\cdot\vec{n})^2v_2^2-9\vec{S}_1\cdot\vec{S}_2(\vec{v}_2\cdot\vec{n})^2v_1^2\nn\\
&&~~~~~-36\vec{S}_1\cdot\vec{S}_2\vec{v}_1\cdot\vec{v}_2\vec{v}_1\cdot\vec{n}\vec{v}_2\cdot\vec{n}+6\vec{S}_1\cdot\vec{v}_1\vec{S}_2\cdot\vec{v}_1(\vec{v}_2\cdot\vec{n})^2+12\vec{S}_1\cdot\vec{v}_1\vec{S}_2\cdot\vec{v}_2\vec{v}_1\cdot\vec{n}\vec{v}_2\cdot\vec{n}\nn\\
&&~~~~~+12\vec{S}_1\cdot\vec{v}_2\vec{S}_2\cdot\vec{v}_1\vec{v}_1\cdot\vec{n}\vec{v}_2\cdot\vec{n}+6\vec{S}_1\cdot\vec{v}_2\vec{S}_2\cdot\vec{v}_2(\vec{v}_1\cdot\vec{n})^2+6\vec{S}_1\cdot\vec{v}_1\vec{S}_2\cdot\vec{n}\vec{v}_1\cdot\vec{n}v_2^2\nn\\
&&~~~~~+12\vec{S}_1\cdot\vec{v}_1\vec{S}_2\cdot\vec{n}\vec{v}_1\cdot\vec{v}_2\vec{v}_2\cdot\vec{n}+12\vec{S}_1\cdot\vec{v}_2\vec{S}_2\cdot\vec{n}\vec{v}_1\cdot\vec{v}_2\vec{v}_1\cdot\vec{n}+6\vec{S}_1\cdot\vec{v}_2\vec{S}_2\cdot\vec{n}\vec{v}_2\cdot\vec{n}v_1^2\nn\\
&&~~~~~+6\vec{S}_1\cdot\vec{n}\vec{S}_2\cdot\vec{v}_1\vec{v}_1\cdot\vec{n}v_2^2+12\vec{S}_1\cdot\vec{n}\vec{S}_2\cdot\vec{v}_1\vec{v}_1\cdot\vec{v}_2\vec{v}_2\cdot\vec{n}+12\vec{S}_1\cdot\vec{n}\vec{S}_2\cdot\vec{v}_2\vec{v}_1\cdot\vec{v}_2\vec{v}_1\cdot\vec{n}\nn\\
&&~~~~~+6\vec{S}_1\cdot\vec{n}\vec{S}_2\cdot\vec{v}_2\vec{v}_2\cdot\vec{n}v_1^2+3\vec{S}_1\cdot\vec{n}\vec{S}_2\cdot\vec{n}v_1^2v_2^2+6\vec{S}_1\cdot\vec{n}\vec{S}_2\cdot\vec{n}(\vec{v}_1\cdot\vec{v}_2)^2+45\vec{S}_1\cdot\vec{S}_2(\vec{v}_1\cdot\vec{n})^2(\vec{v}_2\cdot\vec{n})^2\nn\\
&&~~~~~-30\vec{S}_1\cdot\vec{v}_1\vec{S}_2\cdot\vec{n}\vec{v}_1\cdot\vec{n}(\vec{v}_2\cdot\vec{n})^2-30\vec{S}_1\cdot\vec{v}_2\vec{S}_2\cdot\vec{n}\vec{v}_2\cdot\vec{n}(\vec{v}_1\cdot\vec{n})^2-30\vec{S}_1\cdot\vec{n}\vec{S}_2\cdot\vec{v}_1\vec{v}_1\cdot\vec{n}(\vec{v}_2\cdot\vec{n})^2\nn\\
&&~~~~~-30\vec{S}_1\cdot\vec{n}\vec{S}_2\cdot\vec{v}_2\vec{v}_2\cdot\vec{n}(\vec{v}_1\cdot\vec{n})^2-15\vec{S}_1\cdot\vec{n}\vec{S}_2\cdot\vec{n}(\vec{v}_1\cdot\vec{n})^2v_2^2-15\vec{S}_1\cdot\vec{n}\vec{S}_2\cdot\vec{n}(\vec{v}_2\cdot\vec{n})^2v_1^2\nn\\
&&\left.~~~~~-60\vec{S}_1\cdot\vec{n}\vec{S}_2\cdot\vec{n}\vec{v}_1\cdot\vec{v}_2\vec{v}_1\cdot\vec{n}\vec{v}_2\cdot\vec{n}+105\vec{S}_1\cdot\vec{n}\vec{S}_2\cdot\vec{n}(\vec{v}_1\cdot\vec{n})^2(\vec{v}_2\cdot\vec{n})^2\right],\\
Fig.~1(d)&=& \frac{G}{r^2}\left[S_1^{0i}v_1^i\vec{S}_2\times\vec{n}\cdot\vec{a}_2-S_2^{0i}v_2^i\vec{S}_1\times\vec{n}\cdot\vec{a}_1-S_1^{0i}v_1^i\dot{S}_2^{0j}n^j+S_2^{0i}v_2^i\dot{S}_1^{0j}n^j\right]\nn\\
&&+\frac{G}{r^3}\left[S_1^{0i}S_2^{0j}\left(2v_1^iv_2^j-3v_1^in^j\vec{v}_2\cdot\vec{n}-3n^iv_2^j\vec{v}_1\cdot\vec{n}\right)\right.\nn\\
&&\left.~~~~~-3S_1^{0i}v_1^i\vec{S}_2\times\vec{v}_2\cdot\vec{n}\vec{v}_2\cdot\vec{n}-3S_2^{0i}v_2^i\vec{S}_1\times\vec{v}_1\cdot\vec{n}\vec{v}_1\cdot\vec{n}\right],\\
Fig.~1(e)&=& \frac{G}{2r}\left[\vec{S}_1\cdot\vec{S}_2\vec{a}_1\cdot\vec{a}_2-\vec{S}_1\cdot\vec{a}_2\vec{S}_2\cdot\vec{a}_1-\vec{S}_1\times\vec{n}\cdot\vec{a}_1\vec{S}_2\times\vec{n}\cdot\vec{a}_2\right.\nn\\
&&\left.~~~~~+\dot{S}_1^{0i}\dot{S}_2^{0i}-\dot{S}_1^{0i}n^i\dot{S}_2^{0j}n^j+\dot{S}_1^{0i}\left((\vec{S}_2\times\vec{a}_2)^i+n^i\vec{S}_2\times\vec{n}\cdot\vec{a}_2\right)+\dot{S}_2^{0i}\left((\vec{S}_1\times\vec{a}_1)^i+n^i\vec{S}_1\times\vec{n}\cdot\vec{a}_1\right)\right]\nn\\
&&-\frac{G}{2r^2}\left[\vec{S}_1\cdot\vec{S}_2\vec{a}_2\cdot\vec{v}_1\vec{v}_1\cdot\vec{n}-\vec{S}_1\cdot\vec{S}_2\vec{a}_1\cdot\vec{v}_2\vec{v}_2\cdot\vec{n}+\vec{S}_1\cdot\vec{v}_2\vec{S}_2\cdot\vec{a}_1\vec{v}_2\cdot\vec{n}-\vec{S}_1\cdot\vec{a}_2\vec{S}_2\cdot\vec{v}_1\vec{v}_1\cdot\vec{n}\right.\nn\\
&&~~~~~-\vec{S}_1\times\vec{v}_1\cdot\vec{n}\vec{S}_2\times\vec{v}_1\cdot\vec{a}_2+\vec{S}_1\times\vec{v}_2\cdot\vec{a}_1\vec{S}_2\times\vec{v}_2\cdot\vec{n}\nn\\
&&~~~~~+3\vec{S}_1\times\vec{v}_1\cdot\vec{n}\vec{S}_2\times\vec{n}\cdot\vec{a}_2\vec{v}_1\cdot\vec{n}-3\vec{S}_1\times\vec{n}\cdot\vec{a}_1\vec{S}_2\times\vec{v}_2\cdot\vec{n}\vec{v}_2\cdot\vec{n}\nn\\
&&~~~~~+S_1^{0i}\left(\dot{S}_2^{0i}\vec{v}_1\cdot\vec{n}+v_1^i\dot{S}_2^{0j}n^j+n^i\dot{S}_2^{0j}v_1^j-3n^i\dot{S}_2^{0j}n^j\vec{v}_1\cdot\vec{n}+(\vec{S}_2\times\vec{a}_2)^i\vec{v}_1\cdot\vec{n}-v_1^i\vec{S}_2\times\vec{n}\cdot\vec{a}_2\right.\nn\\
&&~~~~~\left.-n^i\vec{S}_2\times\vec{v}_1\cdot\vec{a}_2+3n^i\vec{S}_2\times\vec{n}\cdot\vec{a}_2\vec{v}_1\cdot\vec{n}\right)-\dot{S}_1^{0i}\left((\vec{S}_2\times\vec{v}_2)^i\vec{v}_2\cdot\vec{n}+v_2^i\vec{S}_2\times\vec{v}_2\cdot\vec{n}\right.\nn\\ 
&&~~~~~\left.-3n^i\vec{S}_2\times\vec{v}_2\cdot\vec{n}\vec{v}_2\cdot\vec{n}\right)-S_2^{0i}\left(\dot{S}_1^{0i}\vec{v}_2\cdot\vec{n}+v_2^i\dot{S}_1^{0j}n^j+n^i\dot{S}_1^{0j}v_2^j-3n^i\dot{S}_1^{0j}n^j\vec{v}_2\cdot\vec{n}\right.\nn\\
&&~~~~~\left.+(\vec{S}_1\times\vec{a}_1)^i\vec{v}_2\cdot\vec{n}-v_2^i\vec{S}_1\times\vec{n}\cdot\vec{a}_1-n^i\vec{S}_1\times\vec{v}_2\cdot\vec{a}_1+3n^i\vec{S}_1\times\vec{n}\cdot\vec{a}_1\vec{v}_2\cdot\vec{n}\right)\nn\\
&&~~~~~\left.+\dot{S}_2^{0i}\left((\vec{S}_1\times\vec{v}_1)^i\vec{v}_1\cdot\vec{n}+v_1^i\vec{S}_1\times\vec{v}_1\cdot\vec{n}-3n^i\vec{S}_1\times\vec{v}_1\cdot\vec{n}\vec{v}_1\cdot\vec{n}\right)\right]\nn
\eea

\bea
&&+\frac{G}{2r^3}\left[\vec{S}_1\cdot\vec{S}_2(\vec{v}_1\cdot\vec{v}_2)^2-\vec{S}_1\cdot\vec{v}_2\vec{S}_2\cdot\vec{v}_1\vec{v}_1\cdot\vec{v}_2+\vec{S}_1\times\vec{v}_1\cdot\vec{v}_2\vec{S}_2\times\vec{v}_2\cdot\vec{v}_1-3\vec{S}_1\cdot\vec{S}_2\vec{v}_1\cdot\vec{v}_2\vec{v}_1\cdot\vec{n}\vec{v}_2\cdot\vec{n}\right.\nn\\
&&~~~~~+3\vec{S}_1\cdot\vec{v}_2\vec{S}_2\cdot\vec{v}_1\vec{v}_1\cdot\vec{n}\vec{v}_2\cdot\vec{n}-3\vec{S}_1\times\vec{v}_1\cdot\vec{v}_2\vec{S}_2\times\vec{v}_2\cdot\vec{n}\vec{v}_1\cdot\vec{n}-3\vec{S}_1\times\vec{v}_1\cdot\vec{n}\vec{S}_2\times\vec{v}_2\cdot\vec{v}_1\vec{v}_2\cdot\vec{n}\nn\\
&&~~~~~-3\vec{S}_1\times\vec{v}_1\cdot\vec{n}\vec{S}_2\times\vec{v}_2\cdot\vec{n}\vec{v}_1\cdot\vec{v}_2+15\vec{S}_1\times\vec{v}_1\cdot\vec{n}\vec{S}_2\times\vec{v}_2\cdot\vec{n}\vec{v}_1\cdot\vec{n}\vec{v}_2\cdot\vec{n}\nn\\
&&~~~~~+S_1^{0i}\left((\vec{S}_2\times\vec{v}_2)^i\vec{v}_1\cdot\vec{v}_2+v_2^i\vec{S}_2\times\vec{v}_2\cdot\vec{v}_1-3(\vec{S}_2\times\vec{v}_2)^i\vec{v}_1\cdot\vec{n}\vec{v}_2\cdot\vec{n}-3v_1^i\vec{S}_2\times\vec{v}_2\cdot\vec{n}\vec{v}_2\cdot\vec{n}\right.\nn\\
&&\left.~~~~~-3v_2^i\vec{S}_2\times\vec{v}_2\cdot\vec{n}\vec{v}_1\cdot\vec{n}-3n^i\vec{S}_2\times\vec{v}_2\cdot\vec{v}_1\vec{v}_2\cdot\vec{n}-3n^i\vec{S}_2\times\vec{v}_2\cdot\vec{n}\vec{v}_1\cdot\vec{v}_2+15n^i\vec{S}_2\times\vec{v}_2\cdot\vec{n}\vec{v}_1\cdot\vec{n}\vec{v}_2\cdot\vec{n}\right)\nn\\
&&~~~~~+S_2^{0i}\left((\vec{S}_1\times\vec{v}_1)^i\vec{v}_1\cdot\vec{v}_2+v_1^i\vec{S}_1\times\vec{v}_1\cdot\vec{v}_2-3(\vec{S}_1\times\vec{v}_1)^i\vec{v}_1\cdot\vec{n}\vec{v}_2\cdot\vec{n}-3v_2^i\vec{S}_1\times\vec{v}_1\cdot\vec{n}\vec{v}_1\cdot\vec{n}\right.\nn\\
&&~~~~~-3v_1^i\vec{S}_1\times\vec{v}_1\cdot\vec{n}\vec{v}_2\cdot\vec{n}-3n^i\vec{S}_1\times\vec{v}_1\cdot\vec{v}_2\vec{v}_1\cdot\vec{n}-3n^i\vec{S}_1\times\vec{v}_1\cdot\vec{n}\vec{v}_1\cdot\vec{v}_2\nn\\
&&\left.~~~~~+15n^i\vec{S}_1\times\vec{v}_1\cdot\vec{n}\vec{v}_1\cdot\vec{n}\vec{v}_2\cdot\vec{n}\right)
+S_1^{0i}S_2^{0i} \vec{v}_1\cdot\vec{v}_2 + S_1^{0i}v_1^i S_2^{0j}v_2^j + S_1^{0i}v_2^i S_2^{0j}v_1^j
-3S_1^{0i}S_2^{0i} \vec{v}_1\cdot\vec{n} \vec{v}_2\cdot\vec{n}
\nn\\
&&~~~~~
-3S_1^{0i}v_1^i S_2^{0j}n^j\vec{v}_2\cdot\vec{n}-3S_1^{0i}n^i S_2^{0j}v_1^j\vec{v}_2\cdot\vec{n}
-3S_1^{0i}v_2^i S_2^{0j}n^j\vec{v}_1\cdot\vec{n}-3S_1^{0i}n^i S_2^{0j}v_2^j\vec{v}_1\cdot\vec{n}
\nn\\
&&\left.~~~~~
-3S_1^{0i}n^i S_2^{0j}n^j\vec{v}_1\cdot\vec{v}_2
+15S_1^{0i}n^i S_2^{0j}n^j\vec{v}_1\cdot\vec{n}\vec{v}_2\cdot\vec{n}\right],\\
Fig.~1(f)&=& \frac{G}{r^2}\left[S_1^{0i}\left((\vec{S}_2\times\vec{n})^i\vec{a}_2\cdot\vec{v}_1-2v_1^i\vec{S}_2\times\vec{n}\cdot\vec{a}_2-a_2^i\vec{S}_2\times\vec{v}_1\cdot\vec{n}\right)\right.\nn\\
&&\left.~~~~~-S_2^{0i}\left((\vec{S}_1\times\vec{n})^i\vec{a}_1\cdot\vec{v}_2-2v_2^i\vec{S}_1\times\vec{n}\cdot\vec{a}_1-a_1^i\vec{S}_1\times\vec{v}_2\cdot\vec{n}\right)\right]\nn\\
&&-\frac{G}{r^3}\left[S_1^{0i}\left((\vec{S}_2\times\vec{v}_2)^i\vec{v}_1\cdot\vec{v}_2+v_2^i\vec{S}_2\times\vec{v}_2\cdot\vec{v}_1\right.\right.\nn\\
&&\left.~~~~~-3(\vec{S}_2\times\vec{n})^i\vec{v}_1\cdot\vec{v}_2\vec{v}_2\cdot\vec{n}-6v_1^i\vec{S}_2\times\vec{v}_2\cdot\vec{n}\vec{v}_2\cdot\vec{n}+3v_2^i\vec{S}_2\times\vec{v}_1\cdot\vec{n}\vec{v}_2\cdot\vec{n}\right)\nn\\
&&~~~~~+S_2^{0i}\left((\vec{S}_1\times\vec{v}_1)^i\vec{v}_1\cdot\vec{v}_2+v_1^i\vec{S}_1\times\vec{v}_1\cdot\vec{v}_2\right.\nn\\
&&\left.\left.~~~~~-3(\vec{S}_1\times\vec{n})^i\vec{v}_1\cdot\vec{v}_2\vec{v}_1\cdot\vec{n}-6v_2^i\vec{S}_1\times\vec{v}_1\cdot\vec{n}\vec{v}_1\cdot\vec{n}+3v_1^i\vec{S}_1\times\vec{v}_2\cdot\vec{n}\vec{v}_1\cdot\vec{n}\right)\right],\\
Fig.~1(g)&=& -\frac{G}{2r}\left[\vec{S}_1\cdot\vec{a}_1\vec{S}_2\cdot\vec{a}_2-2\vec{S}_1\cdot\vec{a}_2\vec{S}_2\cdot\vec{a}_1-\vec{S}_1\cdot\vec{n}\vec{S}_2\cdot\vec{n}\vec{a}_1\cdot\vec{a}_2\right.\nn\\
&&\left.~~~~~-2\vec{S}_1\times\vec{n}\cdot\vec{a}_1\vec{S}_2\times\vec{n}\cdot\vec{a}_2+\vec{S}_1\times\vec{n}\cdot\vec{a}_2\vec{S}_2\times\vec{n}\cdot\vec{a}_1\right]\nn\\ &&+\frac{G}{2r^2}\left[\vec{S}_1\cdot\vec{v}_1\vec{S}_2\cdot\vec{a}_2\vec{v}_1\cdot\vec{n}+\vec{S}_1\cdot\vec{v}_1\vec{S}_2\cdot\vec{n}\vec{a}_2\cdot\vec{v}_1+\vec{S}_1\cdot\vec{n}\vec{S}_2\cdot\vec{v}_1\vec{a}_2\cdot\vec{v}_1-2\vec{S}_1\cdot\vec{a}_2\vec{S}_2\cdot\vec{v}_1\vec{v}_1\cdot\vec{n}\right.\nn\\
&&~~~~~-2\vec{S}_1\times\vec{v}_1\cdot\vec{n}\vec{S}_2\times\vec{v}_1\cdot\vec{a}_2+\vec{S}_1\times\vec{v}_1\cdot\vec{a}_2\vec{S}_2\times\vec{v}_1\cdot\vec{n}-3\vec{S}_1\cdot\vec{n}\vec{S}_2\cdot\vec{n}\vec{a}_2\cdot\vec{v}_1\vec{v}_1\cdot\vec{n}\nonumber\\
&&~~~~~+6\vec{S}_1\times\vec{v}_1\cdot\vec{n}\vec{S}_2\times\vec{n}\cdot\vec{a}_2\vec{v}_1\cdot\vec{n}-3\vec{S}_1\times\vec{n}\cdot\vec{a}_2\vec{S}_2\times\vec{v}_1\cdot\vec{n}\vec{v}_1\cdot\vec{n}\nn\\
&&~~~~~-\vec{S}_1\cdot\vec{a}_1\vec{S}_2\cdot\vec{v}_2\vec{v}_2\cdot\vec{n}-\vec{S}_1\cdot\vec{n}\vec{S}_2\cdot\vec{v}_2\vec{a}_1\cdot\vec{v}_2-\vec{S}_1\cdot\vec{v}_2\vec{S}_2\cdot\vec{n}\vec{a}_1\cdot\vec{v}_2+2\vec{S}_1\cdot\vec{v}_2\vec{S}_2\cdot\vec{a}_1\vec{v}_2\cdot\vec{n}\nn\\
&&~~~~~+2\vec{S}_1\times\vec{v}_2\cdot\vec{a}_1\vec{S}_2\times\vec{v}_2\cdot\vec{n}-\vec{S}_1\times\vec{v}_2\cdot\vec{n}\vec{S}_2\times\vec{v}_2\cdot\vec{a}_1+3\vec{S}_1\cdot\vec{n}\vec{S}_2\cdot\vec{n}\vec{a}_1\cdot\vec{v}_2\vec{v}_2\cdot\vec{n}\nonumber\\
&&\left.~~~~~-6\vec{S}_1\times\vec{n}\cdot\vec{a}_1\vec{S}_2\times\vec{v}_2\cdot\vec{n}\vec{v}_2\cdot\vec{n}+3\vec{S}_1\times\vec{v}_2\cdot\vec{n}\vec{S}_2\times\vec{n}\cdot\vec{a}_1\vec{v}_2\cdot\vec{n}\right]\nn\\
&&-\frac{G}{2r^3}\left[2\vec{S}_1\cdot\vec{v}_1\vec{S}_2\cdot\vec{v}_2\vec{v}_1\cdot\vec{v}_2-\vec{S}_1\cdot\vec{v}_2\vec{S}_2\cdot\vec{v}_1\vec{v}_1\cdot\vec{v}_2+\vec{S}_1\times\vec{v}_1\cdot\vec{v}_2\vec{S}_2\times\vec{v}_2\cdot\vec{v}_1\right.\nn\\
&&~~~~~-3\vec{S}_1\cdot\vec{v}_1\vec{S}_2\cdot\vec{v}_2\vec{v}_1\cdot\vec{n}\vec{v}_2\cdot\vec{n}+6\vec{S}_1\cdot\vec{v}_2\vec{S}_2\cdot\vec{v}_1\vec{v}_1\cdot\vec{n}\vec{v}_2\cdot\vec{n}-3\vec{S}_1\cdot\vec{v}_1\vec{S}_2\cdot\vec{n}\vec{v}_1\cdot\vec{v}_2\vec{v}_2\cdot\vec{n}\nn\\
&&~~~~~-3\vec{S}_1\cdot\vec{v}_2\vec{S}_2\cdot\vec{n}\vec{v}_1\cdot\vec{v}_2\vec{v}_1\cdot\vec{n}-3\vec{S}_1\cdot\vec{n}\vec{S}_2\cdot\vec{v}_1\vec{v}_1\cdot\vec{v}_2\vec{v}_2\cdot\vec{n}-3\vec{S}_1\cdot\vec{n}\vec{S}_2\cdot\vec{v}_2\vec{v}_1\cdot\vec{v}_2\vec{v}_1\cdot\vec{n}\nn\\
&&~~~~~-3\vec{S}_1\cdot\vec{n}\vec{S}_2\cdot\vec{n}(\vec{v}_1\cdot\vec{v}_2)^2-3\vec{S}_1\times\vec{v}_1\cdot\vec{v}_2\vec{S}_2\times\vec{v}_1\cdot\vec{n}\vec{v}_2\cdot\vec{n}-6\vec{S}_1\times\vec{v}_1\cdot\vec{v}_2\vec{S}_2\times\vec{v}_2\cdot\vec{n}\vec{v}_1\cdot\vec{n}\nn\\
&&~~~~~-6\vec{S}_1\times\vec{v}_1\cdot\vec{n}\vec{S}_2\times\vec{v}_2\cdot\vec{v}_1\vec{v}_2\cdot\vec{n}-3\vec{S}_1\times\vec{v}_2\cdot\vec{n}\vec{S}_2\times\vec{v}_2\cdot\vec{v}_1\vec{v}_1\cdot\vec{n}-6\vec{S}_1\times\vec{v}_1\cdot\vec{n}\vec{S}_2\times\vec{v}_2\cdot\vec{n}\vec{v}_1\cdot\vec{v}_2\nn\\
&&~~~~~+3\vec{S}_1\times\vec{v}_2\cdot\vec{n}\vec{S}_2\times\vec{v}_1\cdot\vec{n}\vec{v}_1\cdot\vec{v}_2+15\vec{S}_1\cdot\vec{n}\vec{S}_2\cdot\vec{n}\vec{v}_1\cdot\vec{v}_2\vec{v}_1\cdot\vec{n}\vec{v}_2\cdot\vec{n}\nn\\
&&\left.~~~~~+30\vec{S}_1\times\vec{v}_1\cdot\vec{n}\vec{S}_2\times\vec{v}_2\cdot\vec{n}\vec{v}_1\cdot\vec{n}\vec{v}_2\cdot\vec{n}-15\vec{S}_1\times\vec{v}_2\cdot\vec{n}\vec{S}_2\times\vec{v}_1\cdot\vec{n}\vec{v}_1\cdot\vec{n}\vec{v}_2\cdot\vec{n}\right].
\eea

The diagrams appearing in Fig.~\ref{fig:4PNGeom} already appeared in the NLO spin-orbit interaction and the 2PN orbital interaction. Diagrams (a1), (a2), and (b), correspond to diagrams (a3), (b1), and (b2), respectively, in Fig.~2 of \cite{Levi:2010zu}. They yield acceleration and precession terms. It can be easily seen, that on substitution of the LO spin-orbit accelerations and LO spin1-spin2 precessions, see Appendix \ref{app:b}, NNLO spin1-spin2 interaction terms are obtained. Diagrams (c), (d1), and (d2) in Fig.~\ref{fig:4PNGeom} also correspond to diagrams (e), (b), and (c), respectively, in Fig.~4 of \cite{Gilmore:2008gq}. Again one can easily get convinced, that since both objects are considered  to be spinning here, on substitution of the LO spin1-spin2 accelerations, see Appendix \ref{app:b}, NNLO spin1-spin2 interaction terms are obtained. 

The values of the one-graviton exchange diagrams, which contribute through the substitution of EOM, are then given by:  
\bea
Fig.~2(a1)&=& 2\frac{Gm_2}{r}\dot{S}_1^{0i}v_2^i,\\
Fig.~2(a2)&=& \frac{Gm_2}{r}\left(\vec{S}_1\times\vec{v}_1\cdot\vec{a}_2-\vec{S}_1\times\vec{n}\cdot\vec{a}_2\vec{v}_1\cdot\vec{n}-\dot{\vec{S}}_1\times\vec{v}_2\cdot\vec{n}\vec{v}_2\cdot\vec{n}\right)+Gm_2\dot{\vec{S}}_1\times\vec{n}\cdot\vec{a}_2,\\
Fig.~2(b)&=& \frac{Gm_2}{2r}\left(\vec{S}_1\times\vec{v}_2\cdot\vec{a}_1-\vec{S}_1\times\vec{n}\cdot\vec{a}_1\vec{v}_2\cdot\vec{n}-\dot{\vec{S}}_1\times\vec{v}_1\cdot\vec{v}_2+\dot{\vec{S}}_1\times\vec{v}_1\cdot\vec{n}\vec{v}_2\cdot\vec{n}-\dot{S}_1^{0i}v_2^i+\dot{S}_1^{0i}n^i\vec{v}_2\cdot\vec{n}\right),\\
Fig.~2(c)&=& -2Gm_1m_2\left(\vec{a}_1\cdot\vec{v}_2\vec{v}_2\cdot\vec{n}-\vec{a}_2\cdot\vec{v}_1\vec{v}_1\cdot\vec{n}\right)+2Gm_1m_2r\vec{a}_1\cdot\vec{a}_2,\\
Fig.~2(d1)&=& \frac{3}{2}Gm_1m_2\left(\vec{a}_1\cdot\vec{v}_1\vec{v}_2\cdot\vec{n}-\vec{a}_2\cdot\vec{v}_2\vec{v}_1\cdot\vec{n}\right),\\
Fig.~2(d2)&=&
\frac{1}{8}Gm_1m_2\left(2\vec{a}_1\cdot\vec{v}_2\vec{v}_2\cdot\vec{n}+\vec{a}_1\cdot\vec{n}\left(v_2^2-(\vec{v}_2\cdot\vec{n})^2\right)-2\vec{a}_2\cdot\vec{v}_1\vec{v}_1\cdot\vec{n}-\vec{a}_2\cdot\vec{n}\left(v_1^2-(\vec{v}_1\cdot\vec{n})^2\right)\right)\nn\\
&&-\frac{1}{8}Gm_1m_2r\left(\vec{a}_1\cdot\vec{a}_2+\vec{a}_1\cdot\vec{n}\vec{a}_2\cdot\vec{n}\right).
\eea

As we noted already, there are several ways to evaluate diagrams including time derivatives, as those appearing here, e.g.~from the NLO spin-orbit interaction. Though these different evaluations lead to physically equivalent potentials, one must consistently use the same NLO spin-orbit potential corresponding to a specific evaluation in all stages of the calculation, e.g.~in the derivation of a corresponding NNLO spin1-spin2 Hamiltonian, where the NLO spin-orbit potential should be taken into account.

We should stress that the SSC dependent parts from the LO and NLO spin-orbit, and NLO spin1-spin2 sectors also contribute upon the insertion of the SSC. In particular, we note that there is a piece, which formally appears at the NLO spin-orbit sector, but it vanishes there. It includes the $S^{0i}$ entry, and so it falls upon the insertion of the SSC at NLO. Yet this piece reappears as a contribution to the NNLO spin1-spin2 sector. It enters from diagrams (a4) and (b2) in Fig.~2 of \cite{Levi:2010zu}, and its value is given by 
\be
\frac{3}{2}\frac{Gm_2}{r^2}\vec{v}_2\cdot\vec{n}S_1^{0i}v_1^i \in \left[Fig.~2(a4)+Fig.~2(b2)\right].\label{formalnloso}
\ee 
Thus there is one further diagram from the spin-orbit sector, Fig.~2(a4) of \cite{Levi:2010zu}, which contributes through its SSC dependent part only, and so we do not repeat this diagram here.

\subsection{Order $G^2$ Feynman diagrams}

\begin{figure}[ht]
\includegraphics{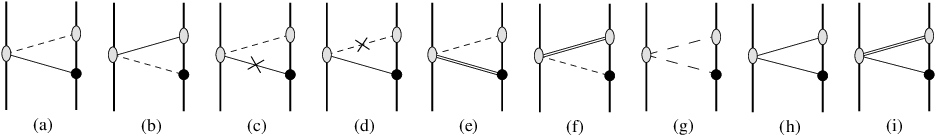}
\caption{NNLO spin1-spin2 Feynman diagrams of order $G^2$: Two-graviton exchange. These diagrams should be included together with their mirror images.}
\label{fig:4PNG22m}
\end{figure}

\begin{figure}[ht]
\includegraphics{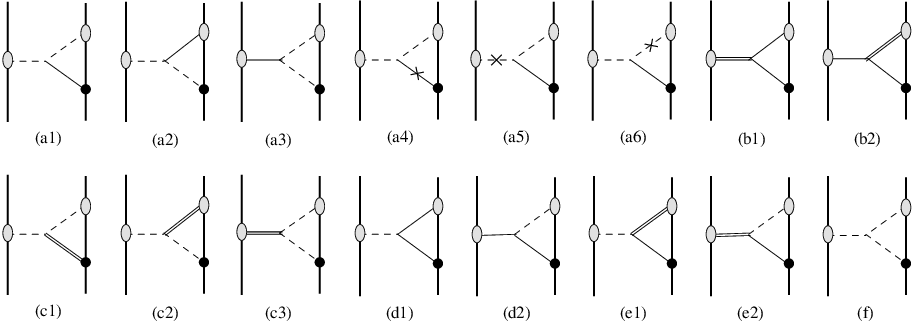}
\caption{NNLO spin1-spin2 Feynman diagrams of order $G^2$: Cubic self-gravitational interaction. These diagrams should be included together with their mirror images.}
\label{fig:4PNG23g}
\end{figure}

For the NNLO spin1-spin2 interaction, we have 25 diagrams at order $G^2$ to evaluate: 9 two-graviton exchange diagrams, and 16 cubic self-gravitational interaction diagrams as shown in Figs.~\ref{fig:4PNG22m} and \ref{fig:4PNG23g}, respectively, (all pure spin1-spin2 diagrams). Here, only diagram (a) of Fig.~\ref{fig:4PNG22m} and diagram (a1) of Fig.~\ref{fig:4PNG23g} appeared already in the NLO spin1-spin2 evaluation: they correspond to diagrams (a) and (b), respectively, of Fig.~3 in \cite{Levi:2008nh}. All the rest are new diagrams. Two further diagrams from the NLO spin-orbit sector, diagrams (a2) and (b2) in Fig.~3 of \cite{Levi:2010zu}, contribute through their SSC dependent parts only, thus we do not repeat them here.

The two-graviton exchange diagrams just factorize into a product of two tensor Fourier integrals, see Appendix~\ref{app:a}. We first encounter here two-graviton exchange diagrams, which involve time derivatives, either from the spin couplings or from propagator correction vertices. Precession terms that arise here are delegated to higher orders.
   
The values of the two-graviton exchange diagrams are given as follows: 
\bea
Fig.~3(a)&=& -\frac{G^2m_2}{2r^4}\left[ 15\vec{S}_1\cdot\vec{S}_2v_2^2-2\vec{S}_1\cdot\vec{S}_2\vec{v}_1\cdot\vec{n}\vec{v}_2\cdot\vec{n}+2\vec{S}_1\cdot\vec{n}\vec{S}_2\cdot\vec{v}_1\vec{v}_2\cdot\vec{n}-39\vec{S}_1\cdot\vec{n}\vec{S}_2\cdot\vec{n}v_2^2\right]\nonumber\\
&&+\frac{G^2m_2}{r^4}\left[S_1^{0i}\left(3(\vec{S}_2\times\vec{v}_1)^i-3(\vec{S}_2\times\vec{v}_2)^i-2(\vec{S}_2\times\vec{n})^i\vec{v}_1\cdot\vec{n}+10(\vec{S}_2\times\vec{n})^i\vec{v}_2\cdot\vec{n} -9n^i\vec{S}_2\times\vec{v}_1\cdot\vec{n}\right)\right.\nonumber\\
&& \left. ~~~~~-S_2^{0i}\left(5(\vec{S}_1\times\vec{v}_1)^i-5(\vec{S}_1\times\vec{v}_2)^i-20(\vec{S}_1\times\vec{n})^i\vec{v}_1\cdot\vec{n}+7(\vec{S}_1\times\vec{n})^i\vec{v}_2\cdot\vec{n}+13n^i\vec{S}_1\times\vec{v}_2\cdot\vec{n}\right) \right],\\
Fig.~3(b)&=& 2\frac{G^2m_2}{r^4}\left[ \vec{S}_1\cdot\vec{S}_2v_2^2-\vec{S}_1\cdot\vec{v}_2\vec{S}_2\cdot\vec{v}_2-7\vec{S}_1\times\vec{v}_2\cdot\vec{n}\,\vec{S}_2\times\vec{v}_2\cdot\vec{n}+S_2^{0i}\left((\vec{S}_1\times\vec{v}_2)^i-7n^i\vec{S}_1\times\vec{v}_2\cdot\vec{n}\right)\right],\\
Fig.~3(c)&=& -\frac{G^2m_2}{2r^4}\left[ \vec{S}_1\cdot\vec{S}_2v_2^2+4\vec{S}_1\cdot\vec{S}_2\vec{v}_1\cdot\vec{v}_2+\vec{S}_1\cdot\vec{v}_1\vec{S}_2\cdot\vec{v}_2-\vec{S}_1\cdot\vec{v}_2\vec{S}_2\cdot\vec{v}_2+11\vec{S}_1\cdot\vec{S}_2(\vec{v}_2\cdot\vec{n})^2\right.\nonumber\\
&&~~~~~-16\vec{S}_1\cdot\vec{S}_2\vec{v}_1\cdot\vec{n}\vec{v}_2\cdot\vec{n}-13\vec{S}_1\cdot\vec{v}_1\vec{S}_2\cdot\vec{n}\vec{v}_2\cdot\vec{n}+13\vec{S}_1\cdot\vec{v}_2\vec{S}_2\cdot\vec{n}\vec{v}_2\cdot\vec{n}-13\vec{S}_1\cdot\vec{n}\vec{S}_2\cdot\vec{v}_1\vec{v}_2\cdot\vec{n}\nonumber\\
&&~~~~~-4\vec{S}_1\cdot\vec{n}\vec{S}_2\cdot\vec{v}_2\vec{v}_1\cdot\vec{n}+15\vec{S}_1\cdot\vec{n}\vec{S}_2\cdot\vec{v}_2\vec{v}_2\cdot\vec{n}-13\vec{S}_1\cdot\vec{n}\vec{S}_2\cdot\vec{n}\vec{v}_1\cdot\vec{v}_2\nonumber\\
&&\left.~~~~~-63\vec{S}_1\cdot\vec{n}\vec{S}_2\cdot\vec{n}(\vec{v}_2\cdot\vec{n})^2+78\vec{S}_1\cdot\vec{n}\vec{S}_2\cdot\vec{n}\vec{v}_1\cdot\vec{n}\vec{v}_2\cdot\vec{n}\right],\\
Fig.~3(d)&=& -\frac{G^2m_2}{2r^4}\left[ \vec{S}_1\cdot\vec{S}_2v_2^2-4\vec{S}_1\cdot\vec{S}_2\vec{v}_1\cdot\vec{v}_2+4\vec{S}_1\cdot\vec{v}_1\vec{S}_2\cdot\vec{v}_2+5\vec{S}_1\cdot\vec{v}_2\vec{S}_2\cdot\vec{v}_1-\vec{S}_1\cdot\vec{v}_2\vec{S}_2\cdot\vec{v}_2-5\vec{S}_1\cdot\vec{S}_2(\vec{v}_2\cdot\vec{n})^2\right.\nonumber\\
&&~~~~~+16\vec{S}_1\cdot\vec{S}_2\vec{v}_1\cdot\vec{n}\vec{v}_2\cdot\vec{n}-13\vec{S}_1\cdot\vec{v}_1\vec{S}_2\cdot\vec{n}\vec{v}_2\cdot\vec{n}-20\vec{S}_1\cdot\vec{v}_2\vec{S}_2\cdot\vec{n}\vec{v}_1\cdot\vec{n}+7\vec{S}_1\cdot\vec{v}_2\vec{S}_2\cdot\vec{n}\vec{v}_2\cdot\vec{n}\nonumber\\
&&~~~~~-13\vec{S}_1\cdot\vec{n}\vec{S}_2\cdot\vec{v}_1\vec{v}_2\cdot\vec{n}-16\vec{S}_1\cdot\vec{n}\vec{S}_2\cdot\vec{v}_2\vec{v}_1\cdot\vec{n}+5\vec{S}_1\cdot\vec{n}\vec{S}_2\cdot\vec{v}_2\vec{v}_2\cdot\vec{n}-13\vec{S}_1\cdot\vec{n}\vec{S}_2\cdot\vec{n}\vec{v}_1\cdot\vec{v}_2\nonumber\\
&&\left.~~~~~-15\vec{S}_1\cdot\vec{n}\vec{S}_2\cdot\vec{n}(\vec{v}_2\cdot\vec{n})^2+78\vec{S}_1\cdot\vec{n}\vec{S}_2\cdot\vec{n}\vec{v}_1\cdot\vec{n}\vec{v}_2\cdot\vec{n}\right],\\
Fig.~3(e)&=& 2\frac{G^2m_2}{r^4}\left[ 4\vec{S}_1\cdot\vec{S}_2v_2^2-2\vec{S}_1\cdot\vec{v}_2\vec{S}_2\cdot\vec{v}_2-3\vec{S}_1\cdot\vec{S}_2(\vec{v}_2\cdot\vec{n})^2+3\vec{S}_1\cdot\vec{n}\vec{S}_2\cdot\vec{v}_2\vec{v}_2\cdot\vec{n}-6\vec{S}_1\cdot\vec{n}\vec{S}_2\cdot\vec{n}v_2^2\right.\nonumber\\
&&\left.~~~~~-3\vec{S}_1\times\vec{v}_2\cdot\vec{n}\,\vec{S}_2\times\vec{v}_2\cdot\vec{n}\right],\\
Fig.~3(f)&=& -2\frac{G^2m_2}{r^4}\left[ \vec{S}_1\cdot\vec{S}_2v_2^2-\vec{S}_1\cdot\vec{S}_2(\vec{v}_2\cdot\vec{n})^2+\vec{S}_1\cdot\vec{n}\vec{S}_2\cdot\vec{v}_2\vec{v}_2\cdot\vec{n}-\vec{S}_1\cdot\vec{n}\vec{S}_2\cdot\vec{n}v_2^2\right.\nonumber\\
&&\left.~~~~~-3\vec{S}_1\times\vec{v}_2\cdot\vec{n}\,\vec{S}_2\times\vec{v}_2\cdot\vec{n}\right],\\
Fig.~3(g)&=& -2\frac{G^2m_2}{r^4}\left[ \vec{S}_1\cdot\vec{S}_2v_2^2-2\vec{S}_1\cdot\vec{S}_2\vec{v}_1\cdot\vec{v}_2-\vec{S}_1\cdot\vec{v}_1\vec{S}_2\cdot\vec{v}_2-\vec{S}_1\cdot\vec{v}_2\vec{S}_2\cdot\vec{v}_2-3\vec{S}_1\cdot\vec{S}_2(\vec{v}_2\cdot\vec{n})^2\right.\nonumber\\
&&~~~~~+\vec{S}_1\cdot\vec{S}_2\vec{v}_1\cdot\vec{n}\vec{v}_2\cdot\vec{n}-\vec{S}_1\cdot\vec{n}\vec{S}_2\cdot\vec{v}_2\vec{v}_1\cdot\vec{n}+3\vec{S}_1\cdot\vec{n}\vec{S}_2\cdot\vec{v}_2\vec{v}_2\cdot\vec{n}+7\vec{S}_1\cdot\vec{n}\vec{S}_2\cdot\vec{n}\vec{v}_1\cdot\vec{v}_2\nonumber\\
&&\left.~~~~~-5\vec{S}_1\times\vec{v}_2\cdot\vec{n}\,\vec{S}_2\times\vec{v}_1\cdot\vec{n}+S_1^{0i}\left(2(\vec{S}_2\times\vec{v}_2)^i-3(\vec{S}_2\times\vec{n})^i\vec{v}_2\cdot\vec{n}-4n^i\vec{S}_2\times\vec{v}_2\cdot\vec{n}\right)\right],\\
Fig.~3(h)&=& -2\frac{G^2m_2}{r^4}\left[S_1^{0i}\left(S_2^{0i}+(\vec{S}_2\times\vec{v}_2)^i-4n^iS_2^{0j}n^j-4n^i\vec{S}_2\times\vec{v}_2\cdot\vec{n}\right) \right],\\
Fig.~3(i)&=& -\frac{G^2m_2}{r^4}\left[ \vec{S}_1\cdot\vec{S}_2\vec{v}_1\cdot\vec{v}_2+\vec{S}_1\cdot\vec{S}_2\vec{v}_1\cdot\vec{n}\vec{v}_2\cdot\vec{n}-\vec{S}_1\cdot\vec{n}\vec{S}_2\cdot\vec{v}_1\vec{v}_2\cdot\vec{n}-\vec{S}_1\cdot\vec{n}\vec{S}_2\cdot\vec{n}\vec{v}_1\cdot\vec{v}_2\right.\nonumber\\
&&\left.~~~~~+\vec{S}_1\times\vec{v}_2\cdot\vec{n}\,\vec{S}_2\times\vec{v}_1\cdot\vec{n}+S_1^{0i}\left((\vec{S}_2\times\vec{n})^i\vec{v}_2\cdot\vec{n}+2n^i\vec{S}_2\times\vec{v}_2\cdot\vec{n}\right)\right].
\eea

\newpage  %

The cubic self-gravitational interaction diagrams all require the application of one-loop tensor integrals, see Appendix \ref{app:a}. First, one applies the one-loop tensor integrals, which are required here up to order 4, and then the Fourier tensor integrals. Here, we first encounter time derivatives, arising from the spin couplings, propagator correction vertices, and time dependent cubic self-gravitational vertices.

The values of the cubic self-gravitational interaction diagrams are then given by
\bea
Fig.~4(a1)&=& 3\frac{G^2m_2}{r^4}\left[\vec{S}_1\cdot\vec{S}_2-2\vec{S}_1\cdot\vec{n}\,\,\vec{S}_2\cdot\vec{n}\right]v_2^2\nonumber\\
&&-\frac{G^2m_2}{r^4}\left[S_1^{0i}\left((\vec{S}_2\times\vec{v}_1)^i-(\vec{S}_2\times\vec{v}_2)^i+4(\vec{S}_2\times\vec{n})^i\vec{v}_2\cdot\vec{n}-4n^i\vec{S}_2\times\vec{v}_1\cdot\vec{n}\right)\right.\nonumber\\
&&~~~~~-S_2^{0i}\left(4(\vec{S}_1\times\vec{v}_1)^i-4(\vec{S}_1\times\vec{v}_2)^i-16(\vec{S}_1\times\vec{n})^i\vec{v}_1\cdot\vec{n}+4(\vec{S}_1\times\vec{n})^i\vec{v}_2\cdot\vec{n}\right.\nn\\
&&\left.\left.~~~~~+12n^i\vec{S}_1\times\vec{v}_2\cdot\vec{n}\right)\right],\\
Fig.~4(a2)&=& -2\frac{G^2m_2}{r^4}\left[ 3\vec{S}_1\cdot\vec{S}_2v_2^2-3\vec{S}_1\cdot\vec{v}_2\vec{S}_2\cdot\vec{v}_2-4\vec{S}_1\cdot\vec{S}_2(\vec{v}_2\cdot\vec{n})^2+4\vec{S}_1\cdot\vec{v}_2\vec{S}_2\cdot\vec{n}\vec{v}_2\cdot\vec{n}\right.\nonumber\\
&&\left.~~~~~-8\vec{S}_1\times\vec{v}_2\cdot\vec{n}\,\vec{S}_2\times\vec{v}_2\cdot\vec{n}+S_2^{0i}\left(3(\vec{S}_1\times\vec{v}_2)^i-4(\vec{S}_1\times\vec{n})^i\vec{v}_2\cdot\vec{n}-8n^i\vec{S}_1\times\vec{v}_2\cdot\vec{n}\right) \right],\\
Fig.~4(a3)&=& 2\frac{G^2m_2}{r^4}\left[ \vec{S}_1\cdot\vec{S}_2\vec{v}_1\cdot\vec{v}_2-\vec{S}_1\cdot\vec{v}_2\vec{S}_2\cdot\vec{v}_1-4\vec{S}_1\times\vec{v}_1\cdot\vec{n}\,\vec{S}_2\times\vec{v}_2\cdot\vec{n}\right.\nonumber\\
&&\left.~~~~~+S_1^{0i}\left((\vec{S}_2\times\vec{v}_2)^i-4n^i\vec{S}_2\times\vec{v}_2\cdot\vec{n}\right)\right],\\
Fig.~4(a4)&=& \frac{G^2m_2}{2r^4}\left[ 3\vec{S}_1\cdot\vec{S}_2v_2^2+3\vec{S}_1\cdot\vec{S}_2\vec{v}_1\cdot\vec{v}_2+2\vec{S}_1\cdot\vec{v}_1\vec{S}_2\cdot\vec{v}_2-\vec{S}_1\cdot\vec{v}_2\vec{S}_2\cdot\vec{v}_1-3\vec{S}_1\cdot\vec{v}_2\vec{S}_2\cdot\vec{v}_2\right.\nonumber\\
&&~~~~~+4\vec{S}_1\cdot\vec{S}_2(\vec{v}_2\cdot\vec{n})^2-12\vec{S}_1\cdot\vec{S}_2\vec{v}_1\cdot\vec{n}\vec{v}_2\cdot\vec{n}-8\vec{S}_1\cdot\vec{v}_1\vec{S}_2\cdot\vec{n}\vec{v}_2\cdot\vec{n}+4\vec{S}_1\cdot\vec{v}_2\vec{S}_2\cdot\vec{n}\vec{v}_1\cdot\vec{n}\nonumber\\
&&~~~~~+8\vec{S}_1\cdot\vec{v}_2\vec{S}_2\cdot\vec{n}\vec{v}_2\cdot\vec{n}-8\vec{S}_1\cdot\vec{n}\vec{S}_2\cdot\vec{v}_1\vec{v}_2\cdot\vec{n}-8\vec{S}_1\cdot\vec{n}\vec{S}_2\cdot\vec{v}_2\vec{v}_1\cdot\vec{n}+12\vec{S}_1\cdot\vec{n}\vec{S}_2\cdot\vec{v}_2\vec{v}_2\cdot\vec{n}\nonumber\\
&&~~~~~-8\vec{S}_1\cdot\vec{n}\vec{S}_2\cdot\vec{n}\vec{v}_1\cdot\vec{v}_2-4\vec{S}_1\times\vec{v}_2\cdot\vec{n}\,\vec{S}_2\times\vec{v}_2\cdot\vec{n}-36\vec{S}_1\cdot\vec{n}\vec{S}_2\cdot\vec{n}(\vec{v}_2\cdot\vec{n})^2\nonumber\\
&&\left.~~~~~+48\vec{S}_1\cdot\vec{n}\vec{S}_2\cdot\vec{n}\vec{v}_1\cdot\vec{n}\vec{v}_2\cdot\vec{n}\right],\\
Fig.~4(a5)&=& \frac{G^2m_2}{r^4}\left[ \vec{S}_1\cdot\vec{v}_1\vec{S}_2\cdot\vec{v}_2+\vec{S}_1\cdot\vec{v}_2\vec{S}_2\cdot\vec{v}_1-4\vec{S}_1\cdot\vec{v}_1\vec{S}_2\cdot\vec{n}\vec{v}_2\cdot\vec{n}-4\vec{S}_1\cdot\vec{v}_2\vec{S}_2\cdot\vec{n}\vec{v}_1\cdot\vec{n}\right.\nonumber\\
&&~~~~~-4\vec{S}_1\cdot\vec{n}\vec{S}_2\cdot\vec{v}_1\vec{v}_2\cdot\vec{n}-4\vec{S}_1\cdot\vec{n}\vec{S}_2\cdot\vec{v}_2\vec{v}_1\cdot\vec{n}-4\vec{S}_1\cdot\vec{n}\vec{S}_2\cdot\vec{n}\vec{v}_1\cdot\vec{v}_2\nn\\
&&\left.~~~~~+24\vec{S}_1\cdot\vec{n}\vec{S}_2\cdot\vec{n}\vec{v}_1\cdot\vec{n}\vec{v}_2\cdot\vec{n}\right],
\eea

\bea
Fig.~4(a6)&=& \frac{G^2m_2}{2r^4}\left[ \vec{S}_1\cdot\vec{S}_2v_2^2-\vec{S}_1\cdot\vec{S}_2\vec{v}_1\cdot\vec{v}_2+6\vec{S}_1\cdot\vec{v}_1\vec{S}_2\cdot\vec{v}_2+7\vec{S}_1\cdot\vec{v}_2\vec{S}_2\cdot\vec{v}_1-5\vec{S}_1\cdot\vec{v}_2\vec{S}_2\cdot\vec{v}_2\right.\nonumber\\
&&~~~~~-4\vec{S}_1\cdot\vec{S}_2(\vec{v}_2\cdot\vec{n})^2+4\vec{S}_1\cdot\vec{S}_2\vec{v}_1\cdot\vec{n}\vec{v}_2\cdot\vec{n}-24\vec{S}_1\cdot\vec{v}_1\vec{S}_2\cdot\vec{n}\vec{v}_2\cdot\vec{n}-28\vec{S}_1\cdot\vec{v}_2\vec{S}_2\cdot\vec{n}\vec{v}_1\cdot\vec{n}\nonumber\\
&&~~~~~+16\vec{S}_1\cdot\vec{v}_2\vec{S}_2\cdot\vec{n}\vec{v}_2\cdot\vec{n}-24\vec{S}_1\cdot\vec{n}\vec{S}_2\cdot\vec{v}_1\vec{v}_2\cdot\vec{n}-24\vec{S}_1\cdot\vec{n}\vec{S}_2\cdot\vec{v}_2\vec{v}_1\cdot\vec{n}+12\vec{S}_1\cdot\vec{n}\vec{S}_2\cdot\vec{v}_2\vec{v}_2\cdot\vec{n}\nonumber\\
&&~~~~~+8\vec{S}_1\cdot\vec{n}\vec{S}_2\cdot\vec{n}v_2^2-24\vec{S}_1\cdot\vec{n}\vec{S}_2\cdot\vec{n}\vec{v}_1\cdot\vec{v}_2-4\vec{S}_1\times\vec{v}_2\cdot\vec{n}\,\vec{S}_2\times\vec{v}_2\cdot\vec{n}-36\vec{S}_1\cdot\vec{n}\vec{S}_2\cdot\vec{n}(\vec{v}_2\cdot\vec{n})^2\nonumber\\
&&\left.~~~~~+144\vec{S}_1\cdot\vec{n}\vec{S}_2\cdot\vec{n}\vec{v}_1\cdot\vec{n}\vec{v}_2\cdot\vec{n}\right],\\
Fig.~4(b1)&=& -\frac{G^2m_2}{2r^4}\left[ \vec{S}_1\cdot\vec{S}_2\vec{v}_1\cdot\vec{v}_2-\vec{S}_1\cdot\vec{v}_2\vec{S}_2\cdot\vec{v}_1-4\vec{S}_1\times\vec{v}_1\cdot\vec{n}\,\vec{S}_2\times\vec{v}_2\cdot\vec{n}\right.\nonumber\\
&&\left.~~~~~+S_2^{0i}\left((\vec{S}_1\times\vec{v}_1)^i-4n^i\vec{S}_1\times\vec{v}_1\cdot\vec{n}\right)\right],\\
Fig.~4(b2)&=& -\frac{G^2m_2}{2r^4}\left[ \vec{S}_1\cdot\vec{S}_2\vec{v}_1\cdot\vec{v}_2-\vec{S}_1\cdot\vec{v}_2\vec{S}_2\cdot\vec{v}_1-4\vec{S}_1\times\vec{v}_1\cdot\vec{n}\,\vec{S}_2\times\vec{v}_2\cdot\vec{n}\right.\nonumber\\
&&\left.~~~~~+S_1^{0i}\left((\vec{S}_2\times\vec{v}_2)^i-4n^i\vec{S}_2\times\vec{v}_2\cdot\vec{n}\right)\right],\\
Fig.~4(c1)&=& -\frac{G^2m_2}{r^4}\left[ \vec{S}_1\cdot\vec{S}_2v_2^2-5\vec{S}_1\cdot\vec{v}_2\vec{S}_2\cdot\vec{v}_2+8\vec{S}_1\cdot\vec{S}_2(\vec{v}_2\cdot\vec{n})^2+16\vec{S}_1\cdot\vec{v}_2\vec{S}_2\cdot\vec{n}\vec{v}_2\cdot\vec{n}\right.\nonumber\\
&&~~~~~+24\vec{S}_1\cdot\vec{n}\vec{S}_2\cdot\vec{v}_2\vec{v}_2\cdot\vec{n}+8\vec{S}_1\cdot\vec{n}\vec{S}_2\cdot\vec{n}v_2^2-4\vec{S}_1\times\vec{v}_2\cdot\vec{n}\,\vec{S}_2\times\vec{v}_2\cdot\vec{n}\nn\\
&&\left.~~~~~-72\vec{S}_1\cdot\vec{n}\vec{S}_2\cdot\vec{n}(\vec{v}_2\cdot\vec{n})^2\right],\\
Fig.~4(c2)&=& \frac{G^2m_2}{r^4}\left[ 11\vec{S}_1\cdot\vec{S}_2v_2^2-5\vec{S}_1\cdot\vec{v}_2\vec{S}_2\cdot\vec{v}_2-4\vec{S}_1\cdot\vec{S}_2(\vec{v}_2\cdot\vec{n})^2+4\vec{S}_1\cdot\vec{v}_2\vec{S}_2\cdot\vec{n}\vec{v}_2\cdot\vec{n}-12\vec{S}_1\cdot\vec{n}\vec{S}_2\cdot\vec{n}v_2^2\right.\nonumber\\
&&\left.~~~~~-16\vec{S}_1\times\vec{v}_2\cdot\vec{n}\,\vec{S}_2\times\vec{v}_2\cdot\vec{n}\right],\\
Fig.~4(c3)&=& -\frac{G^2m_2}{r^4}\left[ \vec{S}_1\cdot\vec{S}_2\vec{v}_1\cdot\vec{v}_2+\vec{S}_1\cdot\vec{v}_1\vec{S}_2\cdot\vec{v}_2+4\vec{S}_1\cdot\vec{S}_2\vec{v}_1\cdot\vec{n}\vec{v}_2\cdot\vec{n}-4\vec{S}_1\cdot\vec{v}_2\vec{S}_2\cdot\vec{n}\vec{v}_1\cdot\vec{n}\right.\nonumber\\
&&\left.~~~~~-4\vec{S}_1\cdot\vec{n}\vec{S}_2\cdot\vec{n}\vec{v}_1\cdot\vec{v}_2-4\vec{S}_1\times\vec{v}_1\cdot\vec{n}\,\vec{S}_2\times\vec{v}_2\cdot\vec{n}+4\vec{S}_1\times\vec{v}_2\cdot\vec{n}\,\vec{S}_2\times\vec{v}_1\cdot\vec{n}\right],\\
Fig.~4(d1)&=& \frac{G^2m_2}{2r^4}\left[ 2\vec{S}_1\cdot\vec{S}_2v_2^2-\vec{S}_1\cdot\vec{S}_2\vec{v}_1\cdot\vec{v}_2+\vec{S}_1\cdot\vec{v}_2\vec{S}_2\cdot\vec{v}_1-2\vec{S}_1\cdot\vec{v}_2\vec{S}_2\cdot\vec{v}_2-4\vec{S}_1\cdot\vec{S}_2(\vec{v}_2\cdot\vec{n})^2\right.\nonumber\\
&&~~~~~+4\vec{S}_1\cdot\vec{S}_2\vec{v}_1\cdot\vec{n}\vec{v}_2\cdot\vec{n}-4\vec{S}_1\cdot\vec{v}_2\vec{S}_2\cdot\vec{n}\vec{v}_1\cdot\vec{n}+4\vec{S}_1\cdot\vec{v}_2\vec{S}_2\cdot\vec{n}\vec{v}_2\cdot\vec{n}-4\vec{S}_1\times\vec{v}_2\cdot\vec{n}\,\vec{S}_2\times\vec{v}_2\cdot\vec{n}\nonumber\\
&&\left.~~~~~-S_2^{0i}\left((\vec{S}_1\times\vec{v}_1)^i-2(\vec{S}_1\times\vec{v}_2)^i-4(\vec{S}_1\times\vec{n})^i\vec{v}_1\cdot\vec{n}+4(\vec{S}_1\times\vec{n})^i\vec{v}_2\cdot\vec{n}+4n^i\vec{S}_1\times\vec{v}_2\cdot\vec{n}\right)\right],\\
Fig.~4(d2)&=& \frac{G^2m_2}{2r^4}\left[ \vec{S}_1\cdot\vec{S}_2\vec{v}_1\cdot\vec{v}_2-\vec{S}_1\cdot\vec{v}_2\vec{S}_2\cdot\vec{v}_1-4\vec{S}_1\times\vec{v}_1\cdot\vec{n}\,\vec{S}_2\times\vec{v}_2\cdot\vec{n}\right.\nonumber\\
&&\left.~~~~~+S_1^{0i}\left((\vec{S}_2\times\vec{v}_2)^i-4n^i\vec{S}_2\times\vec{v}_2\cdot\vec{n}\right)\right],\\
Fig.~4(e1)&=& -\frac{G^2m_2}{r^4}\left[ \vec{S}_1\cdot\vec{S}_2\vec{v}_1\cdot\vec{v}_2-\vec{S}_1\cdot\vec{v}_1\vec{S}_2\cdot\vec{v}_2-2\vec{S}_1\cdot\vec{v}_2\vec{S}_2\cdot\vec{v}_1+2\vec{S}_1\cdot\vec{v}_2\vec{S}_2\cdot\vec{v}_2+4\vec{S}_1\cdot\vec{S}_2(\vec{v}_2\cdot\vec{n})^2\right.\nonumber\\
&&~~~~~-4\vec{S}_1\cdot\vec{S}_2\vec{v}_1\cdot\vec{n}\vec{v}_2\cdot\vec{n}+4\vec{S}_1\cdot\vec{v}_1\vec{S}_2\cdot\vec{n}\vec{v}_2\cdot\vec{n}+8\vec{S}_1\cdot\vec{v}_2\vec{S}_2\cdot\vec{n}\vec{v}_1\cdot\vec{n}-12\vec{S}_1\cdot\vec{v}_2\vec{S}_2\cdot\vec{n}\vec{v}_2\cdot\vec{n}\nonumber\\
&&~~~~~+4\vec{S}_1\cdot\vec{n}\vec{S}_2\cdot\vec{v}_1\vec{v}_2\cdot\vec{n}+4\vec{S}_1\cdot\vec{n}\vec{S}_2\cdot\vec{v}_2\vec{v}_1\cdot\vec{n}-8\vec{S}_1\cdot\vec{n}\vec{S}_2\cdot\vec{v}_2\vec{v}_2\cdot\vec{n}-4\vec{S}_1\cdot\vec{n}\vec{S}_2\cdot\vec{n}v_2^2\nonumber\\
&&~~~~~+4\vec{S}_1\cdot\vec{n}\vec{S}_2\cdot\vec{n}\vec{v}_1\cdot\vec{v}_2-4\vec{S}_1\times\vec{v}_2\cdot\vec{n}\,\vec{S}_2\times\vec{v}_2\cdot\vec{n}+24\vec{S}_1\cdot\vec{n}\vec{S}_2\cdot\vec{n}(\vec{v}_2\cdot\vec{n})^2\nonumber\\
&&\left.~~~~~-24\vec{S}_1\cdot\vec{n}\vec{S}_2\cdot\vec{n}\vec{v}_1\cdot\vec{n}\vec{v}_2\cdot\vec{n}\right],\\
Fig.~4(e2)&=& -\frac{G^2m_2}{r^4}\left[ \vec{S}_1\cdot\vec{S}_2\vec{v}_1\cdot\vec{v}_2-\vec{S}_1\cdot\vec{v}_2\vec{S}_2\cdot\vec{v}_1-4\vec{S}_1\cdot\vec{S}_2\vec{v}_1\cdot\vec{n}\vec{v}_2\cdot\vec{n}+4\vec{S}_1\cdot\vec{v}_2\vec{S}_2\cdot\vec{n}\vec{v}_1\cdot\vec{n}\right],\\
Fig.~4(f)&=& -\frac{G^2m_2}{r^4}\left[ 2\vec{S}_1\cdot\vec{S}_2v_2^2+2\vec{S}_1\cdot\vec{S}_2\vec{v}_1\cdot\vec{v}_2+3\vec{S}_1\cdot\vec{v}_1\vec{S}_2\cdot\vec{v}_2+\vec{S}_1\cdot\vec{v}_2\vec{S}_2\cdot\vec{v}_1+4\vec{S}_1\cdot\vec{v}_2\vec{S}_2\cdot\vec{v}_2\right.\nonumber\\
&&~~~~~-8\vec{S}_1\cdot\vec{S}_2\vec{v}_1\cdot\vec{n}\vec{v}_2\cdot\vec{n}-12\vec{S}_1\cdot\vec{v}_1\vec{S}_2\cdot\vec{n}\vec{v}_2\cdot\vec{n}-4\vec{S}_1\cdot\vec{v}_2\vec{S}_2\cdot\vec{n}\vec{v}_1\cdot\vec{n}-24\vec{S}_1\cdot\vec{v}_2\vec{S}_2\cdot\vec{n}\vec{v}_2\cdot\vec{n}\nonumber\\
&&~~~~~-12\vec{S}_1\cdot\vec{n}\vec{S}_2\cdot\vec{v}_1\vec{v}_2\cdot\vec{n}-12\vec{S}_1\cdot\vec{n}\vec{S}_2\cdot\vec{v}_2\vec{v}_1\cdot\vec{n}-24\vec{S}_1\cdot\vec{n}\vec{S}_2\cdot\vec{v}_2\vec{v}_2\cdot\vec{n}-12\vec{S}_1\cdot\vec{n}\vec{S}_2\cdot\vec{n}v_2^2\nonumber\\
&&~~~~~-12\vec{S}_1\cdot\vec{n}\vec{S}_2\cdot\vec{n}\vec{v}_1\cdot\vec{v}_2-8\vec{S}_1\times\vec{v}_2\cdot\vec{n}\,\vec{S}_2\times\vec{v}_2\cdot\vec{n}+72\vec{S}_1\cdot\vec{n}\vec{S}_2\cdot\vec{n}(\vec{v}_2\cdot\vec{n})^2\nonumber\\
&&\left.~~~~~+72\vec{S}_1\cdot\vec{n}\vec{S}_2\cdot\vec{n}\vec{v}_1\cdot\vec{n}\vec{v}_2\cdot\vec{n}\right].
\eea

\subsection{Order $G^3$ Feynman diagrams} \label{sec:2loop}

\begin{figure}[ht]
\includegraphics{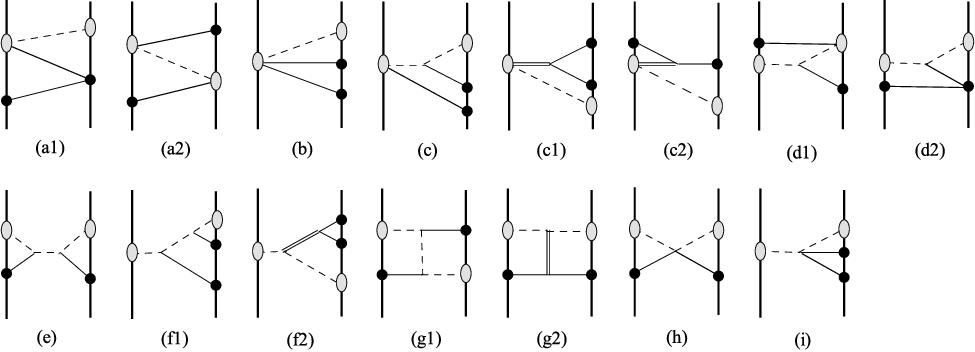}
\caption{NNLO spin1-spin2 Feynman diagrams of order $G^3$. Diagrams (a1), (b), (c), (c1), (c2), (d1), (d2), (f1), (f2), and (i) should be included together with their mirror images.}
\label{fig:4PNG3}
\end{figure}

For the NNLO spin1-spin2 interaction, we also have 15 diagrams at order $G^3$: three are three-graviton exchange diagrams, five are factorizable diagrams with one-loops, and seven are two-loop diagrams, as shown in Fig.~\ref{fig:4PNG3}. All of which are new spin1-spin2 diagrams. The three-graviton exchange diagrams, constructed with either one-, two-, or three-graviton spin couplings, corresponding to diagrams (a1), (a2), and (b), just factorize into a product of three Fourier tensor integrals. We then have diagrams (c), (c1), (c2), (d1), and (d2), which factorize into a product of a one-loop diagram and a one-graviton exchange. 

The values of these simpler diagrams are given by 
\bea
Fig.~5(a1)&=& 2\frac{G^3m_1m_2}{r^5}\left(5\vec{S}_1\cdot\vec{S}_2-13\vec{S}_1\cdot\vec{n}\,\,\vec{S}_2\cdot\vec{n}\right),\\
Fig.~5(a2)&=& \frac{G^3m_1m_2}{r^5}\left(25\vec{S}_1\cdot\vec{S}_2-57\vec{S}_1\cdot\vec{n}\,\,\vec{S}_2\cdot\vec{n}\right),\\
Fig.~5(b)&=& \frac{G^3(m_1^2+m_2^2)}{r^5}\left(11\vec{S}_1\cdot\vec{S}_2-27\vec{S}_1\cdot\vec{n}\,\,\vec{S}_2\cdot\vec{n}\right),\\
Fig.~5(c)&=& -\frac{G^3(m_1^2+m_2^2)}{r^5}\left(9\vec{S}_1\cdot\vec{S}_2-17\vec{S}_1\cdot\vec{n}\,\,\vec{S}_2\cdot\vec{n}\right),\\
Fig.~5(c1)&=& \frac{G^3(m_1^2+m_2^2)}{2r^5}\left(\vec{S}_1\cdot\vec{S}_2-5\vec{S}_1\cdot\vec{n}\,\,\vec{S}_2\cdot\vec{n}\right),\\
Fig.~5(c2)&=& 16\frac{G^3m_1m_2}{r^5}\vec{S}_1\cdot\vec{n}\,\,\vec{S}_2\cdot\vec{n},\\
Fig.~5(d1)&=& -8\frac{G^3m_1m_2}{r^5}\left(3\vec{S}_1\cdot\vec{S}_2-5\vec{S}_1\cdot\vec{n}\,\,\vec{S}_2\cdot\vec{n}\right),\\
Fig.~5(d2)&=& -4\frac{G^3m_1m_2}{r^5}\left(\vec{S}_1\cdot\vec{S}_2-2\vec{S}_1\cdot\vec{n}\,\,\vec{S}_2\cdot\vec{n}\right),
\eea

As for the two-loop diagrams, with two cubic vertices or one quartic vertex, corresponding to diagrams (e) through (i),  they contain two-loop Feynman integrals, where we have to deal with three kinds of two-loop integrals:
\begin{enumerate}
\item \textit{Factorizable} two-loops, as in diagrams (e) and (h), which factorize into a product of two one-loops, such that each one-loop can be performed separately. 
\item \textit{Nested} two-loops, or recursively one-loops, as in diagrams (f1), (f2), and (i), on which a one-loop is nested in a one-loop, so that they should be performed successively, first the nested one-loop, then the outer one-loop.
\item \textit{Irreducible} two-loops, as in diagrams (g1) and (g2), which can be formally reduced, using an integration by parts method \cite{Smirnov:2004ym}, to a sum of integrals of the two previous kinds, i.e.~to a sum of factorizable and nested two-loops. 
\end{enumerate}   
The factorizable two-loop diagrams yield here purely short distance contributions, of the form $\delta^{(2)}(\vec{r})$, which are contact interaction terms that cancel out. For all other two-loop diagrams, calculations should be made, keeping the dimension $d$ general, and the limit $d\to3$ is only taken in the end. For the irreducible two-loop diagrams, which are the most complicated to evaluate here, irreducible two-loop \textit{tensor} integrals of order 4 are encountered, compared to the NNLO orbital interaction, i.e. the 2PN potential, which requires only the irreducible two-loop scalar integral. These are reduced using the integration by parts method \cite{Smirnov:2004ym}, see Appendix \ref{app:a}. The expressions for the irreducible two-loop tensor integrals reductions contain explicit poles in $d=3$, see Appendix \ref{app:a}, but these cancel out in the dimensional regularization.

The values of the two-loop diagrams are then given by
\bea
Fig.~5(e)&=& 0,\\
Fig.~5(f1)&=& \frac{12}{5}\frac{G^3(m_1^2+m_2^2)}{r^5}\left(3\vec{S}_1\cdot\vec{S}_2-5\vec{S}_1\cdot\vec{n}\,\,\vec{S}_2\cdot\vec{n}\right),\\
Fig.~5(f2)&=& \frac{1}{5}\frac{G^3(m_1^2+m_2^2)}{r^5}\left(3\vec{S}_1\cdot\vec{S}_2-5\vec{S}_1\cdot\vec{n}\,\,\vec{S}_2\cdot\vec{n}\right),\\
Fig.~5(g1)&=& 4\frac{G^3m_1m_2}{r^5}\left(\vec{S}_1\cdot\vec{S}_2-3\vec{S}_1\cdot\vec{n}\,\,\vec{S}_2\cdot\vec{n}\right),\\
Fig.~5(g2)&=& 10\frac{G^3m_1m_2}{r^5}\left(\vec{S}_1\cdot\vec{S}_2-5\vec{S}_1\cdot\vec{n}\,\,\vec{S}_2\cdot\vec{n}\right),\\
Fig.~5(h)&=& 0,\\
Fig.~5(i)&=& -\frac{8}{5}\frac{G^3(m_1^2+m_2^2)}{r^5}\left(3\vec{S}_1\cdot\vec{S}_2-5\vec{S}_1\cdot\vec{n}\,\,\vec{S}_2\cdot\vec{n}\right).
\eea

\section{spin1-spin2 Routhian and Hamiltonian at fourth post-Newtonian order}\label{sec:res}

Summing up all of the contributions from the Feynman diagrams, we obtain the NNLO spin1-spin2 interaction Routhian. 
Here, we give the result with the accelerations and precessions dependent terms, and we do not substitute in the $S^{0i}$ dependent terms, using the SSC, so that the $S^{0i}$ entries are left as independent degrees of freedom, and the additional contributions of the field corrections in the $S^{0i}$ entries 
are not taken into account explicitly in this form. We also rearrange terms that should be taken with $1\leftrightarrow2$. Our NNLO spin1-spin2 interaction Routhian for a binary system of compact spinning objects is then given by
\bea
R_{S_1S_2}^{NNLO} &=&-\frac{G}{8r^3}\left[3\vec{S}_1\cdot\vec{S}_2v_1^2v_2^2+2\vec{S}_1\cdot\vec{S}_2(\vec{v}_1\cdot\vec{v}_2)^2-2\vec{S}_1\cdot\vec{v}_1\vec{S}_2\cdot\vec{v}_1v_2^2+4\vec{S}_1\cdot\vec{v}_1\vec{S}_2\cdot\vec{v}_2\vec{v}_1\cdot\vec{v}_2-4\vec{S}_1\cdot\vec{v}_2\vec{S}_2\cdot\vec{v}_1\vec{v}_1\cdot\vec{v}_2\right.\nn\\
&&~~~~~-2\vec{S}_1\cdot\vec{v}_2\vec{S}_2\cdot\vec{v}_2v_1^2-9\vec{S}_1\cdot\vec{S}_2(\vec{v}_1\cdot\vec{n})^2v_2^2-9\vec{S}_1\cdot\vec{S}_2(\vec{v}_2\cdot\vec{n})^2v_1^2-24\vec{S}_1\cdot\vec{S}_2\vec{v}_1\cdot\vec{v}_2\vec{v}_1\cdot\vec{n}\vec{v}_2\cdot\vec{n}\nn\\
&&~~~~~+6\vec{S}_1\cdot\vec{v}_1\vec{S}_2\cdot\vec{v}_1(\vec{v}_2\cdot\vec{n})^2+24\vec{S}_1\cdot\vec{v}_2\vec{S}_2\cdot\vec{v}_1\vec{v}_1\cdot\vec{n}\vec{v}_2\cdot\vec{n}+6\vec{S}_1\cdot\vec{v}_2\vec{S}_2\cdot\vec{v}_2(\vec{v}_1\cdot\vec{n})^2+6\vec{S}_1\cdot\vec{v}_1\vec{S}_2\cdot\vec{n}\vec{v}_1\cdot\vec{n}v_2^2\nn\\
&&~~~~~+6\vec{S}_1\cdot\vec{v}_2\vec{S}_2\cdot\vec{n}\vec{v}_2\cdot\vec{n}v_1^2+6\vec{S}_1\cdot\vec{n}\vec{S}_2\cdot\vec{v}_1\vec{v}_1\cdot\vec{n}v_2^2+6\vec{S}_1\cdot\vec{n}\vec{S}_2\cdot\vec{v}_2\vec{v}_2\cdot\vec{n}v_1^2+3\vec{S}_1\cdot\vec{n}\vec{S}_2\cdot\vec{n}v_1^2v_2^2\nn\\
&&~~~~~-6\vec{S}_1\cdot\vec{n}\vec{S}_2\cdot\vec{n}(\vec{v}_1\cdot\vec{v}_2)^2-12\vec{S}_1\times\vec{v}_1\cdot\vec{v}_2\vec{S}_2\times\vec{v}_1\cdot\vec{n}\vec{v}_2\cdot\vec{n}-12\vec{S}_1\times\vec{v}_1\cdot\vec{v}_2\vec{S}_2\times\vec{v}_2\cdot\vec{n}\vec{v}_1\cdot\vec{n}\nn\\
&&~~~~~-12\vec{S}_1\times\vec{v}_1\cdot\vec{n}\vec{S}_2\times\vec{v}_2\cdot\vec{v}_1\vec{v}_2\cdot\vec{n}-12\vec{S}_1\times\vec{v}_2\cdot\vec{n}\vec{S}_2\times\vec{v}_2\cdot\vec{v}_1\vec{v}_1\cdot\vec{n}-12\vec{S}_1\times\vec{v}_1\cdot\vec{n}\vec{S}_2\times\vec{v}_2\cdot\vec{n}\vec{v}_1\cdot\vec{v}_2\nn\\
&&~~~~~+12\vec{S}_1\times\vec{v}_2\cdot\vec{n}\vec{S}_2\times\vec{v}_1\cdot\vec{n}\vec{v}_1\cdot\vec{v}_2+45\vec{S}_1\cdot\vec{S}_2(\vec{v}_1\cdot\vec{n})^2(\vec{v}_2\cdot\vec{n})^2-30\vec{S}_1\cdot\vec{v}_1\vec{S}_2\cdot\vec{n}\vec{v}_1\cdot\vec{n}(\vec{v}_2\cdot\vec{n})^2\nn\\
&&~~~~~-30\vec{S}_1\cdot\vec{v}_2\vec{S}_2\cdot\vec{n}\vec{v}_2\cdot\vec{n}(\vec{v}_1\cdot\vec{n})^2-30\vec{S}_1\cdot\vec{n}\vec{S}_2\cdot\vec{v}_1\vec{v}_1\cdot\vec{n}(\vec{v}_2\cdot\vec{n})^2-30\vec{S}_1\cdot\vec{n}\vec{S}_2\cdot\vec{v}_2\vec{v}_2\cdot\vec{n}(\vec{v}_1\cdot\vec{n})^2\nn\\
&&~~~~~-15\vec{S}_1\cdot\vec{n}\vec{S}_2\cdot\vec{n}(\vec{v}_1\cdot\vec{n})^2v_2^2-15\vec{S}_1\cdot\vec{n}\vec{S}_2\cdot\vec{n}(\vec{v}_2\cdot\vec{n})^2v_1^2+60\vec{S}_1\times\vec{v}_1\cdot\vec{n}\vec{S}_2\times\vec{v}_2\cdot\vec{n}\vec{v}_1\cdot\vec{n}\vec{v}_2\cdot\vec{n}\nn\\
&&~~~~~-60\vec{S}_1\times\vec{v}_2\cdot\vec{n}\vec{S}_2\times\vec{v}_1\cdot\vec{n}\vec{v}_1\cdot\vec{n}\vec{v}_2\cdot\vec{n}+105\vec{S}_1\cdot\vec{n}\vec{S}_2\cdot\vec{n}(\vec{v}_1\cdot\vec{n})^2(\vec{v}_2\cdot\vec{n})^2\nn\\
&&~~~~~-4\left[S_1^{0i}S_2^{0i}\left(3\vec{v}_1\cdot\vec{v}_2+3\vec{v}_1\cdot\vec{n}\vec{v}_2\cdot\vec{n}\right)+S_1^{0i}S_2^{0j}\left(5v_1^iv_2^j-3v_2^iv_1^j-9v_1^in^j\vec{v}_2\cdot\vec{n}+3v_2^in^j\vec{v}_1\cdot\vec{n}\right.\right.\nn\\
&&~~~~~\left.+3n^iv_1^j\vec{v}_2\cdot\vec{n}-9n^iv_2^j\vec{v}_1\cdot\vec{n}+3n^in^j\vec{v}_1\cdot\vec{v}_2+15n^in^j\vec{v}_1\cdot\vec{n}\vec{v}_2\cdot\vec{n}\right)\nn\\
&&~~~~~-\left(S_1^{0i}\left((\vec{S}_2\times\vec{v}_1)^i\vec{v}_1\cdot\vec{v}_2-(\vec{S}_2\times\vec{v}_1)^iv_2^2-(\vec{S}_2\times\vec{v}_2)^i\vec{v}_1\cdot\vec{v}_2-v_1^i(\vec{S}_2\times\vec{v}_2)\cdot\vec{v}_1+v_2^i(\vec{S}_2\times\vec{v}_2)\cdot\vec{v}_1\right.\right.\nn\\
&&~~~~~-3(\vec{S}_2\times\vec{v}_1)^i\vec{v}_1\cdot\vec{n}\vec{v}_2\cdot\vec{n}+3(\vec{S}_2\times\vec{v}_1)^i(\vec{v}_2\cdot\vec{n})^2+9(\vec{S}_2\times\vec{v}_2)^i\vec{v}_1\cdot\vec{n}\vec{v}_2\cdot\vec{n}+3(\vec{S}_2\times\vec{n})^i\vec{v}_1\cdot\vec{n}v_2^2\nn\\
&&~~~~~-3v_1^i\vec{S}_2\times\vec{v}_1\cdot\vec{n}\vec{v}_2\cdot\vec{n}-3v_1^i\vec{S}_2\times\vec{v}_2\cdot\vec{n}\vec{v}_2\cdot\vec{n}-3v_2^i\vec{S}_2\times\vec{v}_1\cdot\vec{n}\vec{v}_1\cdot\vec{n}+6v_2^i\vec{S}_2\times\vec{v}_1\cdot\vec{n}\vec{v}_2\cdot\vec{n}\nn\\
&&~~~~~+3v_2^i\vec{S}_2\times\vec{v}_2\cdot\vec{n}\vec{v}_1\cdot\vec{n}+3n^i\vec{S}_2\times\vec{v}_2\cdot\vec{v}_1\vec{v}_1\cdot\vec{n}+3n^i\vec{S}_2\times\vec{v}_2\cdot\vec{v}_1\vec{v}_2\cdot\vec{n}-3n^i\vec{S}_2\times\vec{v}_1\cdot\vec{n}\vec{v}_1\cdot\vec{v}_2\nn
\eea
\bea
&&~~~~~+3n^i\vec{S}_2\times\vec{v}_2\cdot\vec{n}\vec{v}_1\cdot\vec{v}_2-15(\vec{S}_2\times\vec{n})^i\vec{v}_1\cdot\vec{n}(\vec{v}_2\cdot\vec{n})^2+15n^i\vec{S}_2\times\vec{v}_1\cdot\vec{n}\vec{v}_1\cdot\vec{n}\vec{v}_2\cdot\vec{n}\nn\\
&&\left.\left.\left.\left.~~~~-15n^i\vec{S}_2\times\vec{v}_2\cdot\vec{n}\vec{v}_1\cdot\vec{n}\vec{v}_2\cdot\vec{n}\right)\right)+(1\longleftrightarrow2)\right]\right]\nn\\
&&-\frac{G}{8r^2}\left[3\vec{S}_1\cdot\vec{S}_2\vec{a}_1\cdot\vec{n}v_2^2+2\vec{S}_1\cdot\vec{S}_2\vec{a}_1\cdot\vec{v}_2\vec{v}_2\cdot\vec{n}-2\vec{S}_1\cdot\vec{v}_2\vec{S}_2\cdot\vec{v}_2\vec{a}_1\cdot\vec{n}+2\vec{S}_1\cdot\vec{a}_1\vec{S}_2\cdot\vec{v}_2\vec{v}_2\cdot\vec{n}+2\vec{S}_1\cdot\vec{n}\vec{S}_2\cdot\vec{v}_2\vec{a}_1\cdot\vec{v}_2\right.\nn\\
&&~~~~~-6\vec{S}_1\cdot\vec{v}_2\vec{S}_2\cdot\vec{a}_1\vec{v}_2\cdot\vec{n}+2\vec{S}_1\cdot\vec{v}_2\vec{S}_2\cdot\vec{n}\vec{a}_1\cdot\vec{v}_2-\vec{S}_1\cdot\vec{a}_1\vec{S}_2\cdot\vec{n}v_2^2-\vec{S}_1\cdot\vec{n}\vec{S}_2\cdot\vec{a}_1v_2^2-4\vec{S}_1\times\vec{v}_2\cdot\vec{a}_1\vec{S}_2\times\vec{v}_2\cdot\vec{n}\nn\\
&&~~~~~+4\vec{S}_1\times\vec{v}_2\cdot\vec{n}\vec{S}_2\times\vec{v}_2\cdot\vec{a}_1-9\vec{S}_1\cdot\vec{S}_2(\vec{v}_2\cdot\vec{n})^2\vec{a}_1\cdot\vec{n}+6\vec{S}_1\cdot\vec{v}_2\vec{S}_2\cdot\vec{n}\vec{v}_2\cdot\vec{n}\vec{a}_1\cdot\vec{n}+6\vec{S}_1\cdot\vec{n}\vec{S}_2\cdot\vec{v}_2\vec{v}_2\cdot\vec{n}\vec{a}_1\cdot\vec{n}\nn\\
&&~~~~~+3\vec{S}_1\cdot\vec{n}\vec{S}_2\cdot\vec{a}_1(\vec{v}_2\cdot\vec{n})^2+3\vec{S}_1\cdot\vec{a}_1\vec{S}_2\cdot\vec{n}(\vec{v}_2\cdot\vec{n})^2+3\vec{S}_1\cdot\vec{n}\vec{S}_2\cdot\vec{n}\vec{a}_1\cdot\vec{n}v_2^2-6\vec{S}_1\cdot\vec{n}\vec{S}_2\cdot\vec{n}\vec{a}_1\cdot\vec{v}_2\vec{v}_2\cdot\vec{n}\nn\\
&&~~~~~-15\vec{S}_1\cdot\vec{n}\vec{S}_2\cdot\vec{n}(\vec{v}_2\cdot\vec{n})^2\vec{a}_1\cdot\vec{n}+12\vec{S}_1\times\vec{n}\cdot\vec{a}_1\vec{S}_2\times\vec{v}_2\cdot\vec{n}\vec{v}_2\cdot\vec{n}-12\vec{S}_1\times\vec{v}_2\cdot\vec{n}\vec{S}_2\times\vec{n}\cdot\vec{a}_1\vec{v}_2\cdot\vec{n}\nn\\
&&~~~~~+4\dot{\vec{S}}_1\cdot\vec{S}_2\vec{v}_2\cdot\vec{n}-4\dot{\vec{S}}_1\cdot\vec{v}_2\vec{S}_2\cdot\vec{n}-4\dot{\vec{S}}_1\cdot\vec{n}\vec{S}_2\cdot\vec{v}_2+12\dot{\vec{S}}_1\cdot\vec{n}\vec{S}_2\cdot\vec{n}\vec{v}_2\cdot\vec{n}\nn\\
&&~~~~~+4\left[S_1^{0i}\left((\vec{S}_2\times\vec{a}_1)^i\vec{v}_2\cdot\vec{n}-v_2^i\vec{S}_2\times\vec{n}\cdot\vec{a}_1-n^i\vec{S}_2\times\vec{v}_2\cdot\vec{a}_1+3n^i\vec{S}_2\times\vec{n}\cdot\vec{a}_1\vec{v}_2\cdot\vec{n}\right)-S_2^{0i}\left((\vec{S}_1\times\vec{v}_2)^i\vec{a}_1\cdot\vec{n}\right.\right.\nn\\
&&~~~~~\left.-(\vec{S}_1\times\vec{n})^i\vec{a}_1\cdot\vec{v}_2+2(\vec{S}_1\times\vec{a}_1)^i\vec{v}_2\cdot\vec{n}+v_2^i\vec{S}_1\times\vec{n}\cdot\vec{a}_1-n^i\vec{S}_1\times\vec{v}_2\cdot\vec{a}_1+2a_1^i\vec{S}_1\times\vec{v}_2\cdot\vec{n}-3(\vec{S}_1\times\vec{n})^i\vec{v}_2\cdot\vec{n}\vec{a}_1\cdot\vec{n}\right.\nn\\
&&~~~~~\left.+3n^i\vec{S}_1\times\vec{n}\cdot\vec{a}_1\vec{v}_2\cdot\vec{n}\right)+\dot{S}_1^{0i}\left(S_2^{0i}\vec{v}_2\cdot\vec{n}+v_2^iS_2^{0j}n^j-3n^iS_2^{0j}v_2^j+3n^iS_2^{0j}n^j\vec{v}_2\cdot\vec{n}-2(\vec{S}_2\times\vec{n})^i+(\vec{S}_2\times\vec{v}_1)^i\vec{v}_2\cdot\vec{n}\right.\nn\\
&&~~~~~-3(\vec{S}_2\times\vec{v}_2)^i\vec{v}_2\cdot\vec{n}-(\vec{S}_2\times\vec{n})^iv_2^2+v_2^i\vec{S}_2\times\vec{v}_1\cdot\vec{n}-v_2^i\vec{S}_2\times\vec{v}_2\cdot\vec{n}-n^i\vec{S}_2\times\vec{v}_2\cdot\vec{v}_1+3(\vec{S}_2\times\vec{n})^i(\vec{v}_2\cdot\vec{n})^2\nn\\
&&\left.\left.\left.~~~~-3n^i\vec{S}_2\times\vec{v}_1\cdot\vec{n}\vec{v}_2\cdot\vec{n}+3n^i\vec{S}_2\times\vec{v}_2\cdot\vec{n}\vec{v}_2\cdot\vec{n}\right)\right]\right]+[1\longleftrightarrow2]\nn\\
&&+\frac{G}{8r}\left[\vec{S}_1\cdot\vec{S}_2\vec{a}_1\cdot\vec{a}_2-3\vec{S}_1\cdot\vec{a}_1\vec{S}_2\cdot\vec{a}_2+5\vec{S}_1\cdot\vec{a}_2\vec{S}_2\cdot\vec{a}_1+3\vec{S}_1\cdot\vec{S}_2\vec{a}_1\cdot\vec{n}\vec{a}_2\cdot\vec{n}-\vec{S}_1\cdot\vec{a}_1\vec{S}_2\cdot\vec{n}\vec{a}_2\cdot\vec{n}-\vec{S}_1\cdot\vec{a}_2\vec{S}_2\cdot\vec{n}\vec{a}_1\cdot\vec{n}\right.\nn\\
&&~~~~~-\vec{S}_1\cdot\vec{n}\vec{S}_2\cdot\vec{a}_1\vec{a}_2\cdot\vec{n}-\vec{S}_1\cdot\vec{n}\vec{S}_2\cdot\vec{a}_2\vec{a}_1\cdot\vec{n}+3\vec{S}_1\cdot\vec{n}\vec{S}_2\cdot\vec{n}\vec{a}_1\cdot\vec{a}_2+4\vec{S}_1\times\vec{n}\cdot\vec{a}_1\vec{S}_2\times\vec{n}\cdot\vec{a}_2-4\vec{S}_1\times\vec{n}\cdot\vec{a}_2\vec{S}_2\times\vec{n}\cdot\vec{a}_1\nn\\
&&~~~~~+3\vec{S}_1\cdot\vec{n}\vec{S}_2\cdot\vec{n}\vec{a}_1\cdot\vec{n}\vec{a}_2\cdot\vec{n}\nn\\
&&~~~~~-4\left[\dot{S}_1^{0i}\dot{S}_2^{0i}+\dot{S}_1^{0i}n^i\dot{S}_2^{0j}n^j\right.\nn\\
&&~~~~~\left.\left.-\dot{S}_1^{0i}\left(2(\vec{S}_2\times\vec{a}_2)^i-(\vec{S}_2\times\vec{n})^i\vec{a}_2\cdot\vec{n}+n^i\vec{S}_2\times\vec{n}\cdot\vec{a}_2\right)-\dot{S}_2^{0i}\left(2(\vec{S}_1\times\vec{a}_1)^i-(\vec{S}_1\times\vec{n})^i\vec{a}_1\cdot\vec{n}+n^i\vec{S}_1\times\vec{n}\cdot\vec{a}_1\right)\right]\right]\nn\\
&&+\frac{Gm_2}{2r}\left[\vec{S}_1\times\vec{v}_2\cdot\vec{a}_1-\vec{S}_1\times\vec{n}\cdot\vec{a}_1\vec{v}_2\cdot\vec{n}+2\frac{m_1}{m_2}\vec{S}_2\times\vec{v}_2\cdot\vec{a}_1-2\frac{m_1}{m_2}\vec{S}_2\times\vec{n}\cdot\vec{a}_1\vec{v}_2\cdot\vec{n}-\dot{\vec{S}}_1\times\vec{v}_1\cdot\vec{v}_2+\dot{\vec{S}}_1\times\vec{v}_1\cdot\vec{n}\vec{v}_2\cdot\vec{n}\right.\nn\\
&&~~~~~\left.-2\dot{\vec{S}}_1\times\vec{v}_2\cdot\vec{n}\vec{v}_2\cdot\vec{n}+3\dot{S}_1^{0i}v_2^i+\dot{S}_1^{0i}n^i\vec{v}_2\cdot\vec{n}\right]+Gm_2\dot{\vec{S}}_1\times\vec{n}\cdot\vec{a}_2+[1\longleftrightarrow2]\nn\\
&&+\frac{1}{8}Gm_1m_2\left[12\vec{a}_1\cdot\vec{v}_1\vec{v}_2\cdot\vec{n}-14\vec{a}_1\cdot\vec{v}_2\vec{v}_2\cdot\vec{n}+\vec{a}_1\cdot\vec{n}v_2^2-\vec{a}_1\cdot\vec{n}(\vec{v}_2\cdot\vec{n})^2-12\vec{a}_2\cdot\vec{v}_2\vec{v}_1\cdot\vec{n}+14\vec{a}_2\cdot\vec{v}_1\vec{v}_1\cdot\vec{n}-\vec{a}_2\cdot\vec{n}v_1^2\right.\nn\\
&&~~~~~\left.+\vec{a}_2\cdot\vec{n}(\vec{v}_1\cdot\vec{n})^2\right]+\frac{1}{8}Gm_1m_2r\left[15\vec{a}_1\cdot\vec{a}_2-\vec{a}_1\cdot\vec{n}\vec{a}_2\cdot\vec{n}\right]\nn\\
&&+\frac{G^2m_2}{2r^4}\left[ 11\vec{S}_1\cdot\vec{S}_2v_2^2+3\vec{S}_1\cdot\vec{v}_1\vec{S}_2\cdot\vec{v}_2+5\vec{S}_1\cdot\vec{v}_2\vec{S}_2\cdot\vec{v}_1-16\vec{S}_1\cdot\vec{v}_2\vec{S}_2\cdot\vec{v}_2-22\vec{S}_1\cdot\vec{S}_2(\vec{v}_2\cdot\vec{n})^2+16\vec{S}_1\cdot\vec{S}_2\vec{v}_1\cdot\vec{n}\vec{v}_2\cdot\vec{n}\right.\nonumber\\
&&~~~~~+2\vec{S}_1\cdot\vec{v}_1\vec{S}_2\cdot\vec{n}\vec{v}_2\cdot\vec{n}-24\vec{S}_1\cdot\vec{v}_2\vec{S}_2\cdot\vec{n}\vec{v}_1\cdot\vec{n}+40\vec{S}_1\cdot\vec{v}_2\vec{S}_2\cdot\vec{n}\vec{v}_2\cdot\vec{n}+2\vec{S}_1\cdot\vec{n}\vec{S}_2\cdot\vec{v}_1\vec{v}_2\cdot\vec{n}
\nonumber\\
&&~~~~~+16\vec{S}_1\cdot\vec{n}\vec{S}_2\cdot\vec{v}_2\vec{v}_2\cdot\vec{n}+7\vec{S}_1\cdot\vec{n}\vec{S}_2\cdot\vec{n}v_2^2-16\vec{S}_1\cdot\vec{n}\vec{S}_2\cdot\vec{n}\vec{v}_1\cdot\vec{v}_2-4\vec{S}_1\times\vec{v}_1\cdot\vec{n}\,\vec{S}_2\times\vec{v}_2\cdot\vec{n}\nonumber\\
&&~~~~~+10\vec{S}_1\times\vec{v}_2\cdot\vec{n}\,\vec{S}_2\times\vec{v}_1\cdot\vec{n}-8\vec{S}_1\times\vec{v}_2\cdot\vec{n}\,\vec{S}_2\times\vec{v}_2\cdot\vec{n}-42\vec{S}_1\cdot\vec{n}\vec{S}_2\cdot\vec{n}(\vec{v}_2\cdot\vec{n})^2-12\vec{S}_1\cdot\vec{n}\vec{S}_2\cdot\vec{n}\vec{v}_1\cdot\vec{n}\vec{v}_2\cdot\vec{n}\nonumber\\
&&~~~~~-4S_1^{0i}S_2^{0i}+16S_1^{0i}n^iS_2^{0j}n^j+2S_1^{0i}\left(2(\vec{S}_2\times\vec{v}_1)^i-6(\vec{S}_2\times\vec{v}_2)^i-2(\vec{S}_2\times\vec{n})^i\vec{v}_1\cdot\vec{n}+11(\vec{S}_2\times\vec{n})^i\vec{v}_2\cdot\vec{n}\right.\nonumber\\
&&~~~~~\left.-5n^i\vec{S}_2\times\vec{v}_1\cdot\vec{n}+6n^i\vec{S}_2\times\vec{v}_2\cdot\vec{n}\right)-2S_2^{0i}\left(2(\vec{S}_1\times\vec{v}_1)^i+2(\vec{S}_1\times\vec{v}_2)^i-6(\vec{S}_1\times\vec{n})^i\vec{v}_1\cdot\vec{n}-3(\vec{S}_1\times\vec{n})^i\vec{v}_2\cdot\vec{n}\right.\nn\\
&&\left.\left.~~~~~-2n^i\vec{S}_1\times\vec{v}_1\cdot\vec{n}+n^i\vec{S}_1\times\vec{v}_2\cdot\vec{n}\right)\right]+[1\longleftrightarrow2]\nn\\ 
&&+\frac{G^3(m_1^2+m_2^2)}{2r^5}\left(11\vec{S}_1\cdot\vec{S}_2-35\vec{S}_1\cdot\vec{n}\,\,\vec{S}_2\cdot\vec{n}\right)+ 3\frac{G^3m_1m_2}{r^5}\left(7\vec{S}_1\cdot\vec{S}_2-27\vec{S}_1\cdot\vec{n}\,\,\vec{S}_2\cdot\vec{n}\right).
\eea

As we already noted, later one should also take into account the SSC dependent parts from the LO and NLO spin-orbit, and NLO spin1-spin2 sectors, appearing in \cite{Levi:2008nh} and \cite{Levi:2010zu}, including in particular also the piece noted in Eq.~(\ref{formalnloso}) here. 

Now we can substitute in the lower order EOM, i.e.~the accelerations and precessions from the orbital and spin interactions, which start contributing at this order, as given in Appendix \ref{app:b}. This is a common procedure for higher-order PN corrections \cite{Schafer:1984, Damour:1985, Damour:1990jh}. 
The time derivatives on the temporal spin components, i.e.~$\dot{S}^{0i}$, are more complicate beyond the leading PN contribution from the spin-orbit sector, since they contain contributions from several sectors at distinct PN orders, as one would expect from just naively differentiating $S^{0i}\sim S v$. Therefore, in a few terms, where they are found, we shifted the time derivatives from them to the other variables, which are physical ones, neglecting total time derivatives. Keeping only possible spin1-spin2 terms up to the PN order considered here, and again rearranging terms that should be taken with $1\leftrightarrow2$, the expression for the NNLO spin1-spin2 Routhian narrows down, and we obtain
\bea \label{eq:fin}
R_{S_1S_2}^{NNLO} 
&=&-\frac{G}{8r^3}\left[3\vec{S}_1\cdot\vec{S}_2v_1^2v_2^2+2\vec{S}_1\cdot\vec{S}_2(\vec{v}_1\cdot\vec{v}_2)^2-2\vec{S}_1\cdot\vec{v}_1\vec{S}_2\cdot\vec{v}_1v_2^2+4\vec{S}_1\cdot\vec{v}_1\vec{S}_2\cdot\vec{v}_2\vec{v}_1\cdot\vec{v}_2-4\vec{S}_1\cdot\vec{v}_2\vec{S}_2\cdot\vec{v}_1\vec{v}_1\cdot\vec{v}_2\right.\nn\\
&&~~~~~-2\vec{S}_1\cdot\vec{v}_2\vec{S}_2\cdot\vec{v}_2v_1^2-9\vec{S}_1\cdot\vec{S}_2(\vec{v}_1\cdot\vec{n})^2v_2^2-9\vec{S}_1\cdot\vec{S}_2(\vec{v}_2\cdot\vec{n})^2v_1^2-24\vec{S}_1\cdot\vec{S}_2\vec{v}_1\cdot\vec{v}_2\vec{v}_1\cdot\vec{n}\vec{v}_2\cdot\vec{n}\nn\\
&&~~~~~+6\vec{S}_1\cdot\vec{v}_1\vec{S}_2\cdot\vec{v}_1(\vec{v}_2\cdot\vec{n})^2+24\vec{S}_1\cdot\vec{v}_2\vec{S}_2\cdot\vec{v}_1\vec{v}_1\cdot\vec{n}\vec{v}_2\cdot\vec{n}+6\vec{S}_1\cdot\vec{v}_2\vec{S}_2\cdot\vec{v}_2(\vec{v}_1\cdot\vec{n})^2+6\vec{S}_1\cdot\vec{v}_1\vec{S}_2\cdot\vec{n}\vec{v}_1\cdot\vec{n}v_2^2\nn\\
&&~~~~~+6\vec{S}_1\cdot\vec{v}_2\vec{S}_2\cdot\vec{n}\vec{v}_2\cdot\vec{n}v_1^2+6\vec{S}_1\cdot\vec{n}\vec{S}_2\cdot\vec{v}_1\vec{v}_1\cdot\vec{n}v_2^2+6\vec{S}_1\cdot\vec{n}\vec{S}_2\cdot\vec{v}_2\vec{v}_2\cdot\vec{n}v_1^2+3\vec{S}_1\cdot\vec{n}\vec{S}_2\cdot\vec{n}v_1^2v_2^2\nn\\
&&~~~~~-6\vec{S}_1\cdot\vec{n}\vec{S}_2\cdot\vec{n}(\vec{v}_1\cdot\vec{v}_2)^2-12\vec{S}_1\times\vec{v}_1\cdot\vec{v}_2\vec{S}_2\times\vec{v}_1\cdot\vec{n}\vec{v}_2\cdot\vec{n}-12\vec{S}_1\times\vec{v}_1\cdot\vec{v}_2\vec{S}_2\times\vec{v}_2\cdot\vec{n}\vec{v}_1\cdot\vec{n}\nn\\
&&~~~~~-12\vec{S}_1\times\vec{v}_1\cdot\vec{n}\vec{S}_2\times\vec{v}_2\cdot\vec{v}_1\vec{v}_2\cdot\vec{n}-12\vec{S}_1\times\vec{v}_2\cdot\vec{n}\vec{S}_2\times\vec{v}_2\cdot\vec{v}_1\vec{v}_1\cdot\vec{n}-12\vec{S}_1\times\vec{v}_1\cdot\vec{n}\vec{S}_2\times\vec{v}_2\cdot\vec{n}\vec{v}_1\cdot\vec{v}_2\nn\\
&&~~~~~+12\vec{S}_1\times\vec{v}_2\cdot\vec{n}\vec{S}_2\times\vec{v}_1\cdot\vec{n}\vec{v}_1\cdot\vec{v}_2+45\vec{S}_1\cdot\vec{S}_2(\vec{v}_1\cdot\vec{n})^2(\vec{v}_2\cdot\vec{n})^2-30\vec{S}_1\cdot\vec{v}_1\vec{S}_2\cdot\vec{n}\vec{v}_1\cdot\vec{n}(\vec{v}_2\cdot\vec{n})^2\nn\\
&&~~~~~-30\vec{S}_1\cdot\vec{v}_2\vec{S}_2\cdot\vec{n}\vec{v}_2\cdot\vec{n}(\vec{v}_1\cdot\vec{n})^2-30\vec{S}_1\cdot\vec{n}\vec{S}_2\cdot\vec{v}_1\vec{v}_1\cdot\vec{n}(\vec{v}_2\cdot\vec{n})^2-30\vec{S}_1\cdot\vec{n}\vec{S}_2\cdot\vec{v}_2\vec{v}_2\cdot\vec{n}(\vec{v}_1\cdot\vec{n})^2\nn\\
&&~~~~~-15\vec{S}_1\cdot\vec{n}\vec{S}_2\cdot\vec{n}(\vec{v}_1\cdot\vec{n})^2v_2^2-15\vec{S}_1\cdot\vec{n}\vec{S}_2\cdot\vec{n}(\vec{v}_2\cdot\vec{n})^2v_1^2+60\vec{S}_1\times\vec{v}_1\cdot\vec{n}\vec{S}_2\times\vec{v}_2\cdot\vec{n}\vec{v}_1\cdot\vec{n}\vec{v}_2\cdot\vec{n}\nn\\
&&~~~~~-60\vec{S}_1\times\vec{v}_2\cdot\vec{n}\vec{S}_2\times\vec{v}_1\cdot\vec{n}\vec{v}_1\cdot\vec{n}\vec{v}_2\cdot\vec{n}+105\vec{S}_1\cdot\vec{n}\vec{S}_2\cdot\vec{n}(\vec{v}_1\cdot\vec{n})^2(\vec{v}_2\cdot\vec{n})^2\nn\\
&&~~~~~-4\left[S_1^{0i}S_2^{0i}\left(3\vec{v}_1\cdot\vec{v}_2+3\vec{v}_1\cdot\vec{n}\vec{v}_2\cdot\vec{n}\right)+S_1^{0i}S_2^{0j}\left(5v_1^iv_2^j-3v_2^iv_1^j-9v_1^in^j\vec{v}_2\cdot\vec{n}\right.\right.\nn\\
&&~~~~~\left.+3v_2^in^j\vec{v}_1\cdot\vec{n}+3n^iv_1^j\vec{v}_2\cdot\vec{n}-9n^iv_2^j\vec{v}_1\cdot\vec{n}+3n^in^j\vec{v}_1\cdot\vec{v}_2+15n^in^j\vec{v}_1\cdot\vec{n}\vec{v}_2\cdot\vec{n}\right)\nn\\
&&~~~~~-\left(S_1^{0i}\left((\vec{S}_2\times\vec{v}_1)^i\vec{v}_1\cdot\vec{v}_2-(\vec{S}_2\times\vec{v}_1)^iv_2^2-(\vec{S}_2\times\vec{v}_2)^i\vec{v}_1\cdot\vec{v}_2-v_1^i(\vec{S}_2\times\vec{v}_2)\cdot\vec{v}_1+v_2^i(\vec{S}_2\times\vec{v}_2)\cdot\vec{v}_1\right.\right.\nn\\
&&~~~~~-3(\vec{S}_2\times\vec{v}_1)^i\vec{v}_1\cdot\vec{n}\vec{v}_2\cdot\vec{n}+3(\vec{S}_2\times\vec{v}_1)^i(\vec{v}_2\cdot\vec{n})^2+9(\vec{S}_2\times\vec{v}_2)^i\vec{v}_1\cdot\vec{n}\vec{v}_2\cdot\vec{n}+3(\vec{S}_2\times\vec{n})^i\vec{v}_1\cdot\vec{n}v_2^2\nn\\
&&~~~~~-3v_1^i\vec{S}_2\times\vec{v}_1\cdot\vec{n}\vec{v}_2\cdot\vec{n}-3v_1^i\vec{S}_2\times\vec{v}_2\cdot\vec{n}\vec{v}_2\cdot\vec{n}-3v_2^i\vec{S}_2\times\vec{v}_1\cdot\vec{n}\vec{v}_1\cdot\vec{n}+6v_2^i\vec{S}_2\times\vec{v}_1\cdot\vec{n}\vec{v}_2\cdot\vec{n}\nn\\
&&~~~~~+3v_2^i\vec{S}_2\times\vec{v}_2\cdot\vec{n}\vec{v}_1\cdot\vec{n}+3n^i\vec{S}_2\times\vec{v}_2\cdot\vec{v}_1\vec{v}_1\cdot\vec{n}+3n^i\vec{S}_2\times\vec{v}_2\cdot\vec{v}_1\vec{v}_2\cdot\vec{n}-3n^i\vec{S}_2\times\vec{v}_1\cdot\vec{n}\vec{v}_1\cdot\vec{v}_2\nn\\
&&~~~~~+3n^i\vec{S}_2\times\vec{v}_2\cdot\vec{n}\vec{v}_1\cdot\vec{v}_2-15(\vec{S}_2\times\vec{n})^i\vec{v}_1\cdot\vec{n}(\vec{v}_2\cdot\vec{n})^2+15n^i\vec{S}_2\times\vec{v}_1\cdot\vec{n}\vec{v}_1\cdot\vec{n}\vec{v}_2\cdot\vec{n}\nn\\
&&\left.\left.\left.\left.~~~~-15n^i\vec{S}_2\times\vec{v}_2\cdot\vec{n}\vec{v}_1\cdot\vec{n}\vec{v}_2\cdot\vec{n}+2(\vec{S}_2\times\vec{v}_1)^i-2(\vec{S}_2\times\vec{v}_2)^i-6(\vec{S}_2\times\vec{n})^i(\vec{v}_1\cdot\vec{n}-\vec{v}_2\cdot\vec{n})\right)\right)+(1\longleftrightarrow2)\right]\right]\nn\\
&&-\frac{Gm_2}{2r^2}S_1^{0i}\left[v_1^i\vec{v}_2\cdot\vec{n}-3v_2^i\vec{v}_1\cdot\vec{n}+2v_2^i\vec{v}_2\cdot\vec{n}+n^i\left(\vec{v}_1\cdot\vec{v}_2-v_2^2-3(\vec{v}_1\cdot\vec{n}\vec{v}_2\cdot\vec{n}-(\vec{v}_2\cdot\vec{n})^2)\right)\right]+[1\longleftrightarrow2]\nn\\
&&+\frac{G^2m_2}{4r^4}\left[ 2\vec{S}_1\cdot\vec{S}_2v_2^2+16\vec{S}_1\cdot\vec{S}_2\vec{v}_1\cdot\vec{v}_2-6\vec{S}_1\cdot\vec{v}_1\vec{S}_2\cdot\vec{v}_2+4\vec{S}_1\cdot\vec{v}_2\vec{S}_2\cdot\vec{v}_1-11\vec{S}_1\cdot\vec{v}_2\vec{S}_2\cdot\vec{v}_2\right.\nonumber\\
&&~~~~~-7\vec{S}_1\cdot\vec{S}_2(\vec{v}_2\cdot\vec{n})^2-42\vec{S}_1\cdot\vec{v}_2\vec{S}_2\cdot\vec{n}\vec{v}_1\cdot\vec{n}+86\vec{S}_1\cdot\vec{v}_2\vec{S}_2\cdot\vec{n}\vec{v}_2\cdot\vec{n}-10\vec{S}_1\cdot\vec{n}\vec{S}_2\cdot\vec{v}_1\vec{v}_2\cdot\vec{n}\nonumber\\
&&~~~~~+14\vec{S}_1\cdot\vec{n}\vec{S}_2\cdot\vec{v}_2\vec{v}_1\cdot\vec{n}+38\vec{S}_1\cdot\vec{n}\vec{S}_2\cdot\vec{v}_2\vec{v}_2\cdot\vec{n}+17\vec{S}_1\cdot\vec{n}\vec{S}_2\cdot\vec{n}v_2^2-30\vec{S}_1\cdot\vec{n}\vec{S}_2\cdot\vec{n}\vec{v}_1\cdot\vec{v}_2\nonumber\\
&&~~~~~-20\vec{S}_1\times\vec{v}_1\cdot\vec{n}\,\vec{S}_2\times\vec{v}_2\cdot\vec{n}+6\vec{S}_1\times\vec{v}_2\cdot\vec{n}\,\vec{S}_2\times\vec{v}_1\cdot\vec{n}+16\vec{S}_1\times\vec{v}_2\cdot\vec{n}\,\vec{S}_2\times\vec{v}_1\cdot\vec{n}\nonumber\\
&&~~~~~-183\vec{S}_1\cdot\vec{n}\vec{S}_2\cdot\vec{n}(\vec{v}_2\cdot\vec{n})^2+42\vec{S}_1\cdot\vec{n}\vec{S}_2\cdot\vec{n}\vec{v}_1\cdot\vec{n}\vec{v}_2\cdot\vec{n}\nn\\
&&~~~~~-2\left[9S_1^{0i}S_2^{0i}-29S_1^{0i}n^iS_2^{0j}n^j-S_1^{0i}\left(4(\vec{S}_2\times\vec{v}_1)^i-10(\vec{S}_2\times\vec{v}_2)^i-4(\vec{S}_2\times\vec{n})^i\vec{v}_1\cdot\vec{n}+22(\vec{S}_2\times\vec{n})^i\vec{v}_2\cdot\vec{n}\right.\right.\nonumber\\
&&~~~~~\left.-10n^i\vec{S}_2\times\vec{v}_1\cdot\vec{n}+5n^i\vec{S}_2\times\vec{v}_2\cdot\vec{n}-(\vec{v}_2\times\vec{n})^i\vec{S}_2\cdot\vec{n}\right)+S_2^{0i}\left(10(\vec{S}_1\times\vec{v}_1)^i-3(\vec{S}_1\times\vec{v}_2)^i\right.\nn\\
&&~~~~~-21(\vec{S}_1\times\vec{n})^i\vec{v}_1\cdot\vec{n}-(\vec{S}_1\times\vec{n})^i\vec{v}_2\cdot\vec{n}-14n^i\vec{S}_1\times\vec{v}_1\cdot\vec{n}+22n^i\vec{S}_1\times\vec{v}_2\cdot\vec{n}-2(\vec{v}_1\times\vec{n})^i\vec{S}_1\cdot\vec{n}\nn\\
&&\left.\left.\left.~~~~~+4(\vec{v}_2\times\vec{n})^i\vec{S}_1\cdot\vec{n}\right)\right]\right]+[1\longleftrightarrow2]\,\,-2\frac{G^2m_1m_2}{r^3}S_1^{0i}n^i+(1\longleftrightarrow2)\nn\\ 
&&+\frac{G^3(m_1^2+m_2^2)}{2r^5}\left(11\vec{S}_1\cdot\vec{S}_2-35\vec{S}_1\cdot\vec{n}\,\,\vec{S}_2\cdot\vec{n}\right)+ \frac{G^3m_1m_2}{2r^5}\left(19\vec{S}_1\cdot\vec{S}_2-99\vec{S}_1\cdot\vec{n}\,\,\vec{S}_2\cdot\vec{n}\right).
\eea
From this, one can derive the NNLO spin1-spin2 Hamiltonian essentially following similar steps to those reported in Sec.~VI of \cite{Levi:2010zu}. These include (1) Legendre transforming with respect to the velocities, (2) substituting in the covariant SSC, including the tetrads corresponding to the metric for a binary of spinning black holes in harmonic coordinates, up to order $G^2$, and (3) mapping the position and spin from ``covariant'' to canonical variables, where these mappings should be extended to higher orders. In fact, it may be possible to eliminate the $S^{i0}$ degrees of freedom using the \textit{covariant} SSC at the level of the action, and get a reduced Routhian result. In a forthcoming paper, 
this derivation will be shown in full detail, and the resulting Hamiltonian will be compared with the NNLO spin1-spin2 Hamiltonian obtained via the ADM canonical formalism in \cite{Hartung:2011ea}. Alternatively, from this result the EOM may be computed directly, using the full spin algebra as in, e.g.~\cite{Yee:1993ya}, see Appendix \ref{app:b}, and then like the steps mentioned above for the obtainment of the Hamiltonian one should (1) substitute in the covariant SSC, including the tetrads up to order $G^2$, and (2) map the position and spin from covariant to canonical variables to get to canonical EOM. 
 
\section{Conclusions}\label{sec:conc}

In this paper, we calculated the NNLO spin1-spin2 potential for a binary of compact spinning objects at the 4PN order. Such high PN orders are required for the successful detection of gravitational radiation. We have performed the calculation using the EFT approach and in terms of the NRG fields. Here, we first demonstrate the ability of the EFT approach to go beyond the NLO in PN corrections of spin effects. The NNLO spin1-spin2 interaction sector includes contributions from 56 diagrams, of which 47 are pure spin1-spin2 diagrams, while a further 9 arise from other sectors, but contribute through the LO spin EOM, that should be taken into account here for the first time, and/or from their SSC dependent parts. Thus, we encounter here several diagrams, that are not pure spin1-spin2 diagrams, but still contribute to the interaction. As for the pure spin1-spin2 diagrams 41 new diagrams appeared here, while 6 others contained new ingredients. 

Two features of the spin couplings present the main difficulties. First, the fact that the spin, formally being a tensor, is derivative-coupled unlike the scalar mass. This calls for higher-order tensor expressions for all  integrals involved in the calculations, which add significantly to the complexity of computations. In particular, for the irreducible two-loop diagrams, which are the most complicated to compute here, irreducible two-loop \textit{tensor} integrals up to order 4 are encountered. Moreover, the derivative-coupling also allows for time derivatives in the worldline spin couplings, which are an additional complication in spin computations. The time derivatives also make the corresponding terms scale at higher PN orders. The second obstacle is the fact that the spin couplings contain $S^{i0}$ entries, which represent the redundant unphysical degrees of freedom related with the spin tensor. These are taken as independent degrees of freedom throughout the calculation, yet eventually, possibly even after the obtainment of the EOM, they are reduced from the final result using some SSC. The $S^{i0}$ entries also yield contributions of higher PN orders with respect to the $S^{ij}$ spin tensor components. Both features make \textit{the PN order of the spin couplings implicit}, compared to the mass couplings, and the power counting, which is essential in the EFT approach, is more difficult with respect to the nonspinning case.

Moreover, unfortunately, not all of the good attributes of the NRG fields pass on to spin interactions. In particular, \textit{all possible diagram topologies are realized at each order of $G$ included}, as was already illustrated in the NLO spin interactions. Also the derivation of the spin couplings is more convenient in terms of the standard Lorentz covariant parametrization, and is not so simple and immediate as for the mass couplings. Still, the NRG fields remain advantageous over other parametrizations, and thus there was no use of automated computations in this work. However, for calculations beyond this order with or without spin effects, it is clear that automated computations utilizing the NRG fields should be implemented, and would be most powerful and efficient. 

Our final result here can be reduced, and a NNLO spin1-spin2 Hamiltonian may be derived from it. In a forthcoming paper, 
this derivation will be shown in full detail, and the resulting Hamiltonian will be compared with the NNLO spin1-spin2 Hamiltonian obtained via the ADM canonical formalism in \cite{Hartung:2011ea}. Alternatively, the EOM may be computed directly from our final result. Future work may shed more light on the equivalence of the ADM canonical formalism and the EFT approach in the treatment of spinning bodies, and may result in improvement in both.

\section*{ACKNOWLEDGMENTS}

I am grateful to Barak Kol for his ongoing support and encouragement. I would like to thank also the other members of the high energy group in the Hebrew University: Shmuel Elitzur, Amit Giveon, and Eliezer Rabinovici, for their special support. I would like to extend special thanks to Adam Schwimmer for his continuous kind hospitality at Weizmann Institute. Further it is with great pleasure that I thank Gerhard Schafer and his group for the warm hospitality, and to Jan Steinhoff and Johannes Hartung for discussions on the results.
This work was supported by the U.S.-Israel Binational Science Foundation, and by the Israel Science Foundation center of excellence.

\appendix
\section{Dimensional regularization and Feynman integrals}\label{app:a}

Throughout the computation of the contributing Feynman diagrams, we encounter in general two types of momentum integrals that need to be evaluated: Fourier integrals that arise from the Fourier transforms of the propagators, and loop integrals, which arise from the cubic and quartic self-gravitational interaction. Both types of integrals are evaluated using dimensional regularization \cite{Smirnov:2004ym}. In order to evaluate the Fourier integrals, one should use the d-dimensional master formula for the scalar integral given by 
\be\label{eq:ft}
I\equiv\int \frac{d^d\bf{k}}{(2\pi)^d}\frac{e^{i\bf{k}\cdot\bf{r}}}{({\bf{k}}^2)^\alpha}=\frac{1}{(4\pi)^{d/2}}\frac{\Gamma(d/2-\alpha)}{\Gamma(\alpha)}\left(\frac{{\bf{r}}^2}{4}\right)^{\alpha-d/2}.
\ee
This formula can easily be derived using Schwinger (or alpha) parameters \cite{Smirnov:2004ym}. 
Differentiating the above with respect to $\bf{r}$, yields the following d-dimensional master formulas for the tensor Fourier integrals: 
\bea
I^i\equiv\int\frac{d^d\bf{k}}{(2\pi)^d}\frac{k^ie^{i\bf{k}\cdot\bf{r}}}{({\bf{k}}^2)^\alpha}&=&\frac{i}{(4\pi)^{d/2}}\frac{\Gamma(d/2-\alpha+1)}{\Gamma(\alpha)}\left(\frac{{\bf{r}}^2}{4}\right)^{\alpha-d/2-1/2}n^i,\\
I^{ij}\equiv\int\frac{d^d\bf{k}}{(2\pi)^d}\frac{k^ik^je^{i\bf{k}\cdot\bf{r}}}{({\bf{k}}^2)^\alpha}&=&\frac{1}{(4\pi)^{d/2}}\frac{\Gamma(d/2-\alpha+1)}{\Gamma(\alpha)}\left(\frac{{\bf{r}}^2}{4}\right)^{\alpha-d/2-1}\left(\frac{1}{2}\delta^{ij}+(\alpha-1-d/2)n^in^j\right),\\
I^{ijl}\equiv\int\frac{d^d\bf{k}}{(2\pi)^d}\frac{k^ik^jk^le^{i\bf{k}\cdot\bf{r}}}{({\bf{k}}^2)^\alpha}&=&\frac{i}{(4\pi)^{d/2}}\frac{\Gamma(d/2-\alpha+2)}{\Gamma(\alpha)}\left(\frac{{\bf{r}}^2}{4}\right)^{\alpha-d/2-3/2}\nn \\
&& \,\,\,\,\,\,\,\,\,\,\,\,\,\,\,\,\,\,\,\,\,\,\,\,\,\,\,\,\,\,\,\,\,\,\,\,\,\,\,\,\,\,\,\,\,\,
\times\left(\frac{1}{2}\left(\delta^{ij}n^l+\delta^{il}n^j+\delta^{jl}n^i\right)+(\alpha-d/2-2)n^in^jn^l\right),\\
I^{ijlm}\equiv\int\frac{d^d\bf{k}}{(2\pi)^d}\frac{k^ik^jk^lk^me^{i\bf{k}\cdot\bf{r}}}{({\bf{k}}^2)^\alpha}&=&\frac{1}{(4\pi)^{d/2}}\frac{\Gamma(d/2-\alpha+2)}{\Gamma(\alpha)}\left(\frac{{\bf{r}}^2}{4}\right)^{\alpha-d/2-2}\left(\frac{1}{4}\left(\delta^{ij}\delta^{lm}+\delta^{il}\delta^{jm}+\delta^{im}\delta^{jl}\right)\right.\nn \\
&& 
+\frac{\alpha-d/2-2}{2}\left(\delta^{ij}n^ln^m+\delta^{il}n^jn^m+\delta^{im}n^jn^l+\delta^{jl}n^in^m+\delta^{jm}n^in^l+\delta^{lm}n^in^j\right)\nn \\
&&\left.\,\,\,\,\,\,\,\,\,\,\,\,\,\,\,\,\,\,\,\,\,\,\,\,\,\,\,\,\,\,\,\,\,\,\,\,\,\,\,\,\,\,\,\,\,\,\,\,\,\,\,\,\,\,\,\,\,\,\,\,\,\,\,\,\,\,\,\,\,\,\,\,\,\,\,\,\,\,\,\,
+(\alpha-d/2-2)(\alpha-d/2-3)n^in^jn^ln^m\right).
\eea
Actually, for this paper, the tensor Fourier integrals of orders 5 and 6 were required. We did not include here the lengthy expressions for them, containing 26 and 76 generic terms, respectively.

In order to evaluate the loop integrals, one should use the d-dimensional master formula for one-loop scalar integrals given by 
\be
J\equiv\int \frac{d^d\bf{k}}{(2\pi)^d}\frac{1}{\left[{\bf{k}}^2\right]^\alpha\left[({\bf{k}-\bf{q}})^2\right]^\beta}=  \frac{1}{(4\pi)^{d/2}}\frac{\Gamma(\alpha+\beta-d/2)}{\Gamma(\alpha)\Gamma(\beta)}\frac{\Gamma(d/2-\alpha)\Gamma(d/2-\beta)}{\Gamma(d-\alpha-\beta)}\left(q^2\right)^{d/2-\alpha-\beta}.\label{eq:1l0}
\ee
This formula can easily be derived using Feynman and Schwinger parameters \cite{Smirnov:2004ym}. 
Similarly, one can also derive the following d-dimensional master formulae for the one-loop tensor integrals: 
\bea
J^i\equiv\int \frac{d^d\bf{k}}{(2\pi)^d}\frac{k^i}{\left[{\bf{k}}^2\right]^\alpha\left[({\bf{k}-\bf{q}})^2\right]^\beta}&=&  \frac{1}{(4\pi)^{d/2}}\frac{\Gamma(\alpha+\beta-d/2)}{\Gamma(\alpha)\Gamma(\beta)}\frac{\Gamma(d/2-\alpha+1)\Gamma(d/2-\beta)}{\Gamma(d-\alpha-\beta+1)}\left(q^2\right)^{d/2-\alpha-\beta}q^i,\\
J^{ij}\equiv\int \frac{d^d\bf{k}}{(2\pi)^d}\frac{k^ik^j}{\left[{\bf{k}}^2\right]^\alpha\left[({\bf{k}-\bf{q}})^2\right]^\beta}&=&  \frac{1}{(4\pi)^{d/2}}\frac{\Gamma(\alpha+\beta-d/2-1)}{\Gamma(\alpha)\Gamma(\beta)}\frac{\Gamma(d/2-\alpha+1)\Gamma(d/2-\beta)}{\Gamma(d-\alpha-\beta+2)}\left(q^2\right)^{d/2-\alpha-\beta}\nn\\
&& \,\,\,\,\,\,\,\,\,\,\,\,\,\,\,\,\,\,\,\,\,\,
\times\left(\frac{d/2-\beta}{2}q^2\delta^{ij}+(\alpha+\beta-d/2-1)(d/2-\alpha+1)q^iq^j\right),\\
J^{ijl}\equiv\int \frac{d^d\bf{k}}{(2\pi)^d}\frac{k^ik^jk^l}{\left[{\bf{k}}^2\right]^\alpha\left[({\bf{k}-\bf{q}})^2\right]^\beta}&=&  \frac{1}{(4\pi)^{d/2}}\frac{\Gamma(\alpha+\beta-d/2-1)}{\Gamma(\alpha)\Gamma(\beta)}\frac{\Gamma(d/2-\alpha+2)\Gamma(d/2-\beta)}{\Gamma(d-\alpha-\beta+3)}\left(q^2\right)^{d/2-\alpha-\beta}\nn\\
&& 
\times\left(\frac{d/2-\beta}{2}q^2\left(\delta^{ij}q^l+\delta^{il}q^j+\delta^{jl}q^i\right)\right.\nn \\
&& \left. \,\,\,\,\,\,\,\,\,
+(\alpha+\beta-d/2-1)(d/2-\alpha+2)q^iq^jq^l\right),
\eea

\bea
J^{ijlm}\equiv\int \frac{d^d\bf{k}}{(2\pi)^d}\frac{k^ik^jk^lk^m}{\left[{\bf{k}}^2\right]^\alpha\left[({\bf{k}-\bf{q}})^2\right]^\beta}&=&  \frac{1}{(4\pi)^{d/2}}\frac{\Gamma(\alpha+\beta-d/2-2)}{\Gamma(\alpha)\Gamma(\beta)}\frac{\Gamma(d/2-\alpha+2)\Gamma(d/2-\beta)}{\Gamma(d-\alpha-\beta+4)}\left(q^2\right)^{d/2-\alpha-\beta}\nn\\
&& 
\times\left(\frac{(d/2-\beta)(d/2-\beta+1)}{4}q^4\left(\delta^{ij}\delta^{lm}+\delta^{il}\delta^{jm}+\delta^{jl}\delta^{im}\right)\right.\nn\\
&& \,\,\,\,\,\,\,\,\,
+(\alpha+\beta-d/2-2)(d/2-\alpha+2)\frac{d/2-\beta}{2}q^2\nn\\
&& \,\,\,\,\,\,\,\,\,\,\,\,\,\,
\times\left(\delta^{ij}q^lq^m+\delta^{il}q^jq^m+\delta^{im}q^jq^l+\delta^{jl}q^iq^m+\delta^{jm}q^iq^l+\delta^{lm}q^iq^j\right)\nn\\
&&\left.\,\,\,\,\,\,\,\,\,
+(\alpha+\beta-d/2-2)(\alpha+\beta-d/2-1)(d/2-\alpha+2)(d/2-\alpha+3)\right.\nn\\
&&\left.\,\,\,\,\,\,\,\,\,\,\,\,\,\,
\times q^iq^jq^lq^m\right).
\eea

In addition, we encounter irreducible two-loop tensor integrals up to order 4. These can be reduced, using an integration by parts method \cite{Smirnov:2004ym}, to a sum of factorizable and nested two-loops, as explained in Sec.~\ref{sec:2loop}. The required irreducible two-loop tensor integrals reductions are given by
\bea
\int_{\bf{k_1}\bf{k_2}} \frac{k_1^ik_2^j}{k_1^2 \left(p-k_1\right)^2 k_2^2 \left(p-k_2\right)^2 \left(k_1-k_2\right)^2}&=&\frac{1}{d-3}\int_{\bf{k_1}\bf{k_2}} \left[\frac{p^ik_2^j}{k_1^4 \left(p-k_1\right)^2 k_2^2 \left(p-k_2\right)^2}-\frac{k_1^ik_2^j}{k_1^4 \left(p-k_1\right)^2 \left(p-k_2\right)^2\left(k_1-k_2\right)^2}\right.\nn\\
&&\left.-\frac{k_1^ik_2^j}{k_1^2 \left(p-k_1\right)^4 k_2^2 \left(k_1-k_2\right)^2}+ \frac{1}{d-4} \left(2\frac{k_2^ik_2^j}{k_1^4 \left(p-k_1\right)^2 k_2^2 \left(p-k_2\right)^2}\right.\right.\nn\\
&&\left.\left.-\frac{k_2^ik_2^j}{k_1^4 \left(p-k_1\right)^2 \left(p-k_2\right)^2 \left(k_1-k_2\right)^2}-\frac{k_2^ik_2^j}{k_1^2 \left(p-k_1\right)^4 k_2^2 \left(k_1-k_2\right)^2}\right)\right],\\
\int_{\bf{k_1}\bf{k_2}} \frac{k_1^ik_1^jk_2^l}{k_1^2 \left(p-k_1\right)^2 k_2^2 \left(p-k_2\right)^2 \left(k_1-k_2\right)^2}&=&\frac{1}{d-3}\int_{\bf{k_1}\bf{k_2}} \left[\frac{p^lk_1^ik_1^j}{k_1^2 \left(p-k_1\right)^2 k_2^4 \left(p-k_2\right)^2}-\frac{k_1^ik_1^jk_2^l}{ \left(p-k_1\right)^2 k_2^4 \left(p-k_2\right)^2\left(k_1-k_2\right)^2}\right.\nn\\
&&\left.-\frac{k_1^ik_1^jk_2^l}{k_1^2 k_2^2 \left(p-k_2\right)^4\left(k_1-k_2\right)^2}+ \frac{1}{d-4} \left(2\frac{k_1^ik_1^jk_1^l}{k_1^2 \left(p-k_1\right)^2 k_2^4 \left(p-k_2\right)^2}\right.\right.\nn\\
&&\left.\left.-\frac{k_1^ik_1^jk_1^l}{\left(p-k_1\right)^2 k_2^4 \left(p-k_2\right)^2 \left(k_1-k_2\right)^2}-\frac{k_1^ik_1^jk_1^l}{k_1^2 k_2^2 \left(p-k_2\right)^4 \left(k_1-k_2\right)^2}\right)\right],\\
\int_{\bf{k_1}\bf{k_2}} \frac{k_1^ik_1^jk_2^lk_2^m}{k_1^2 \left(p-k_1\right)^2 k_2^2 \left(p-k_2\right)^2 \left(k_1-k_2\right)^2}&=&\frac{1}{d-2}\int_{\bf{k_1}\bf{k_2}} \left[\frac{k_1^ik_1^jk_2^lk_2^m}{k_1^4 \left(p-k_1\right)^2 k_2^2 \left(p-k_2\right)^2}+\frac{k_1^ik_1^jk_2^lk_2^m}{k_1^2 \left(p-k_1\right)^4 k_2^2 \left(p-k_2\right)^2}\right.\nn\\
&&-\frac{k_1^ik_1^jk_2^lk_2^m}{k_1^4 \left(p-k_1\right)^2 \left(p-k_2\right)^2\left(k_1-k_2\right)^2}-\frac{k_1^ik_1^jk_2^lk_2^m}{k_1^2 \left(p-k_1\right)^4 k_2^2 \left(k_1-k_2\right)^2}\nn\\
&&+\frac{1}{d-3}\left(\frac{p^ik_2^jk_2^lk_2^m}{k_1^4 \left(p-k_1\right)^2 k_2^2 \left(p-k_2\right)^2}+\frac{p^jk_2^ik_2^lk_2^m}{k_1^4 \left(p-k_1\right)^2 k_2^2 \left(p-k_2\right)^2}\right.\nn\\
&&-\frac{k_1^ik_2^jk_2^lk_2^m}{k_1^4 \left(p-k_1\right)^2 \left(p-k_2\right)^2\left(k_1-k_2\right)^2}-\frac{k_1^jk_2^ik_2^lk_2^m}{k_1^4 \left(p-k_1\right)^2 \left(p-k_2\right)^2\left(k_1-k_2\right)^2}\nn\\
&&-\frac{k_1^ik_2^jk_2^lk_2^m}{k_1^2 \left(p-k_1\right)^4 k_2^2 \left(k_1-k_2\right)^2}-\frac{k_1^jk_2^ik_2^lk_2^m}{k_1^2 \left(p-k_1\right)^4 k_2^2 \left(k_1-k_2\right)^2}\nn\\
&&\left.+ \frac{2}{d-4} \left(2\frac{k_2^ik_2^jk_2^lk_2^m}{k_1^4 \left(p-k_1\right)^2 k_2^2 \left(p-k_2\right)^2}-\frac{k_2^ik_2^jk_2^lk_2^m}{k_1^4 \left(p-k_1\right)^2 \left(p-k_2\right)^2 \left(k_1-k_2\right)^2}\right.\right.\nn\\
&&\left.\left.\left.-\frac{k_2^ik_2^jk_2^lk_2^m}{k_1^2 \left(p-k_1\right)^4 k_2^2 \left(k_1-k_2\right)^2}\right)\right)\right].
\eea
It should be noted that these expressions contain explicit poles in $d=3$, but these cancel out in the dimensional regularization. 

\section{Accelerations and Precessions}\label{app:b}

In the diagrams of Figs.~\ref{fig:4PNG} and \ref{fig:4PNGeom}, there appear spin-orbit accelerations $\vec{a}_{(SO)}$ and precessions $\dot{\vec{S}}_{(SO)}$, $\dot{S}_{(SO)}^{0i}$, and spin1-spin2 precessions $\dot{\vec{S}}_{(S_1S_2)}$,  $\dot{S}_{(S_1S_2)}^{0i}$, that should be substituted in the spin1-spin2, and spin-orbit diagram values, respectively. LO spin1-spin2 accelerations $\vec{a}_{(S_1S_2)}$ should also be substituted in the orbital interaction diagrams. This is in addition to the required substitution of the Newtonian acceleration $\vec{a}_{(N)}$ (in principle, the 1PN acceleration $\vec{a}_{(1PN)}$ is also required here). We note that the time derivatives on the temporal spin components, i.e.~$\dot{S}^{0i}$, are more complicate beyond the leading PN contribution from the spin-orbit sector, since they contain contributions from several sectors at distinct PN orders, as one would expect from just naively differentiating $S^{0i}\sim S v$. Therefore, in a few terms where they are found, we shifted the time derivatives from them to the other variables, which are physical ones, neglecting total time derivatives. Thus, the required EOM of spin effects are derived here from the LO spin-orbit Routhian in \cite{Levi:2010zu}, and the LO spin1-spin2 Routhian in \cite{Levi:2008nh}. The accelerations are derived from the Euler-Lagrange equations or by applying the variation principle on the Routhian, namely,
\be
\frac{\delta \int dt\,\cal{R}}{\delta x^\mu}=0.\nn
\ee
The accelerations required for the NNLO spin1-spin2 potential are then:
\bea
\vec{a}_{1(N)}&=&-\frac{Gm_2}{r^2}\vec{n},\\
\vec{a}_{1(S_1S_2)}&=&-\frac{3~G}{m_1r^4}\left[\left(\vec{S}_1\cdot\vec{S}_2-5\vec{S}_1\cdot\vec{n}\vec{S}_2\cdot\vec{n}\right)\vec{n}+\vec{S}_1\cdot\vec{n}\vec{S}_2+\vec{S}_2\cdot\vec{n}\vec{S}_1\right],\\
a_{1(SO)}^i=&&\frac{G}{r^3}~\frac{m_2}{m_1}\left[2(\vec{S}_1\times\vec{v}_1)^i-3(\vec{S}_1\times\vec{v}_2)^i-3(\vec{S}_1\times\vec{n})^i\left(\vec{v}_1\cdot\vec{n}-\vec{v}_2\cdot\vec{n}\right)-3n^i\left(\vec{S}_1\times\vec{v}_1\cdot\vec{n}-2\vec{S}_1\times\vec{v}_2\cdot\vec{n}\right)\right.\nn\\
&&~~~~~\left.+S_1^{0i}-3n^iS_1^{0j}n^j\right]\nn\\
&&+\frac{G}{r^3}\left[4(\vec{S}_2\times\vec{v}_1)^i-3(\vec{S}_2\times\vec{v}_2)^i-6(\vec{S}_2\times\vec{n})^i\left(\vec{v}_1\cdot\vec{n}-\vec{v}_2\cdot\vec{n}\right)-3n^i\left(2\vec{S}_2\times\vec{v}_1\cdot\vec{n}-\vec{S}_2\times\vec{v}_2\cdot\vec{n}\right)\right.\nn\\
&&~~~~~\left.-S_2^{0i}+3n^iS_2^{0j}n^j\right],
\eea
where we have written them for particle 1, and for particle 2, the equations should just be taken with the particle labels exchanged, i.e.~$1\leftrightarrow2$. For the last equation here the vanishing of the spin precession ${\dot{\vec{S}}}$ at Newtonian order is used.

The precessions are obtained using Hamilton's equations for the Routhian \cite{Yee:1993ya}, namely:
\bdm
\begin{array}{cc}
\dot{\vec{S}}=\{\vec{S},{\cal{R}}\}, 
& \dot{S}^{0i}=\{S^{0i},{\cal{R}}\},
\end{array}
\edm
where the reduced spin algebra is just obtained from the full spin algebra, e.g.~in \cite{Yee:1993ya}, such that:
\bea
\{S^i,S^j\}\,\,&=&-\epsilon^{ijk}S^k,\nn\\
\{S^i,S^{0j}\}\,&=&-\epsilon^{ijk}S^{0k},\nn\\
\{S^{0i},S^{0j}\}&=&\,\,\,\,\,\epsilon^{ijk}S^{k}.
\eea

The precessions required for the NNLO spin1-spin2 potential are (also given for particle 1, with $1\leftrightarrow2$ for particle 2):
\bea
\dot{\vec{S}}_{1(S_1S_2)}&=&\frac{G}{r^3}\left(\vec{S}_1\times\vec{S}_2-3\vec{S}_1\times\vec{n}\vec{S}_2\cdot\vec{n}\right),\\
\dot{S}_{1(SO)}^i&=&\frac{Gm_2}{r^2}\left[\vec{S}_1\cdot\vec{n}\left(\vec{v}_1-2\vec{v}_2\right)^i-\vec{S}_1\cdot\left(\vec{v}_1-2\vec{v}_2\right)n^i+\epsilon_{ijk}S_1^{0j}n^k\right],\\
\dot{S}_{1(SO)}^{0i}&=&-\frac{Gm_2}{r^2}(\vec{S}_1\times\vec{n})^i,
\eea
where in the last equation for $\dot{S}_{1(SO)}^{0i}$ only the leading PN order contribution is considered here.


\begin{thebibliography}{99} 

\bibitem{ligo}
  LIGO webpage
	\url{http://www.ligo.caltech.edu/}. 
	
\bibitem{virgo}
  Virgo webpage
	\url{http://www.virgo.infn.it/}. 
	
\bibitem{geo600}
  GEO 600 webpage
	\url{http://www.geo600.org}. 
	
\bibitem{lcgt}
  LCGT webpage
	\url{http://gw.icrr.u-tokyo.ac.jp/lcgt}. 

\bibitem{lisa}
	ESA LISA webpage,
	\url{http://sci.esa.int/science-e/www/area/index.cfm?fareaid=27};\\
	NASA LISA webpage
	\url{http://lisa.jpl.nasa.gov/}. 
		
\bibitem{Blanchet:2002av}
  L.~Blanchet,
  ``Gravitational radiation from post-Newtonian sources and inspiralling compact binaries,''
  Living Rev.\ Rel.\  {\bf 9}, 4 (2006)
  \url{http://www.livingreviews.org/lrr-2006-4}.  
  
\bibitem{Foffa:2011ub}
  S.~Foffa and R.~Sturani,
  ``Effective field theory calculation of conservative binary dynamics at third post-Newtonian order,''
  Phys.\ Rev.\  D {\bf 84}, 044031 (2011)
  [arXiv:1104.1122 [gr-qc]].
  
\bibitem{Reisswig:2009vc}
  C.~Reisswig, S.~Husa, L.~Rezzolla, E.~N.~Dorband, D.~Pollney, J.~Seiler,
  ``Gravitational-wave detectability of equal-mass black-hole binaries with aligned spins,''
  Phys.\ Rev.\  {\bf D80}, 124026 (2009).
  [arXiv:0907.0462 [gr-qc]].   
      
\bibitem{Goldberger:2004jt}
  W.~D.~Goldberger and I.~Z.~Rothstein,
  ``An effective field theory of gravity for extended objects,''
  Phys.\ Rev.\  D {\bf 73}, 104029 (2006)
  [arXiv:hep-th/0409156].

\bibitem{Goldberger:2007hy}
  W.~D.~Goldberger,
  ``Les Houches lectures on effective field theories and gravitational radiation,''
	arXiv:hep-ph/0701129.
     
\bibitem{Gilmore:2008gq}
  J.~B.~Gilmore and A.~Ross,
  ``Effective field theory calculation of second post-Newtonian binary dynamics,''
  Phys.\ Rev.\  D {\bf 78}, 124021 (2008)
  [arXiv:0810.1328 [gr-qc]].
  
\bibitem{Chu:2008xm}
  Y.~Z.~Chu,
  ``The n-body problem in General Relativity up to the second post-Newtonian
  order from perturbative field theory,''
  Phys.\ Rev.\  D {\bf 79}, 044031 (2009)
  [arXiv:0812.0012 [gr-qc]].
  
\bibitem{Porto:2005ac}
  R.~A.~Porto,
  ``Post-Newtonian corrections to the motion of spinning bodies in NRGR,''
  Phys.\ Rev.\  D {\bf 73}, 104031 (2006)
  [arXiv:gr-qc/0511061].

\bibitem{Porto:2006bt}
  R.~A.~Porto and I.~Z.~Rothstein,
  ``The hyperfine Einstein-Infeld-Hoffmann potential,''
  Phys.\ Rev.\ Lett.\  {\bf 97}, 021101 (2006)
  [arXiv:gr-qc/0604099].

\bibitem{Porto:2007tt}
  R.~A.~Porto and I.~Z.~Rothstein,
  ``Comment on `On the next-to-leading order gravitational spin(1)-spin(2) dynamics' by J. Steinhoff et al,''
  arXiv:0712.2032 [gr-qc].
  
\bibitem{Levi:2008nh}
  M.~Levi,
  ``Next-to-leading order gravitational spin1-spin2 coupling with Kaluza-Klein reduction,''
  Phys.\ Rev.\  D {\bf 82}, 064029 (2010)
  [arXiv:0802.1508 [gr-qc]].
  
\bibitem{Porto:2008tb}
  R.~A.~Porto and I.~Z.~Rothstein,
  ``Spin(1)Spin(2) Effects in the Motion of Inspiralling Compact Binaries at
  Third Order in the Post-Newtonian Expansion,''
  Phys.\ Rev.\  D {\bf 78}, 044012 (2008)
  [Erratum-ibid.\  D {\bf 81}, 029904 (2010)]
  [arXiv:0802.0720 [gr-qc]].
  
\bibitem{Steinhoff:2007mb}
  J.~Steinhoff, S.~Hergt and G.~Schaefer,
  ``On the next-to-leading order gravitational spin(1)-spin(2) dynamics,''
  Phys.\ Rev.\  D {\bf 77}, 081501 (2008)
  [arXiv:0712.1716 [gr-qc]].
    
\bibitem{Porto:2008jj}
  R.~A.~Porto and I.~Z.~Rothstein,
  ``Next to Leading Order Spin(1)Spin(1) Effects in the Motion of Inspiralling Compact Binaries,''
  Phys.\ Rev.\  D {\bf 78}, 044013 (2008)
  [Erratum-ibid.\  D {\bf 81}, 029905 (2010)]
  [arXiv:0804.0260 [gr-qc]].
  
\bibitem{Hergt:2008jn}
  S.~Hergt and G.~Schafer,
  ``Higher-order-in-spin interaction Hamiltonians for binary black holes from Poincar\'e invariance,''
  Phys.\ Rev.\  D {\bf 78}, 124004 (2008)
  [arXiv:0809.2208 [gr-qc]].

\bibitem{Steinhoff:2008ji}
  J.~Steinhoff, S.~Hergt and G.~Schafer,
  ``Spin-squared Hamiltonian of next-to-leading order gravitational interaction,''
  Phys.\ Rev.\  D {\bf 78}, 101503 (2008)
  [arXiv:0809.2200 [gr-qc]].
  
\bibitem{Hergt:2010pa}
  S.~Hergt, J.~Steinhoff and G.~Schaefer,
  ``Reduced Hamiltonian for next-to-leading order Spin-Squared Dynamics of General Compact Binaries,''
  Class.\ Quant.\ Grav.\  {\bf 27}, 135007 (2010)
  [arXiv:1002.2093 [gr-qc]].
  
\bibitem{Perrodin:2010dy}
  D.~L.~Perrodin,
  ``Subleading Spin-Orbit Correction to the Newtonian Potential in Effective Field Theory Formalism,''
  arXiv:1005.0634 [gr-qc].
     
\bibitem{Porto:2010tr}
  R.~A.~Porto,
  ``Next to leading order spin-orbit effects in the motion of inspiralling compact binaries,''
  Class.\ Quant.\ Grav.\  {\bf 27}, 205001 (2010)
  [arXiv:1005.5730 [gr-qc]].
  
\bibitem{Levi:2010zu}
  M.~Levi,
  ``Next-to-leading order gravitational spin-orbit coupling in an effective field theory approach,''
  Phys.\ Rev.\  D {\bf 82}, 104004 (2010) 
  [arXiv:1006.4139 [gr-qc]].            
  
\bibitem{Tagoshi:2000zg}
  H.~Tagoshi, A.~Ohashi and B.~J.~Owen,
  ``Gravitational field and equations of motion of spinning compact binaries to 2.5 post Newtonian order,''
  Phys.\ Rev.\  D {\bf 63}, 044006 (2001)
  [arXiv:gr-qc/0010014].
  
\bibitem{Faye:2006gx}
  G.~Faye, L.~Blanchet and A.~Buonanno,
  ``Higher-order spin effects in the dynamics of compact binaries. I: Equations of motion,''
  Phys.\ Rev.\  D {\bf 74}, 104033 (2006)
  [arXiv:gr-qc/0605139].
  
\bibitem{Blanchet:2006gy}
  L.~Blanchet, A.~Buonanno and G.~Faye,
  ``Higher-order spin effects in the dynamics of compact binaries II. Radiation field,''
  Phys.\ Rev.\  D {\bf 74}, 104034 (2006)
  [Erratum-ibid.\  D {\bf 75}, 049903 (2007), Erratum-ibid.\  D {\bf 81}, 089901 (2010).]
  [arXiv:gr-qc/0605140].
  
\bibitem{Damour:2007nc}
  T.~Damour, P.~Jaranowski and G.~Schafer,
  ``Hamiltonian of two spinning compact bodies with next-to-leading order
  gravitational spin-orbit coupling,''
  Phys.\ Rev.\  D {\bf 77}, 064032 (2008)
  [arXiv:0711.1048 [gr-qc]].

\bibitem{Steinhoff:2009ei}
  J.~Steinhoff and G.~Schafer,
  ``Canonical formulation of self-gravitating spinning-object systems,''
  Europhys.\ Lett.\  {\bf 87}, 50004 (2009)
  [arXiv:0907.1967 [gr-qc]].
      
\bibitem{Steinhoff:2009hx}
  J.~Steinhoff and H.~Wang,
  ``Canonical formulation of gravitating spinning objects at 3.5PN,''
  Phys.\ Rev.\  D {\bf 81}, 024022 (2010)
  [arXiv:0910.1008 [gr-qc]].
             
\bibitem{Hartung:2010jg}
  J.~Hartung and J.~Steinhoff,
  ``Next-to-leading order spin-orbit and spin(a)-spin(b) Hamiltonians for n
  gravitating spinning compact objects,''
  Phys.\ Rev.\  D {\bf 83}, 044008 (2011)
  [arXiv:1011.1179 [gr-qc]].
    
\bibitem{Hartung:2011te}
  J.~Hartung and J.~Steinhoff,
  ``Next-to-next-to-leading order post-Newtonian spin-orbit Hamiltonian for self-gravitating binaries,''
  Annalen Phys.\  {\bf 523}, 783 (2011)
  [arXiv:1104.3079 [gr-qc]].
    
\bibitem{Goldberger:2009qd}
  W.~D.~Goldberger and A.~Ross,
  ``Gravitational radiative corrections from effective field theory,''
  Phys.\ Rev.\  D {\bf 81}, 124015 (2010)
  [arXiv:0912.4254 [gr-qc]].
  
\bibitem{Porto:2010zg}
  R.~A.~Porto, A.~Ross and I.~Z.~Rothstein,
  ``Spin induced multipole moments for the gravitational wave flux from binary
  inspirals to third Post-Newtonian order,''
  JCAP {\bf 1103}, 009 (2011)
  [arXiv:1007.1312 [gr-qc]].
  
\bibitem{Blanchet:2011zv}
  L.~Blanchet, A.~Buonanno and G.~Faye,
  ``Tail-induced spin-orbit effect in the gravitational radiation of compact binaries,''
  Phys.\ Rev.\  D {\bf 84}, 064041 (2011)
  [arXiv:1104.5659 [gr-qc]].        
                     
\bibitem{Galley:2009px}
  C.~R.~Galley and M.~Tiglio,
  ``Radiation reaction and gravitational waves in the effective field theory approach,''
  Phys.\ Rev.\  D {\bf 79}, 124027 (2009)
  [arXiv:0903.1122 [gr-qc]].
  
\bibitem{Galley:2008ih}
  C.~R.~Galley and B.~L.~Hu,
  ``Self-force on extreme mass ratio inspirals via curved spacetime effective field theory,''
  Phys.\ Rev.\  D {\bf 79} (2009) 064002
  [arXiv:0801.0900 [gr-qc]].
  
\bibitem{Galley:2010xn} 
  C.~R.~Galley,
  ``A nonlinear scalar model of extreme mass ratio inspirals in effective field theory 
  I. Self force through third order,''
  Class.\ Quant.\ Grav.\  {\bf 29}, 015010 (2012)
  [arXiv:1012.4488 [gr-qc]].

\bibitem{Galley:2011te} 
  C.~R.~Galley,
  ``A Nonlinear scalar model of extreme mass ratio inspirals in effective field theory 
  II. Scalar perturbations and a master source,''
  Class.\ Quant.\ Grav.\  {\bf 29}, 015011 (2012)
  [arXiv:1107.0766 [gr-qc]].
    
\bibitem{Kol:2011pj}
  B.~Kol,
  ``Classical Effective Field Theory for Weak Ultra Relativistic Scattering,''
  JHEP {\bf 1107}, 062 (2011)
  [arXiv:1103.5741 [hep-th]].
  
\bibitem{Kol:2007bc}
  B.~Kol and M.~Smolkin,
  ``Non-Relativistic Gravitation: From Newton to Einstein and Back,''
  Class.\ Quant.\ Grav.\  {\bf 25}, 145011 (2008)
  [arXiv:0712.4116 [hep-th]].

\bibitem{Kol:2010si}
  B.~Kol and M.~Smolkin,
  ``Einstein's action and the harmonic gauge in terms of Newtonian fields,''
  arXiv:1009.1876 [hep-th].            
  
\bibitem{Kol:2010ze}
  B.~Kol, M.~Levi and M.~Smolkin,
  ``Comparing space+time decompositions in the post-Newtonian limit,''
  Class.\ Quant.\ Grav.\  {\bf 28}, 145021 (2011)
  [arXiv:1011.6024 [gr-qc]].                       
                
\bibitem{Hartung:2011ea}
  J.~Hartung and J.~Steinhoff,
  ``Next-to-next-to-leading order post-Newtonian spin(1)-spin(2) Hamiltonian
  for self-gravitating binaries,''
  Annalen Phys.\  {\bf 523}, 919 (2011)
  [arXiv:1107.4294 [gr-qc]].

\bibitem{Yee:1993ya}
  K.~Yee and M.~Bander,
  ``Equations Of Motion For Spinning Particles In External Electromagnetic And Gravitational Fields,''
  Phys.\ Rev.\  D {\bf 48}, 2797 (1993)
  [arXiv:hep-th/9302117].
      
\bibitem{Schafer:1984}  
  G.~Schafer, 
  ``Acceleration-dependent lagrangians in general relativity,''
  Phys.\ Lett.\ A {\bf 100}, 128 (1984).  
  
\bibitem{Damour:1985}
  T.~Damour and G.~Schafer,
  ``Lagrangians for n point masses at the second post-Newtonian approximation of general relativity,''
  Gen.\ Rel.\ Grav.\  {\bf 17}, 879 (1985).  

\bibitem{Damour:1990jh}
  T.~Damour and G.~Schafer,
  ``Redefinition of position variables and the reduction of higher order Lagrangians,''
  J.\ Math.\ Phys.\  {\bf 32}, 127 (1991).
  
\bibitem{Smirnov:2004ym}
  V.~A.~Smirnov,
  ``Evaluating Feynman integrals,''
  Springer Tracts Mod.\ Phys.\  {\bf 211}, 1-244 (2004).
                                                                 	                                             	  
\end{thebibliography}
\end{document}